\pdfoutput=1

\documentclass{aa}

\usepackage{graphicx}
\usepackage{txfonts}
\usepackage{longtable}
\newcommand{\ks}{K$_\mathrm{s}$}

\newcommand{\um}{$\mu$m}                                 
\newcommand{\lsun}{L$_{\odot}$}               
\newcommand{\msun}{M$_{\odot}$}

\newcommand{\lsim}{\;\lower.6ex\hbox{$\sim$}\kern-7.75pt\raise.65ex\hbox $<$\;}
\newcommand{\gsim}{\;\lower.6ex\hbox{$\sim$}\kern-7.75pt\raise.65ex\hbox $>$\;}

\newcommand{\pam}{.\hskip-2pt$^{\prime}$}
\newcommand{\pas}{.\hskip-2pt$^{\prime\prime}$}

\newcommand{\amin}{$^{\prime}$}                   
\newcommand{\asec}{$^{\prime \prime}$}

\begin{document}

   \title{Properties of Stellar Clusters around High-Mass Young
       Stars\thanks{Based on observations obtained at the Palomar Observatory and
       at the ESO La Silla Observatory (Chile), programme 65.I-0310(A)}}


   \author{F. Faustini \inst{1}, S. Molinari \inst{1}, L. Testi \inst{2,3} \and J. Brand \inst{4}}

   \offprints{faustini@ifsi-roma.inaf.it}

   \institute{Istituto di Fisica dello Spazio Interplanetario -
   INAF, via Fosso del Cavaliere 100, I-00133, Rome
              (\email{faustini, molinari@ifsi-roma.inaf.it})
            \and
            European Southern Observatory, Karl Schwarzschild str. 2, Garching bei Muenchen, Germany
            (\email{ltesti@eso.org})
            \and
             Osservatorio Astronomico di Arcetri - INAF, via Enrico Fermi 5, I-50125 Firenze
         \and
            Istituto di Radioastronomia - INAF, Via Gobetti 101, Bologna (\email{brand@ira.inaf.it})
             }

   \date{Received ; accepted}


  \abstract
{Twenty-six high-luminosity IRAS sources believed to be collection of stars 
in the early phases
of high-mass star formation have been observed in the Near-IR (J, H, \ks) to
characterize the clustering properties of their young stellar population and
compare them with those of more evolved objects (e.g., Herbig Ae/Be stars) of
comparable mass. All the observed sources possess strong continuum and/or line 
emission in the millimeter, being therefore associated with gas and dust
envelopes. Nine sources have far-IR colors characteristic of UCHII regions
while the other 17 are likely being experiencing an evolutionary phase that
precedes the Hot-Cores, as suggested by a variety of evidence collected in
the past decade.}
{To gain insight into the initial conditions of star formation in these
clusters (Initial Mass Function [IMF], Star Formation History [SFH]), and to
deduce mean values for cluster ages.}
{For each cluster we carry out aperture photometry.
We derive stellar density profiles, color-color
and color-magnitude diagrams, and color (HKCF) and luminosity (KLF) functions.
These two functions are compared with simulated KLFs and HKCFs from a
model that generates populations of synthetic clusters starting from
assumptions on the IMF, the SFH, and the Pre-MS evolution, and using the
average properties of the observed clusters as boundary conditions (bolometric
luminosity, dust distribution, infrared excess, extinction).}
{Twenty-two sources show evidence of clustering with a stellar richness indicator
that varies from a few up to several tens of objects, and a median cluster radius 
of 0.7~pc. A considerable number of cluster members present an infrared excess 
characteristic of young Pre-Main-Sequence objects. For a subset of 9 detected clusters, we could perform a statistically significant comparison of the observed KLFs with those resulting from synthetic 
cluster models; for these clusters we find that the median stellar age ranges  
between $2.5\cdot 10^5$ and $5\cdot 10^6$ years, with evidence of an age spread of the same entity within each cluster. We also find evidence 
that older clusters tend to be smaller in size, in line with the fact that our 
clusters are on average larger than those around relatively older Herbig Ae/Be stars.
Our models allow us to explore the relationship of the mass of the most massive
star in the cluster with both the clusters richness and their total stellar mass. Although such relationships are predicted by several classes of cluster formation models, their detailed analysis suggests that our modeled clusters may not be consistent with them resulting from random sampling of the IMF.}
{Our results are consistent with a star
formation which takes place continuously over a period of time which is longer than a typical crossing time. }


\maketitle
%

\section{Introduction}

The last few  decades have been characterized by a large effort to improve
our understanding of how stars form both from a theoretical and
from an observational point of view. As a result,
today we have reached a good understanding of how isolated low-mass stars
form (Klein at al. \cite{Klein}). The widely accepted scenario is
that low mass stars form through the gravitational collapse of a
prestellar core followed at later stages by disk accretion.

Extending this theory to high-mass stars is not trivial.
High mass (proto-)stars reach the Zero Age Main Sequence while still accreting.
When the central protostar reaches about 10~$M_\odot$ hydrogen
fusion ignites in the core and the star's radiation pressure and wind should
prevent further accretion. This obviously is a paradox given that more
massive stars do form.
Several theories have been put forward to solve this
dilemma (Zinnecker \& Yorke \cite{ZinYor}): accretion rates as high as 
three orders of magnitude larger than in the case of low-mass stars 
(Cesaroni~\cite{ces05}), and non-spherical accretion geometries 
(Nakano~\cite{TN89}, Yorke~\cite{Yorke}, Keto~\cite{keto}), or 
coalescence in dense (proto-)stellar clusters (Bonnell et al.~\cite{bonnell}).

All these theories have predictions that, in principle, could be tested
observationally. In the last decade a large effort was made in trying to
detect massive accretion disks (Cesaroni et al.~\cite{Ces06}), powerful
outflows (Beuther et al. \cite{beu02}, Cesaroni et al.~\cite{cesaroni})
and dense protostellar clusters (Testi et al.~\cite{testi3}, de Wit et
al.~\cite{dewit}),
all of these phenomena are predicted by one or the other formation theory.
None of these efforts have provided conclusive arguments in favour or against
any of the theories.

In this paper we explore the properties of embedded clusters associated with
high-mass protostellar candidates. Our sample was selected from a larger
sample of candidate high-mass protostars selected and analyzed, in the past
decade, by Molinari et al. (\cite{Metal96}, \cite{Metal98}, \cite{Metal00},
\cite{Metal02}) and Brand et al. (\cite{Brandal01}). In Sect.~\ref{obs} we 
present the observations and data analysis
(source extraction, photometry), in Sect.~\ref{results} we discuss data
elaboration and interpretation. In Sect.~\ref{IMFSFH} we present our Synthetic 
Cluster Generation model and the method of comparison between synthetic and 
observed clusters and the results of using
this technique. Finally in Sect.~\ref{disc} we compare our
objects with more evolved ones and present our conclusions.


\section{Observations and Data Analysis}
\label{obs}

Program fields are listed in Table \ref{journal} and were imaged in
J, H, and \ks\ bands. A total of 15 fields were observed in three nights in
November 1998 at the Palomar 60-inch telescope equipped with a
256$\times$256 NICMOS-3 array, with a pixel scale of 0\pas62/pix and
total FOV of 2\pam6$\times$2\pam6. The remaining 11 fields were
observed in 3 nights in August 2000 at the ESO-NTT using the
1024$\times$1024 SOFI camera, with a pixel scale of 0\pas29/pix and a
total FOV of 4\pam9$\times$4\pam9. Standard dithering techniques were
used to minimize impact of bad pixels and optimize flat-fielding,
allowing to achieve for each field a total of 5min integration time per 
band (in the central portion of the observed field) for a covered area 
of 3\pam5$\times$3\pam5 for Palomar observations, and 20min (10min for 
the band \ks\ ) at NTT with a total covered area of 6\pam5$\times$6\pam5.
Suitable calibration sources from the list of Hunt
et al. (\cite{Hetal98}) were observed regularly during the
observations to track atmospheric variations for different
airmasses. Standard stars and target fields were observed at airmasses no
greater than 1.7 at NTT, and 1.3 at Palomar; we determined average zero-point
magnitudes for each night and used them to calibrate our photometry. For each
field the images in the three bands were registered and astrometric solutions
were determined using a few bright optically visible sources.

\begin{table}[ht]
\setlength{\tabcolsep}{0.085in}
\caption{Journal of Observations} 
\label{journal}
\begin{tabular}{lcccc} \hline\hline
Source & IRAS Name & $\alpha$(J2000) & $\delta$(J2000) & Tel. \\
Mol$^a$ &  &  &  &  \\
\hline
3   & 00420+5530  & 00:44:57.6 & +55:46:52 &  Pal \\
8   & 05137+3919 & 05:17:13.3 & +39:22:14 & Pal \\
9   & 05168+3634 & 05:20:16.2 & +36:37:21 &  Pal \\
11 & 05345+3157  & 05:37:47.8 & +31:59:24 &  Pal \\
12 & 05373+2349  & 05:40:24.4 & +23:50:54 &  NTT \\
15 & 06056+2131 & 06:08:41.0 & +21:31:01 &  Pal \\
28 & 06584$-$0852 & 07:00:51.0 & $-$08:56:29 & Pal \\
30 & 17450$-$1742 & 17:48:09.3 & $-$27:43:21 & NTT \\
38 & 18024$-$2119 & 18:06:18.0 & $-$21:42:00 & NTT \\
45 & 18144$-$1723 & 18:17:24.2 & $-$17:22:13 & NTT \\
50 & 18162$-$1612 & 18:19:07.5 & $-$16:11:21 & NTT \\
59 & 18278$-$1009 & 18:30:35.2 & $-$10:07:12 & Pal \\
75 & 18511+0146 & 18:53:38.1 & +01:50:27 & Pal \\
82 & 18565+0349 & 18:59:03.4 & +03:53:22 & NTT \\
84 & 18567+0700 & 18:59:13.6 & +07:04:47 & NTT \\
98 & 19092+0841 & 19:11:37.4 & +08:46:30.0 & NTT \\
99 & 19094+0944 & 19:11:52.0 & +09:49:46 & Pal \\
103 & 19213+1723 & 19:23:37.0 & +17:28:59 & NTT \\
109 & 19374+2352 & 19:39:33.2 & +23:59:55 & NTT$^b$ \\
110 & 19388+2357 & 19:40:59.4 & +24:04:39 & NTT$^b$ \\
136 & 21307+5049  & 21:32:31.5 & +51:02:22.0 & Pal \\
139 & 21519+5613 & 21:53:38.8 & +56:27:53.0 & Pal \\
143 & 22172+5549  & 22:19:09.0 & +56:04:45.0 & Pal \\
148 & 22305+5803 & 22:32:24.3 & +58:18:58.2 & Pal \\
151 & 22506+5944 & 22:52:38.6 & +60:00:56.0 & Pal \\
160 & 23385+6053  & 23:40:53.3 & +61:10:19.1 & Pal \\
\hline
\end{tabular}
\parbox{8.5cm}{$^a$ Source running number from Molinari et al. \cite{Metal96}.}\\
\parbox{8.5cm}{$^b$ Imaged only in \ks.}\\
\end{table}

The K$_s$ images for all observed fields, with superimposed submillimeter continuum emission distribution when available (Molinari et al. \cite{Moli08a}) are presented in Appendix \ref{app_images} and available online at http://galatea.ifsi-roma.inaf.it/faustini/K-images.


\subsection{Point Source Extraction and Photometry}
\label{phot}

The extraction and photometry of point sources for all images was
carried using the IRAF package. The r.m.s  of the background signal
and the FWHM of point sources were measured throughout the images to
characterize the image noise and PSF properties; these parameters
were fed to the DAOFIND task for source extraction, where a
detection threshold of 3$\sigma$ was used for all images.
Sources with saturated pixels were excluded from the analysis; the linearity
of the system response was checked {\it a posteriori} comparing, both for the
Palomar and the NTT data, the magnitudes obtained to those from 2MASS using a
few stars with magnitudes reaching up to the maximum values found in our
photometry files; the relations between the 2MASS magnitudes and ours in
the three bands were found to be linear over the entire magnitude range of the
detected sources. There were clearly brighter objects in the various fields,
but their peaks were already flagged as saturated and were excluded from the
detection process.

The photometry of sources is critical in very dense stellar fields
like the inner Galactic Plane, where all of our target fields lie, and where
the crowding is such that more than one source can enter any
plausible aperture that can be chosen, or any annulus used for
background estimation. This problem is of course magnified in the
clustered environments that are detected in sites of massive star
formation (see \S\ref{resclust} below).

The first approach we tried
to follow is the PSF-fitting photometry which should be
less prone to these problems. We chose a sub-sample of test fields with 
different levels of stellar crowding. An important aspect
in this procedure is the
modelling of the PSF; we made several trials selecting a variable
number of point-like sources (from 3 to 30) of different brightness
levels and different position in the field. We find that the
resulting PSF model is not particularly sensitive to the choice of
numbers and/or brightness of the stars; on the other hand the
results are quite different depending on the mean stellar density of
the field. The photometry was carried out using the ALLSTAR task, particularly
suited for crowded fields. However, we also tested the other two tasks (PEAK
and NSTAR) and they produce comparable results for most of the sources.
We note, however, that especially in the most crowded areas the subtraction of
the PSF-fitted sources from the image introduces two spurious effects: an
unacceptably high level of residuals with brightness levels well above the
detection threshold used and a great number of negative holes which show that
the psf-fit includes a portion of background in the extimation of the source
flux, therefore overestimating its value.
This is due both to the limited accuracy of the PSF model that
can be built in very crowded fields, where faint neighboring stars can
enter in the area where the PSF model is estimated, and to the presence of a
significant and variable background which is quite common and expected in the
Galactic Plane. A similar
conclusion was reached by Hillenbrand \& Carpenter (\cite{HC00}) in
their study of the inner Orion Nebula Cluster.

The second approach we followed is standard aperture photometry. The
choice of the radii for the aperture and for the background annuli
is of course extremely important. The optimum aperture should not be too 
large to include nearby sources and not too small to significantly cut 
the PSF and severely underestimate the flux. We did
several attempts on one of the most crowded fields (Mol30, observed at
NTT) with three different aperture radii equal to the PSF FWHM (typically 
0\pas7 at NTT and 1\pas4 at Palomar in K$_s$),
twice and thrice this value. For each photometry run we analyze the
sources' flux distribution and, as expected, the median flux is found
to increase with increasing aperture radius. Increasing the aperture
from one to two PSF FWHMs raises the median source flux by an amount
compatible with the inclusion of the first ring of an Airy
diffraction pattern; instead, an aperture radius equal to three
times the PSF FWHM produces a flux increase much too large compared
to the additional fraction of the Airy profile entering the aperture,
therefore being caused by the inclusion of nearby sources. We adopted
an aperture radius equal to the PSF FWHM to
minimize neighbour contamination, and then applied an aperture
correction factor for the fraction of the PSF cut out of the
aperture; this was estimated via multi-aperture photometry (starting from a
1 FWHM size) on relatively isolated stars in the target fields.

A further effect to be corrected for, also given the crowding of our fields, 
is the possible contamination arising from the tails of the brightness
profiles of neighbouring stars. To quantify this contamination we created a
grid of simulations with two symmetric Gaussians with a wide
variety of peak contrast and at different reciprocal distances.
We computed the fraction of the Gaussian profile of the neighbouring source
that falls into the photometry aperture centered on the main source,
and hence generated a matrix of photometry corrections for different
source distances and peak contrasts. At this point we run throught
the magnitude file produced by the aperture photometry task and for
each source we apply a magnitude correction depending on the
presence, distance and contrast ratios with other neighbouring
stars.

In spite of the various issues discussed above, the photometry obtained with
the two methods are in a good agreement with each other, except at faint
magnitudes.
For these faint objects we always find that the PSF photometry tends to
produce lower magnitudes (hence stronger sources) than the aperture
photometry; this effect is easily understood given our finding (see above)
that the subtraction of PSF-fitted sources always leaves negative holes in the
residual image, and this effect is much more important for faint stars. We
thus decided to adopt the magnitudes determined from aperture photometry.

For each target field we estimated the limiting magnitude (LM) using
artificial star experiments. The fields were populated using the IRAF task
ADDSTAR with 400 fake stars with magnitudes distributed in bins of 0.25~mag
between values of 15 and 21; the percentage of recovered stars as a function
of magnitude gives an estimate of the completeness level of our photometry.
The star recovery percentage does not decrease monotonically with increasing
magnitude because fake stars can be placed also very close to bright real
stars and then go undetected by the finding algorithm. However, we find that
the limit of 85-90\% recovery fraction is reached on average around J=18.7,
H=17.7 and K$_s$=17.4 for NTT images, and J=18.0, H=17.3 and K$_s$=16.6 for
Palomar images. We find that the typical photometric uncertainty is below 0.1mag
close to the limiting magnitude.

To verify the integrity of our photometry we compared our magnitudes with
those extracted from 2MASS point source catalog for all the fields of our
sample. Considering the different spatial resolutions between 2MASS and the
telescopes used for our observations, this comparison was limited to 
those 2MASS point-like sources associated with a single source in
the Palomar or NTT images. The median differences with respect to 2MASS for
the various fields are of the order of $-$0.1, $-$0.2 and $-$0.3~mag for J, H,
and K$_s$ bands, respectively. Within each field, the scatter around these
median values is $\sim$ 0.1~mag in all three bands, confirming the
internal
consistency of our photometry. Noticeable departures ($\sim$0.5~mag) of the
median difference with 2MASS from the above values are observed for the field
of source Mol11 (Palomar), and for sources Mol103, Mol109 and Mol110 (NTT).
However, the latter sources were observed on the same night, which our log
registered as not good due to sky variations which are not tracked by
night-averaged zero-points. We stress again, however, that these are
systematic differences with respect to 2MASS in this limited number of cases;
the r.m.s. scatter around these median differences are $\sim$ 0.1~mag in all
bands and this should give confidence that the internal consistency of the
photometry in each field is preserved. We then decided to rescale our
photometry to the 2MASS photometric system to remove these systematic effects.
The (J-H) and (H-K) color differences between 2MASS and our photometry are
not correlated with the magnitude, so that no magnitude-dependent color effect
is introduced in this rescaling.

\section{Results}
\label{results}

\subsection{Cluster Identification}
\label{resclust}

The identification of a cluster results from the analysis of stellar
density in the field. Since our target fields are sites of massive
star formation associated with local peaks of dust column densities
and hence of visual extinction, the \ks\ images are clearly more
suited for this type of analysis. 

Stellar density maps were built for each field counting stars in a running 
boxcar of size equal to 20\asec. The box size was determined empirically to enhance 
the statistical significance of local stellar density peaks and to maximize the 
ability to detect the clusters. Larger boxes tend to smear the cluster into the 
background stellar density field decreasing the statistical significance of the 
peak, which may lead to non-detection of a clearly evident cluster, particularly 
in the rich inner Galaxy fields (this happens, e.g., for source Mol103, see 
Fig.~\ref{map_mol103} in the Appendix). Smaller boxes produce noisy density 
maps  where the number of sources in each bin starts to be comparable to the 
fluctuations of the background density field due either to intrinsic variations 
of the field star density or to variable extinction from diffuse foreground ISM 
in the Galactic Plane (where all of our sources are located). For most of our 
objects in the outer Galaxy this analysis is just used to locate the position 
of the peak stellar density, since the clusters are obvious already from the 
visual inspection (Mol3 to Mol28, and Mol143 to Mol151, see Appendix). For 
the rest of the fields the density maps are used to ascertain the presence 
of a cluster; especially toward the inner Galaxy the density maps tend to 
show more than one peak at comparable levels. It is important to remember, 
however, that this is a search for stellar clusters toward regions where 
indications of active star formation are already available, and this 
information can be used. In particular, the coincidence of these peaks 
with cold dust clumps as traced by intense submillimeter and millimeter 
emission (Beltr\'an et al.~\cite{beltran06}, Molinari et al.~\cite{Moli08a}) 
is critical to consider the density peak as a real feature associated with 
the star formation region. Casual association is ruled out by the high number 
of positive associations (see Table~\ref{clusters}).

\begin{figure}
  \centering
  \resizebox{\hsize}{!}{\includegraphics[angle=90]{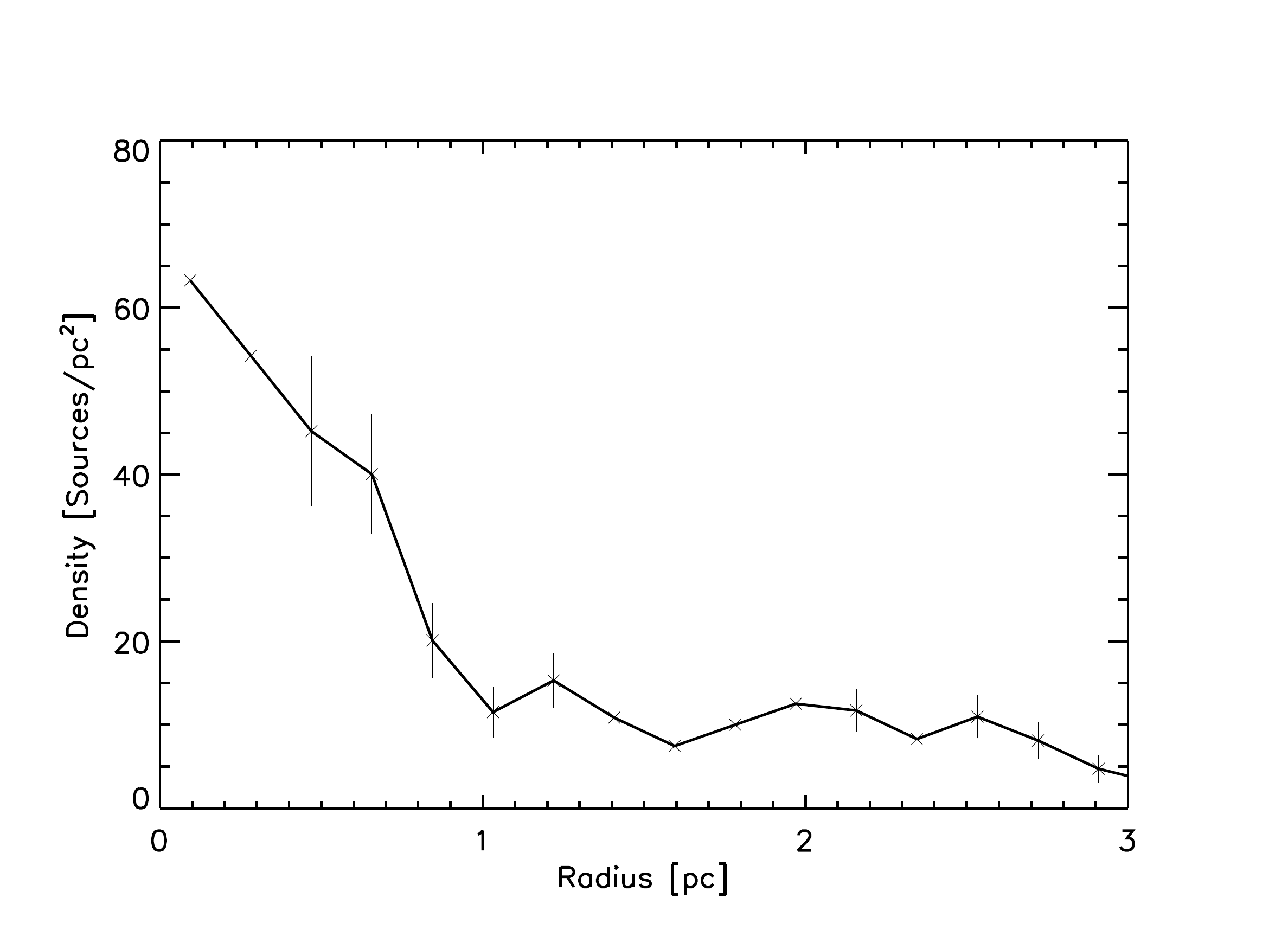}}
  \caption{Stellar density (in stars/pc$^2$), for Mol28, as a function of the radial
distance (in parsecs) from cluster center. Error-bars are computed as the
Poissonian fluctuations of source counts in each bin.}
  \label{dens}
\end{figure}

As a further confirmation step for the positive detection of a cluster 
we build radial stellar density profiles where stars are counted in annuli of 
increasing internal radius and constant width and then ratioed to the area of 
the annuli (Testi, Palla \& Natta~\cite{testi2}); uncertainties are assigned 
assuming Poisson statistics on the number of stars in each annulus. We then 
assign a positive cluster identification if the radial profile exhibits at 
least two annuli with values above the background. In order to refine the 
location of the density peak, we repeat the radial density profile analysis 
starting from several locations within 10\asec\ of the peak derived from the 
density maps; the location which maximizes the overall statistical significance 
of the annuli is then assigned as the cluster center. Figure \ref{dens} shows 
the typical footprint of a cluster, where
the stellar density is plotted as a function of distance \textit{r}
from the start location; the density has a maximum at \textit{r}=0
and decreases until it reaches a constant value which is the average
background/foreground stellar density.

There are two exceptions in this analysis. The first is for source 
Mol160. The K$_s$-band image shows clear stellar density enhancement in a 
semi-circular annulus which surrounds the northern side of the dense millimeter 
core (see Figure~\ref{map_mol160} in the Appendix), which appears devoid of stars. 
This stellar density enhancement is coincident with the emission patterns visible 
in the mid-IR (Molinari et al.~\cite{Moli08b}) so it is clearly a stellar 
population associated with the star forming region. Since the millimeter peak 
is at the center of symmetry of the semi-circular stellar distribution, we will 
consider this as the center of the cluster. This is just for completeness of 
reporting, since we cannot say if the low density of stars at the millimeter 
peak is an effect of extreme visual extinction or reflects an intrinsic paucity 
of NIR-visible forming stars, as the proposed extreme youth of the massive YSO 
accreting in its depth would seem to suggest (Molinari et al.~\cite{Moli08b}).

The second exception is for source Mol8. The stellar density analysis 
shows two peaks which are coincident with two distinct dust cores (see 
Fig.~\ref{map_mol008}); we then assume the presence of two distinct clusters, 
rather than a sub-clustering feature within the same cluster. The radial density 
profile analysis cannot be used here, so we fit elliptical gaussians to the peaks 
in the density maps, allowing for an underlying constant level representing the 
background stellar density. The resulting cluster richness is obtained integrating 
the fitted gaussian, and the cluster radius is taken equal to the fitted FWHM 
(the fitted gaussians are nearly circular).

Always following Testi et al. (\cite{testi2}), we
determine the richness indicator of the cluster I$_c$ by integrating the
background-subtracted density profile; the cluster radius is taken as the 
radial distance from start location where the density profile reaches a constant 
value. This richness indicator is a very convenient figure to use in these cases 
where no detailed information is available for each single star in the region 
and the membership of the cluster cannot then be established for each single star.  
These values are reported in Col.~3 of Table~\ref{clusters} for all fields
where a cluster has been clearly revealed. Col.~1 gives the target name (cf. 
Table~\ref{journal}); it's kinematic distance is listed in Col.~2. The parameter
N$_{obs}$ (Col.~4) is the number of cluster members derived (see
Sect.~\ref{klf} below) from the integration of the
background-subtracted \ks\ Luminosity Function (hereafter KLF, see Sect.~\ref{klf}). 
Also reported in Col.~8 is the
mass of the hosting molecular clump; this is derived from the cold dust emission 
as reported in Molinari et al.~\cite{Moli08a}, \cite{Metal00}), integrated over 
the entire spatial extent of the cluster; conversion into masses is done under optically thin assumption 
assuming T=30~K, $\beta=1.5$ (Molinari et al.~\cite{Moli08a}) and a mass opacity 
$\kappa_{230GHz}=0.005 cm^2 g^{-1}$ which incorporates a gas/dust weight ratio of 100 (Preibisch et al.~\cite{P93}). The IRAS source 
bolometric luminosity, Col.~9, is taken from Molinari et al.
(\cite{Metal96}, \cite{Metal00}, \cite{Metal02}, \cite{Moli08a}); in Col.~10 
we list the A$_V$ on the peak cluster position as estimated from submm
observations (Molinari et al.~\cite{Moli08a}, \cite{Metal00}). In 
Col.~11 and Col.~12 the coordinates of centers of the identified clusters 
are reported. Columns 6 and 7 contain parameters that will be described later 
in the text (see Sect.~\ref{clust_nature}).

\begin{table*}
\setlength{\tabcolsep}{0.12in}
\caption{Results for Cluster Detection}
\label{clusters}
\begin{tabular}{lccccccccccc} \\ \hline\hline
 Sou. & d$^a$ & I$_c$ & N$_{obs}$ & R$_{clu}$ & \multicolumn{2}{c}{Pre-MS} & M$_{gas}$ & L$_{bol}$ & A$_V$ $_{peak}$ & \multicolumn{2}{c}{Cluster Center} \\
 Mol & kpc & & & pc & \% & \% & \msun & 10$^3$\lsun & mag & $\alpha$(J2000) & $\delta$(J2000) \\
 &  & & &  & \textit{CC} & \textit{CM} &  &  & \\ \hline
3    & 2.17 & 78   & 78   & 1.7 & 34 & 99 & 910 & 12.4 & 18 &	00:44:57.4	&	+55:47:20.0 \\
8A    & 11.5 & 25  & 30   & 1.3 & 37 & 27 & 1650 & 57.0 & 18 &	05:17:13.8	&	+39:22:29.7 \\
8B    & 11.5 & 27  & 24   & 1.3 & 9 & 4 & 1780 & 5.5 & 8 &	05:17:12.0	&	+39:21:51.8 \\
9    & 6.2 & 7   & 7   & 0.6 & 16 & $-^g$ & $-$ & 24 & $-$ &	05:20:16.9	&	+36:37:22.0 \\
11   & 2.1 & 51   & 48   & 0.5 & 12 & 41 & 360 & 4.6 & 35 &	05:37:47.7	&	+31:59:24.0 \\
12   & 1.6 & 12   & 13   & 0.3 & 0    & 30 & 72 & 1.6 & 46 &	05:40:24.4	&	+23:51:54.8 \\
15  & 1.5 & 64   & 61   & 0.3 & 34  & $-^g$ & $-$ & 5.8 & $-$ &	06:08:41.0	&	+21:31:00.0 \\
28   & 4.5 & 75   & 75   & 1.0 & 58 & 95  & 220 & 9.1 & 4 &	07:00:51.5	&	$-$08:56:18.2 \\
30   & 0.3 &  200$^e$ & \multicolumn{4}{c}{no cluster detected$^b$} & - & 0.14 & $-$ & $-$ & $-$ \\
38   & 0.5 & $-$5 & \multicolumn{4}{c}{no cluster detected$^c$} & -$^h$ & 0.19 & 40 & $-$ & $-$ \\
45   & 4.3 & 28   & 27   & 0.5 & 68 & 19   & 1340 & 21.2 & 77 &	18:17:24.1	&	$-$17:22:12.3 \\
50   & 4.9 & 46   & 43 & 0.6 & 36 & 41 & 80 & 17.3 & 25 & 18:19:07.6 & $-$16:11:21.0 \\
59   & 5.7 & 0 & \multicolumn{4}{c}{no cluster detected$^c$} & -$^h$ & 11 & 29 & $-$ & $-$ \\
75   & 3.9 & 8     & 7     & 0.7 & 37 & 24 & 1310 & 13.3 & 38 &	18:53:38.1	&	+01:50:26.5 \\
82   & 6.8 & 8 & 10 & 0.3 & 0 & 52 & 590 & 15.4 & 52 &	18:59:03.2	&	+03:53:16.7 \\
84   & 2.2 & 21   & 21   & 0.3 & 0 & 54 & 28 & 4.3 & 15 &	18:59:14.3	&	+07:04:52.3 \\
98   & 4.5 & 6 & \multicolumn{4}{c}{no cluster detected$^f$} & -$^h$ & 9.2 & 68 &	$-$ & $-$  \\
99  & 6.1 & 45   & 38   & 0.9 & 49 & $-^g$ & $-$ & 37.3 & $-$ &	19:11:51.4	&	+09:49:35.4 \\
103 & 4.1 & 105 & 107 & 0.7 & 46 & 87 & 510 & 28.2 & 42 &	19:23:36.2	&	+17:28:58.1 \\
109 & 4.3 & 19   & 17   & 0.5 & $^d$ & $^d$ & 1030 & 26.7 & 90 &	19:39:33.0	&	+24:00:21.3 \\
110 & 4.3 & 23   & 20   & 0.5 & $^d$ & $^d$ & 400 & 14.8 & 55 &	19:40:58.5	&	+24:04:36.3 \\
136  & 3.6 & 21   & 19   & 0.6 & 18 & 52 & 230 & 4 & 21 &	21:32:31.4	&	+51:02:23.1 \\
139 & 7.3 & 25   & 24   & 1.2 & 10 & 31 & 1870 & 1.35 & 20 &	21:53:39.2	&	+56:27:50.7 \\
143  & 5.0 & 25   & 22   & 0.8 & 10 & 76 & 630 & 7.8 & 26 &	22:19:09.0	&	+56:04:58.7 \\
148 & 5.1 & 43 & 41 & 0.9 & 43 & 64 & 22 & 7.8 & 13 &	22:32:23.4	&	+58:19:01.3 \\
151 & 5.4 & 15   & 14   & 0.9 & 12 & 30 & 2020 & 25 & 40 &	22:52:38.3	&	+60:00:44.6 \\
160  & 5.0 & 36   & 34   & 1.3 & 30 & 76 & 1830 & 16 & 32 &	23:40:53.1	&	+61:10:21.0 \\ \hline
\end{tabular}\\
\parbox{16cm}{$^a$ Kinetic distance using the rotation curve from Brand \& Blitz \cite{BrBl}.}\\
\parbox{16cm}{$^b$ Stellar density analysis inconclusive due to extreme
crowdedness of this field.}\\
\parbox{16cm}{$^c$ Stellar density reveales no peaks close to the IRAS
position or the submm peak.}\\
\parbox{16cm}{$^d$ Only observed in \ks.}\\
\parbox{16cm}{$^e$ Detection refused due to extreme field complexity (see text).}\\
\parbox{16cm}{$^f$ Detection refused because only 1 annulus in the radial density profile is above background (see text).}\\
\parbox{16cm}{$^g$ No extinction estimate is available due to lack of submm information to evaluate de-reddening correction.}\\
\parbox{16cm}{$^h$ Extinction estimate is available from single-pointing submillimeter data (Molinari et al. \cite{Metal00}) but not from maps, so that a reliable clump mass estimate is not possible.}

\setlength{\tabcolsep}{0.07in}
\end{table*}

Following the described procedure, a cluster is detected within 1\amin\ of the IRAS position for 22
out of the 26 observed fields (85\% detection rate). In two
cases (Mol38 and Mol59) the stellar density map does not show a clear peak above the fluctuations of 
the field stellar density. For Mol 98 the radial density profile only shows one annulus above the 
background, so they fail the criterion that the stellar density enhancement should be 
significantly resolved above the background in two annuli. In one case (Mol30) several stellar density 
peaks are found in proximity of the IRAS source, but the lack of information of 
submillimeter/millimeter continuum prevents any firm conclusion.

\begin{figure}[ht]
  \centering
  \resizebox{\hsize}{!}{\includegraphics[angle=90]{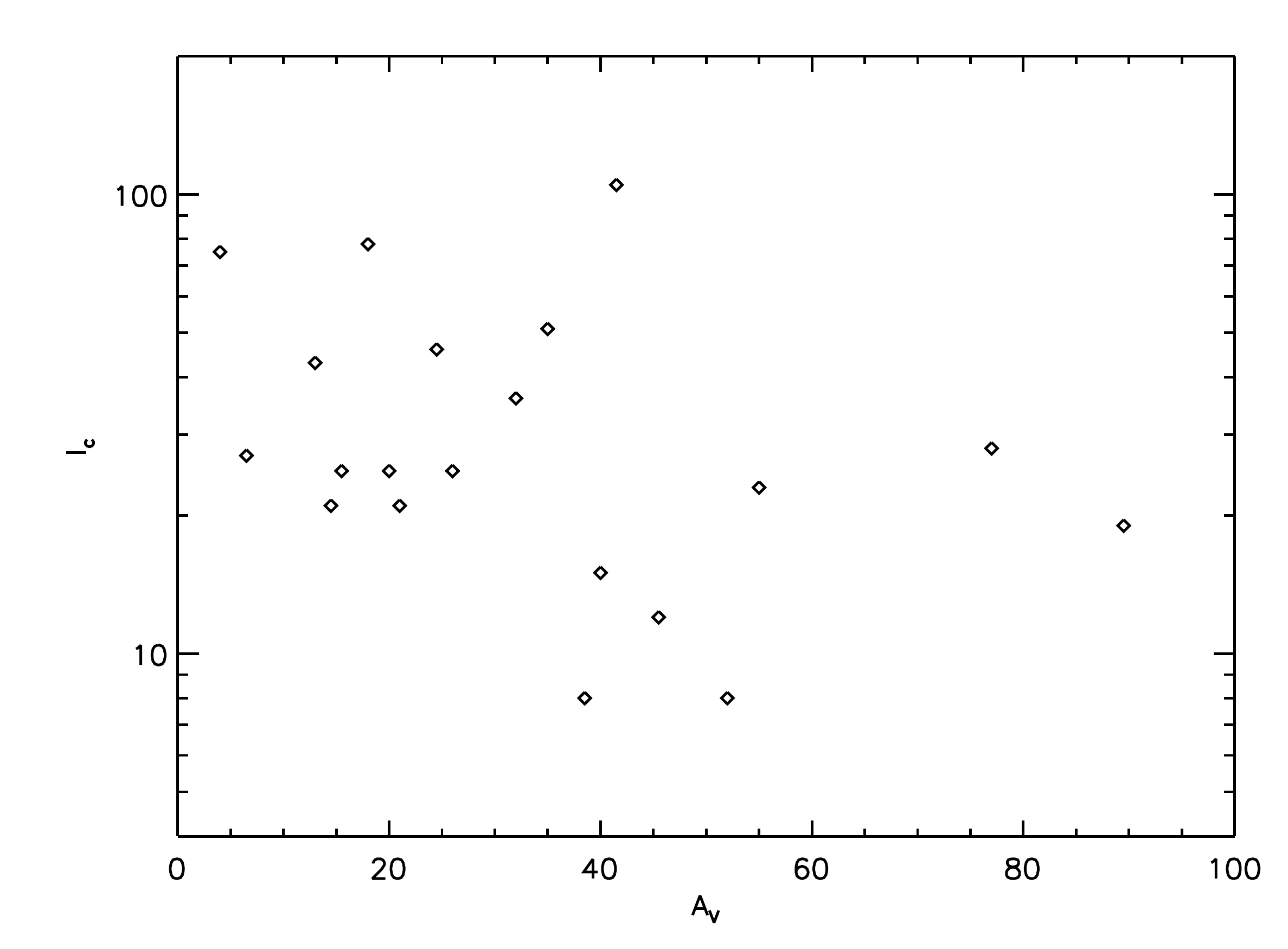}}
  \caption{Cluster richness indicator I$_c$ as a function of A$_V$ on cluster
center; for a few detected cluster we do not have an estimate of A$_V$).}
  \label{ic_av}
\end{figure}

Figure~\ref{ic_av} shows the run of I$_c$ as a function of peak A$_V$ and suggests that with larger 
dust extinction it may be more difficult, or less likely, to detect a cluster at 2.2~\um. 

Our detection rate is quite high and tells us that young stellar clusters
in sites of intermediate and massive star formation are essentially ubiquitous. While this
evidence was established for relatively old Pre-MS systems like Herbig Ae/Be
stars (Testi et al.~\cite{testi3}), we hereby verify that this is the case
also in much younger systems where the most massive stars may even be in a
pre-Hot Core stage (Molinari et al. \cite{Moli08a}).

Our detection rate is higher compared to other similar searches of
stellar clusters toward high-mass YSOs. For example Kumar et al.
(\cite{Kumar06}) use the 2MASS archive and report a rate of 
25\% (rising to 60\% neglecting the inner Galaxy regions) toward a larger sample
which also includes the sources of this work; in particular we detect all
clusters also detected by Kumar et al. and in addition we reveal clusters
toward 13 objects for which Kumar et al. report no detection. The
reason for this discrepancy may be due to the fact that we obtained dedicated
observations while Kumar et al. used data from the 2MASS archive; the
diffraction-limited spatial resolution of our data is between a factor of 4
and a factor of 10 better with respect to 2MASS, and this certainly
facilitates cluster detection especially in particularly crowded areas like
the inner Galactic plane. To test this hypothesis we degraded the NTT K$_s$ 
image of Mol103, also considered in Kumar et al., to the 2MASS resolution; 
extraction and photometry were performed as outlined above but the search 
for a cluster based on the stellar radial density profiles revealed no cluster. 
Besides, the estimated number of  members (corrected for the contribution of
fore/background stars) for 7 out of the 10 clusters detected both by us and by
Kumar et al. is at least a factor of two less in the latter study. 

Kumar \& Grave (\cite{Kumar08}) conducted a similar study on a large 
sample of high-mass YSOs, that include a certain number of our sources, using 
this time data from the GLIMPSE survey (Benjamin et al. \cite{benjamin}). In 
this work they detect no significant cluster around any targets in a sample 
of 509 objects. As the authors say in the paper, however, GLIMPSE data are 
sensitive to 2-4 \msun\ pre-main sequence stars at the distance of 3 kpc. 
Based on color-magnitude analysis (see later below) our mass sensitivity 
is of the order of 1 \msun\ at a distance of 3.6 kpc and $\sim$0.6 \msun\ 
at a distance of 2.1 Kpc. Probing longer wavelengths, GLIMPSE is likely to 
be more sensitive to younger sources compared to the classical J, H, K 
range which also samples relatively older pre-MS objects. The combination 
of sampling higher-mass (and hence more rare stars due to the shape of the 
IMF) and relatively younger stars (which, as indeed our analysis finds, 
may not be the majority in a young cluster) may plausibly be the reason 
of the negative cluster detection results in Kumar \& Grave.

\begin{figure}[ht]
  \centering
  \resizebox{\hsize}{!}{\includegraphics[angle=90]{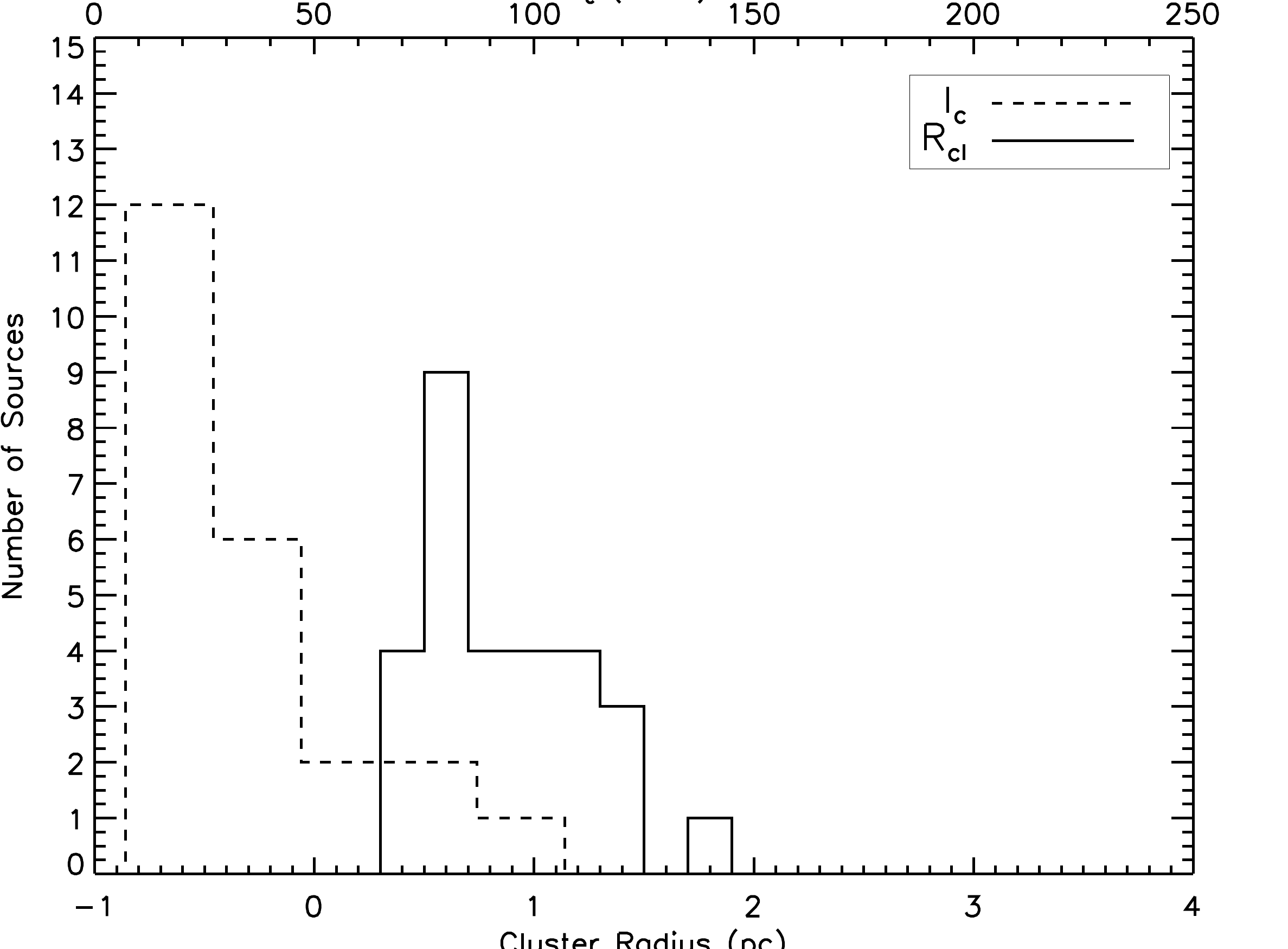}}
  \caption{Distribution of the cluster's radii in parsecs (full line), and of
the cluster richness indicator I$_c$ in number of stars (dashed lines); the
median values for the two distributions are 0.6~pc and 37 stars, respectively}
  \label{rclu_ic_hist}
\end{figure}

The distribution of the radii of the detected clusters is shown with the full
line in Fig.~\ref{rclu_ic_hist}; the median value is 0.7~pc. 
The dashed histogram (which refers to the
upper X-axis) shows the distribution of the cluster richness indicator I$_c$,
with a median number of stars of 27. We note that the value of I$_c$ 
for many of our clusters is less than the limit of 35 suggested by Lada \& Lada 
(\cite{lala}) to be a \textit{bona fide} cluster. This definition stems from the 
argument that a less rich agglomerate may not survive the formation process as an 
entity. Our interest, however, is to investigate the spatial properties of the 
young stellar population in a star forming region at the time of active formation, 
without worrying about its possible persistence as a cluster at the end of the 
formation phase. However, we prefer not to introduce a new term to identify 
the structures we see and still use the term cluster, although in a \asec 
weaker\asec\ sense compared to Lada \& Lada.

\subsection{Properties of Identified Clusters}
\label{clust_nature}

We will first derive qualitative indications concerning the nature of the identified
clusters using simple diagnostic tools like color-color and color-magnitude
diagrams. These diagrams have been drawn for all detected clusters and are
available in electronic form; we here illustrate the particular case for
Mol28.

\subsubsection{Color-Color Analysis}
\label{cc_anal}

\begin{figure}[ht]
  \centering
  \resizebox{\hsize}{!}{\includegraphics[angle=90]{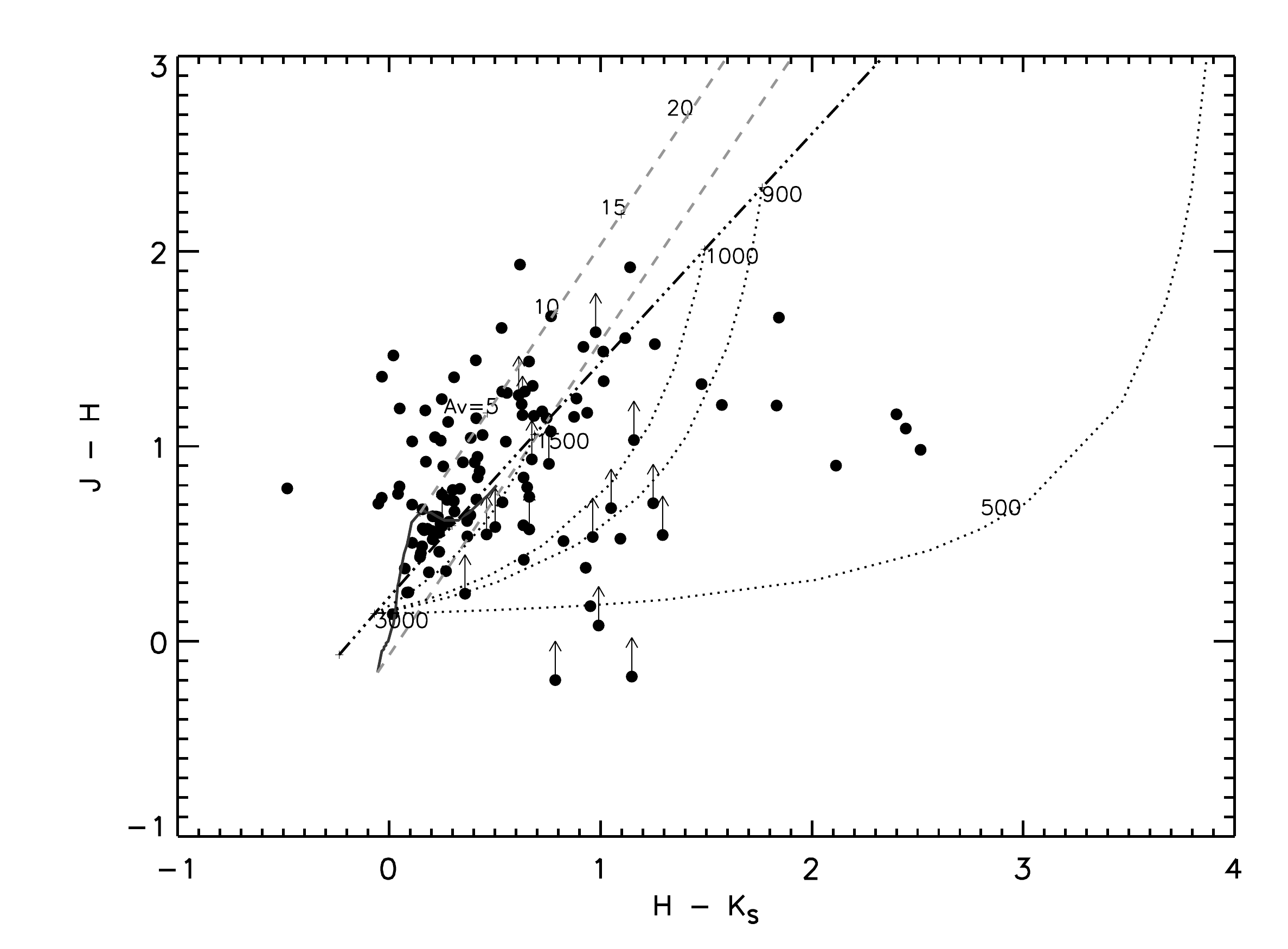}}
  \caption{[J-H]\textit{vs}[H-\ks] diagram for Mol28. [J-H] is obviously a
lower limit for sources not detected in J. The continuous curve at the bottom-left 
represents the Main Sequence, while the dashed grey lines represent the effect of reddening (Rieke \& Lebofsky \cite{RL85}) for variable amounts of extinction as indicted along the lines. The dashed-dotted 
black line is the
Black-Body curve, and the dotted curves are two-component Black-Body with
varying relative contribution (respectively, from the inner to the outer
curve, 3000-1500K, 3000-1000K, 3000-900K and  3000-500K). }
  \label{colcol}
\end{figure}

Fig. \ref{colcol} shows the [J-H]\textit{vs}[H-\ks] diagram for all sources
detected within a distance equal to R$_{cl}$ centered on the stellar density 
peak. The full circles represent all sources detected in all three bands,
the arrows represent sources with lower limits (in magnitude) in the J band.
The plot shows more stars than the I$_c$ value reported in
Table~\ref{clusters} because we also include the fore/background stars that
cannot be individually identified against the true cluster members. A
significant fraction of the sources have colors compatible with main-sequence
stars with a variable amount of extinction reddening (computed adopting the
Rieke \& Lebofsky (\cite{RL85}) extinction curve), but many sources show
colors typical of young pre-MS objects with an intrinsic IR excess arising
from warm circumstellar dust distributed in disks (Lada \& Adams \cite{LA92}).
The set of dotted curves represents the locus
of two-component black-bodies with temperatures as indicated at the start 
and end of each dotted line; along each curve the relative contribution of the the two
blackbodies is varied. These curves mimic the effect of a temperature stratification in
the dusty circumstellar envelopes, and the presence of sources in the area
covered by these curves is an indication of the presence of warm
circumstellar dust.

A straightforward indication of the youth of the cluster may be
offered by the fraction of sources which are not compatible with
reddened MS stars, i.e. those with IR excess. The number of stars 
with IR excess is normalized to the total number of stars
revealed in the cluster area corrected for the expected number of
fore/background stars estimated from the areas surrounding the cluster (but
still in the same imaged field).
To be conservative we extend the region of the MS by 0.2~magnitudes to the
right corresponding to about 2$\sigma$ uncertainty on measured
magnitudes. This ratio is reported as a percentage value in Col.~6 of
Table~\ref{clusters}).

\subsubsection{Color-Magnitude Analysis}
\label{cm_anal}

\begin{figure}[ht]
  \centering
  \resizebox{\hsize}{!}{\includegraphics[angle=90]{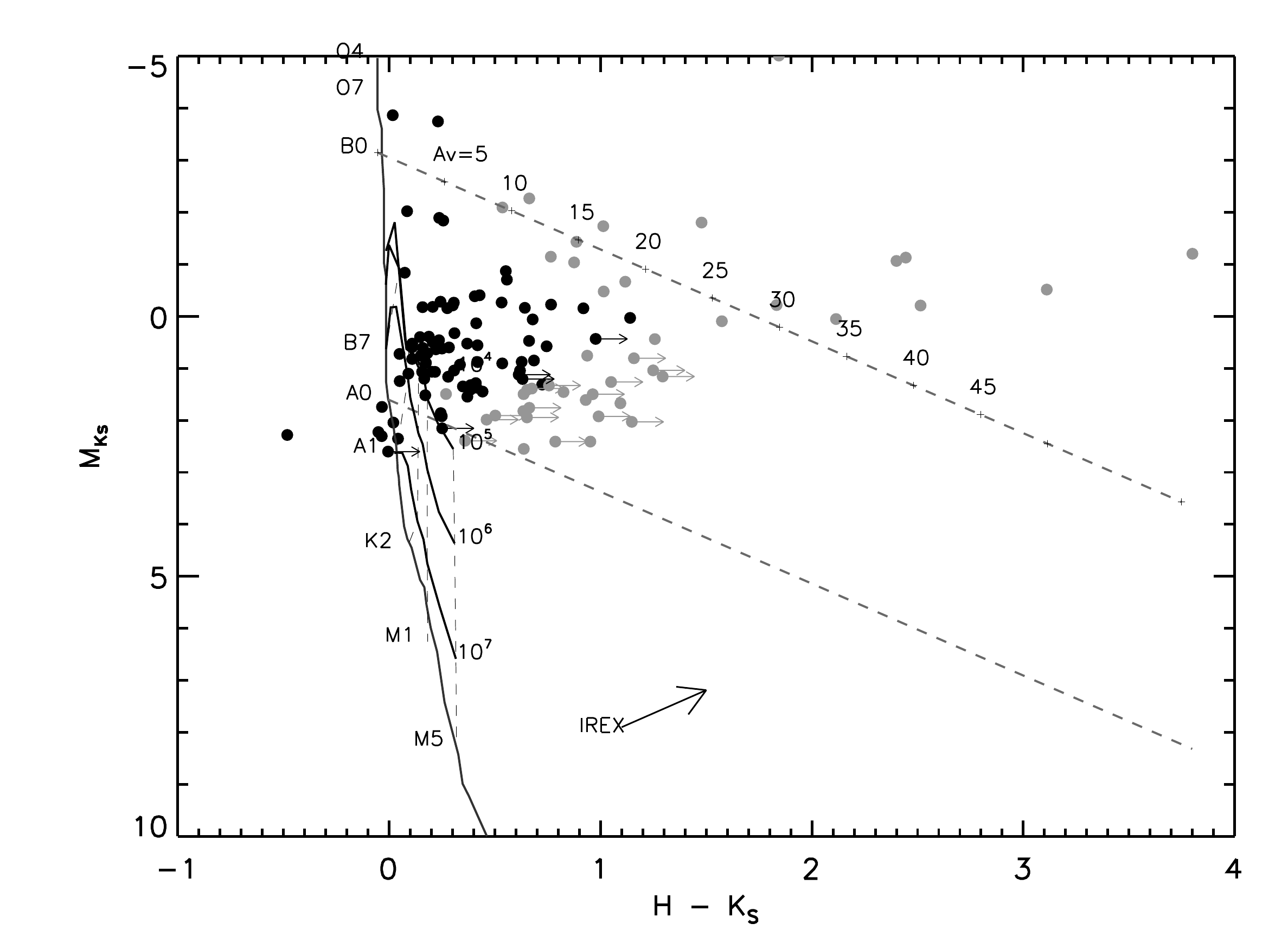}}
  \caption{\ks \textit{vs}[H-\ks] diagram for Mol28. The leftmost curve
represents the Main Sequence, while the dashed lines represent the effect of
reddening for variable amounts of extinction. Isochrones from Palla \& Stahler
(\cite{PS99}) are also indicated with full lines for different Pre-MS ages.
The arrow labeled IREX indicates the direction of change due to IR-Excess 
(see Sect.~\ref{gmdg}). 
[H-K] is obviously a lower limit for those sources not
detected in H. Symbols in grey color indicate sources with IR excess as
determined from the color-color diagrams (see fig. \ref{colcol}).}
  \label{colmag}
\end{figure}

Additional evolutionary indications for the detected clusters may be
derived from the \ks-[H-\ks] diagram, reported for Mol28 in Fig.~\ref{colmag}. 
Compared to the main sequence (the leftmost almost
vertical curve in the figure) a significant fraction of the
sources are on its right, where the evolutionary tracks for Pre-MS
sources (Palla \& Stahler~\cite{PS99}) can also be found, and
could be therefore interpreted as very young pre-MS objects. The
distribution of sources in the diagram spans a much larger region
that the one covered by the Pre-MS isochrones, and this is due to
the combined effect of extinction reddening and IR excess.
Extinction effects can be appreciated looking at the dotted lines
originating on the main sequence and extending toward
bottom-right for increasing values of A$_V$; on the other hand
the presence of a warm dusty circumstellar envelope implies an
increase in absolute emission and in SED steepness, therefore
shifting a pure photosphere toward top-right of the diagram (as
shown by the arrow labeled 'IREX' in Fig.~\ref{colmag}). Similar
to the color-color analysis, it is impossible to try and estimate
the age of individual stellar sources based on their location on
the pre-MS isochrones, because we do not know the amount of A$_V$
by which we should de-redden each object. We follow a
conservative approach by dereddening each object using half of the 
exctinction estimated for each location from millimeter maps; this 
corresponds to putting each object midway through the clump.

A further correction is to remove the IR excess for those sources in which
this is apparent in the color-color diagram (see fig. \ref{colcol}), as 
estimated using the formulation suggested by
Hillenbrand \& Carpenter (\cite{HC00}), and that will also be used
later in this work (see Sect.~\ref{gmdg}). The fraction of pre-MS stars over 
the total in each cluster area will still be contaminated by fore/background 
stars; to estimate this contamination we choose an off-cluster area in the 
same imaged field and we simply compute the ratio of sources with pre-MS colors 
over the total (in these off-cluster regions there is no significant reddening 
to correct for). Col.~7 of Table~\ref{clusters} reports for each cluster the 
fraction of stars (detected in the cluster area in all three bands)
situated more than 0.2~mag to the right of the MS after the
various corrections have been applied.

\section{Initial Mass Functions and Star Formation Histories}
\label{IMFSFH}

As it is apparent from the qualitative analysis presented in the
previous paragraphs, the diagnostic power of our observations is
limited because we do not know which objects in the cluster area are
real cluster members and we do not know the exact amount of dust
extinction (originating within the hosting clump) and IR excess
(originating in the immediate circumstellar environment) pertaining to
each source. Lacking the detailed knowledge on individual stars
in the clusters, fundamental quantities like the Initial Mass
Function (IMF) and the Star Formation History (SFH) cannot be
directly derived from, e.g., the \ks\ luminosity function (KLF). We are
then forced to obtain these using statistical simulations of
clusters based on different input parameters and performing a
statistical comparison between synthetic and observed KLFs and HKCFs.

We will first derive the observed KLFs from the observations; we
will then illustrate in detail the model used for the cluster
simulations, exploring the sensitivity of the results to a wide
range of input parameters; finally, modeled and observed KLFs will be 
compared to infer statistical indications for the IMF and SFH for our clusters.

\subsection{Observed \ks\ Luminosity Functions}
\label{klf}

The KLF for each cluster is obtained simply counting all detected
sources within the cluster area as identified from the cluster
density profile (see \S\ref{resclust}). Similar to the other diagnistic
tools (\S\ref{cc_anal} and \ref{cm_anal}), the KLF will be
contaminated by field stars that cannot be individually identified.
To account for the field star contamination in a statistical
way we subtract from the KLF built on the cluster area, the KLF built in a
region outside the cluster area but still in the same imaged field, after
normalising for the different areas. Regions where the field star KLF is
built have a lower extinction respect to the cluster one, so the background
contribution to the cluster KLF is likely to be overestimated. Field-subtracted KLFs 
for all clusters are presented in Appendix \ref{klf_plots}, and are also available online\footnote{at http://galatea.ifsi-roma.inaf.it/faustini/KLF/}.

\begin{figure}
  \centering
  \resizebox{\hsize}{!}{\includegraphics[angle=90]{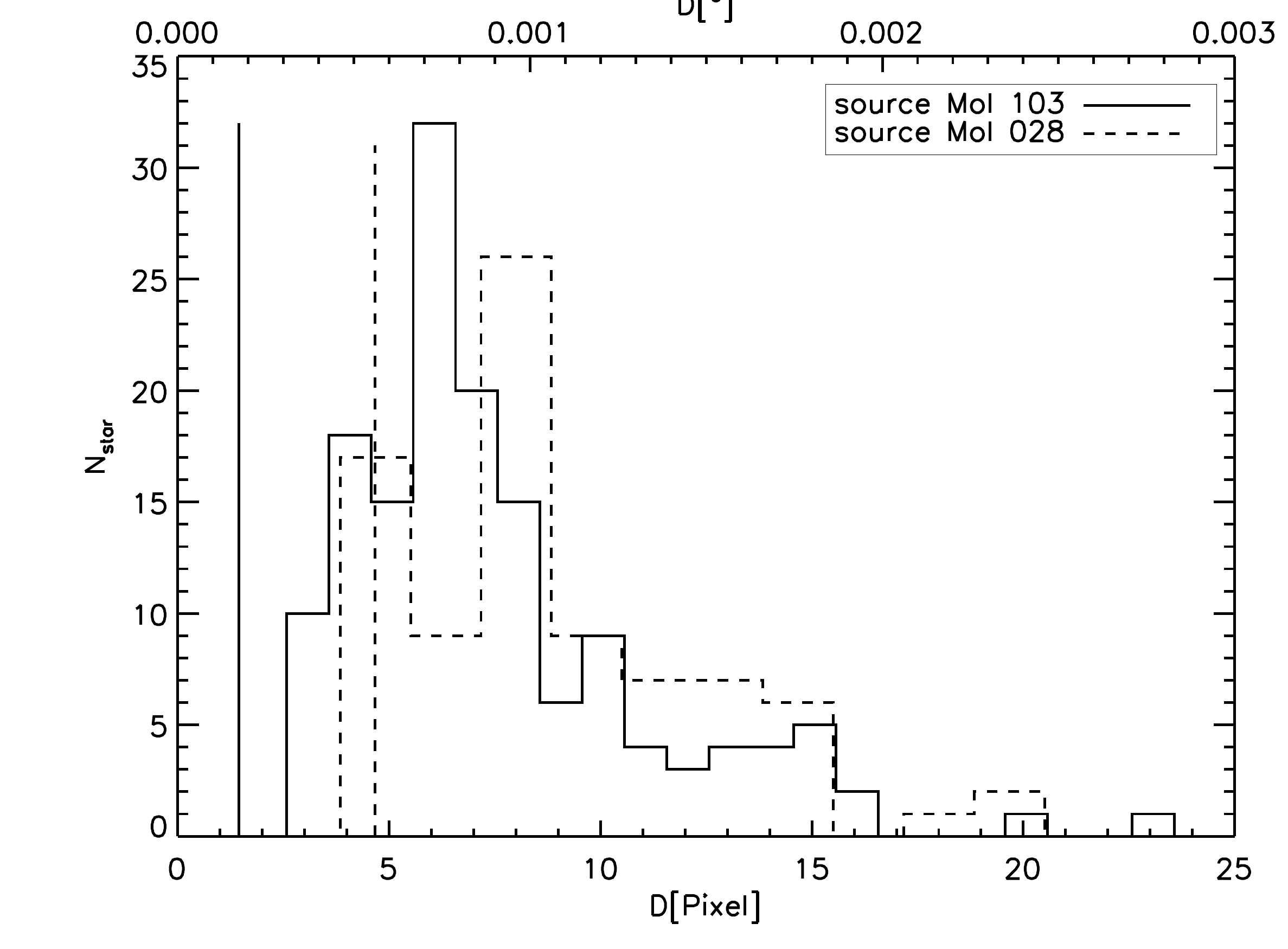}}
  \caption{Distribution of identified sources as function of nearest-neighbour
distance (D) for two of examined fields (Mol28 dashed line and Mol103 full
line).}
  \label{mindist}
\end{figure}

The integral of the KLF gives an independent estimate of the number
of cluster members; these values are reported as N$_{obs}$ in Table
\ref{clusters}; their agreement with richness indicator I$_c$ confirms
the consistency of the analysis. All KLFs show a dominant peak which
always lies close to the completeness limit, showing that our
observations are not sensitive enough to trace the low-mass stellar
component of our clusters. Many of the KLFs present a separate small 
peak at low magnitudes (one or two sources at most, on
average). Can this be due to confusion arising from source crowding and insufficient spatial resolution ? We studied for each cluster the distribution of distances of each star
from its nearest neighbour and we find that there are essentially
two types of distributions, reported in Fig.~\ref{mindist}. In the first type (full line in figure) the distribution has a peak
corresponding to an inter-star distance significantly higher than the value
corresponding to half the PSF FWHM (the full vertical line); in this
case the suggestion is that all cluster members have been resolved
from their neighbour. In the second type (dashed line in the figure)
the distribution has its peak very close to half the PSF's FWHM (the
dashed vertical line), indicating that source blending should be
certainly considered possible. We verified that all clusters
exhibiting a distance distribution of the second type do show a
second faint peak at high brightness in their KLFs, therefore
confirming that this feature is an artifact of the relatively low
spatial resolution which in some cases is not sufficient to resolve
all cluster members.

\subsection{Synthetic KLF. Synthetic Cluster Generator: a Near-IR Cluster Simulator}
\label{gmdg}

As already mentioned, from our data alone we cannot derive masses and ages.
We thus developed a model to create statistically significant cluster
simulations obtained for different assumptions of IMF and SFH
(source ages and their distribution), and compare the synthetic KLFs
with the observed field-subtracted KLFs. This model we call
Synthetic cluster Generator (SCG).

\subsubsection{SCG: Model Description}

A cluster is created by adding stars whose masses and ages are assigned via a Monte-Carlo extraction 
according to the chosen IMF and SFH; the pre-MS evolutionary tracks of Palla \& Stahler (\cite{PS99}) are then 
used to convert them into J, H and \ks\ magnitudes. The 3D distribution of stars is obtained by randomly 
choosing for
each star a set of \textit{{x},\textit{y},\textit{z}} coordinates using the observed stellar density profile
(see Sect.~\ref{resclust}), approximated as a radially symmetric Gaussian, as weight-function; this is 
needed to assign, using submm continuum images, the proper column of cold dust to extinguish the near-IR 
radiation. Other analytical functions could have been used, e.g. a King profile, but the statistics 
of our clusters are not sufficiently high to try and explore the effect of different radial profile assumptions.

To convert the submm flux into dust
column density we used the dust temperature and emissivity exponent $\beta$ as determined in  Molinari et
al.~(\cite{Metal00}); mean values from the latter work were adopted for those fields not covered as part of that work. 

To properly simulate the 
pre-MS stars, we also need to include the effect of an IR excess
due to warm dust in the circumstellar envelopes and disks. We used 
the distribution (modeled as a Gaussian) of [H-\ks]$_{ex}$
color excesses as measured for a sample of Pre-MS stars in
Taurus, as used by Hillenbrand \& Carpenter (\cite{HC00}), as a weight-function to randomly assign a 
[H-\ks]$_{ex}$ to each simulated star in our model; the \ks \textit{vs} [H-\ks]$_{ex}$ relationship adopted in the above mentioned work was then used to derive the H and \ks\ excess-corrected magnitudes. The K$_s$ magnitude of the synthetic star was then compared with the limiting magnitude typical of the cluster being simulated to determine if the star could have been detected in our observations. This procedure is repeated until the number of synthetic detectable stars equal the value of I$_c$ as determined in our observations; at this point the cluster generation process is complete. 

Since the simulation is based on Monte-Carlo extraction for stellar
mass, age and position in the cluster, each independent run for a
fixed set of input parameters can in principle result in very
different outputs in terms of cluster luminosity, total stellar
mass, maximum stellar mass and synthetic KLF. To provide statistical significance, the model is run 200 times for any given set of input parameters, and the median KLF is later adopted for the comparison with the observed one. Clearly, the
predictive power of this simulation model resides in its capability
to characterize the cluster properties for any given parameter set.
In other words, the distribution of the resulting quantities should
not be uniform but peaked around characteristic values. We will come back to
this in section Sect.~\ref{gmdgpp}

\subsubsection{SCG: Input Assumptions}

We tested three different assumptions for star formation histories
in our cluster simulations. The first choice is to assume that stars
in the cluster formed in a single burst-like event (hereafter SB)
some t$_1$ years ago. The explored range in the
simulations is $10^3 \leq \mathrm{t}_1 \leq 10^8$~yrs. The second
choice is to have the formation of stars proceed at a constant rate
(hereafter CR) from a time t$_1$~years ago to a time t$_2$~years
ago. The ranges explored in the simulations are $10^4
\leq \mathrm{t}_1 \leq 10^8$~yrs and $10^3 \leq \mathrm{t}_2 \leq
10^7$~yrs, where always $\mathrm{t}_1 >  \mathrm{t}_2$. The
third possibility we explored is a variation of the previous one,
where the star formation rate is not constant but varies with time
in a Gaussian fashion (hereafter GR). Within the boundaries for
start and end of the star formation process, t$_1$ and t$_2$ varied
as above, we also varied both the time t$_c$ for the peak of the
Gaussian in the range $10^{3.7} \leq \mathrm{t}_c \leq 10^{7.7}$;
the Log$_{10}$($\sigma$) of the Gaussian-like SFH was allowed to
assume the two values 0.1 and 0.5.

As for the IMFs, we allowed three different choices from Kroupa
et al. (\cite{kroupa}), Scalo (\cite{scalo}) and Salpeter
(\cite{salpeter}), with the latter modified introducing a different slope
for M$<$1~\msun\ that coincides with that of the Scalo (\cite{scalo}) IMF; the
three IMFs will be coded as IMF1, IMF2, and IMF3
respectively. The IMF from Kroupa et al. favors the low-mass end of the
distribution, while the classical Salpeter IMF is flatter at low
mass but heavier at intermediate and high masses (above 1 \msun).
The IMF from Scalo is kind of intermediate between the two,
resembling Salpeter's one below 1~\msun\ and above 10~\msun, and
Kroupa's for 1~\msun$<$M$<$10~\msun.


\subsubsection{SCG: Predictive Power}
\label{gmdgpp}

\begin{figure}[ht]
  \centering
  \hspace{-1cm}
  \resizebox{\hsize}{!}{\includegraphics[angle=90]{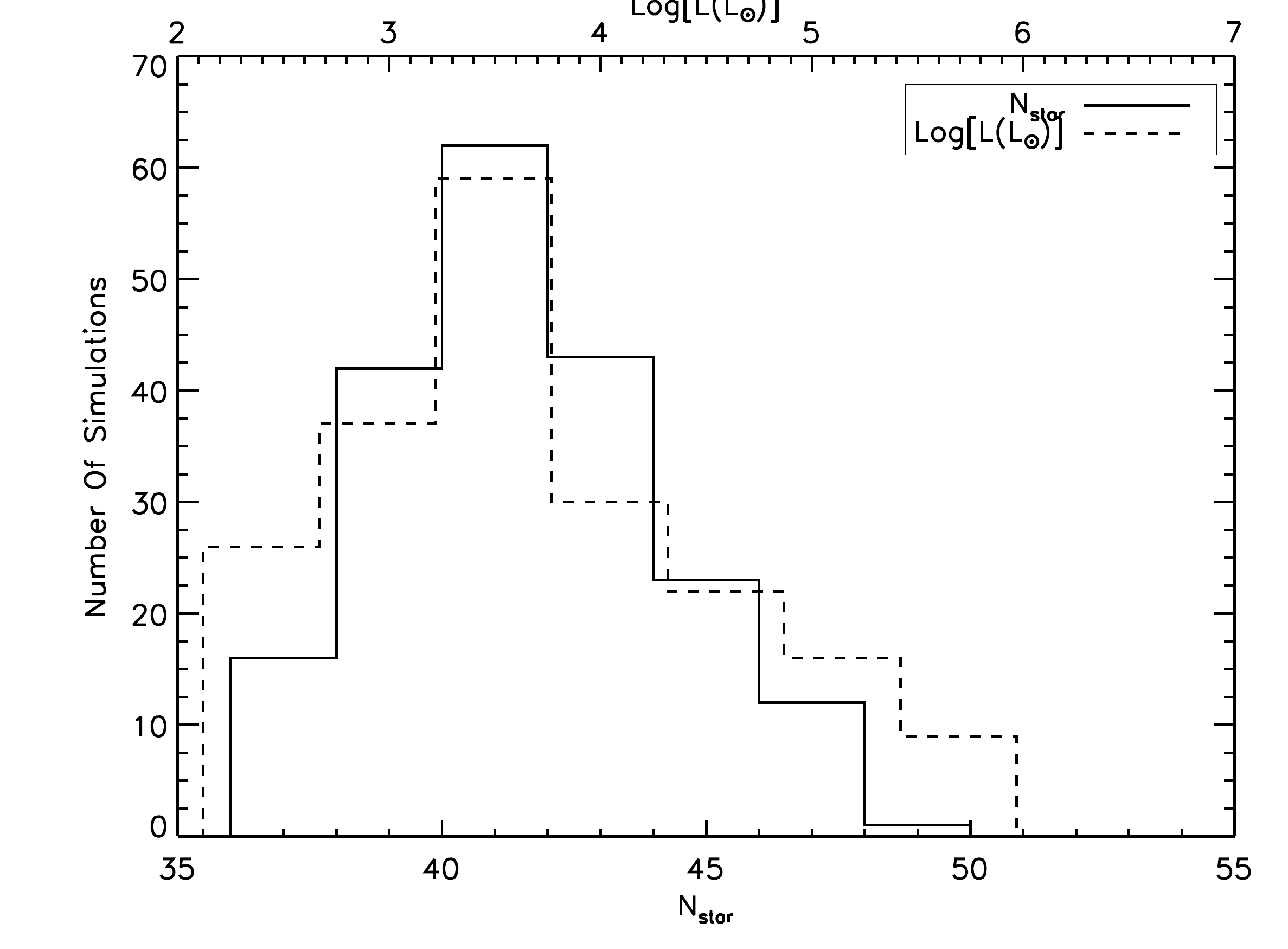}}
  \caption{Distribution of the predicted number of cluster members (full line)
and total luminosity (dashed line) for 200 SCG runs for Mol160 with a Salpeter
IMF and a constant star formation rate with t$_1$=10$^6$~yrs and
t$_2$=10$^4$~yrs.}
  \label{dist_lumnum}
\end{figure}

In order to verify our model's predictive power,
we ran a set of 200 simulations for a cluster with
a Salpeter IMF and a constant star formation rate with
t$_1$=10$^6$~yrs and t$_2$=10$^4$~yrs. Figure~\ref{dist_lumnum} shows
the distribution of the predicted number of stars and the total luminosity
for the 200 simulations. The number of cluster members shows very
little variations, as expected since the number of detectable stars
is the parameter we use to stop the simulation; on the other hand
the distribution of the total luminosity is not particularly peaked,
as the central 3 bins containing about 60\% of the simulations 
span almost two decades in luminosity. 

\begin{figure}[ht]
  \centering
  \hspace{-1cm}
  \resizebox{\hsize}{!}{\includegraphics[angle=90]{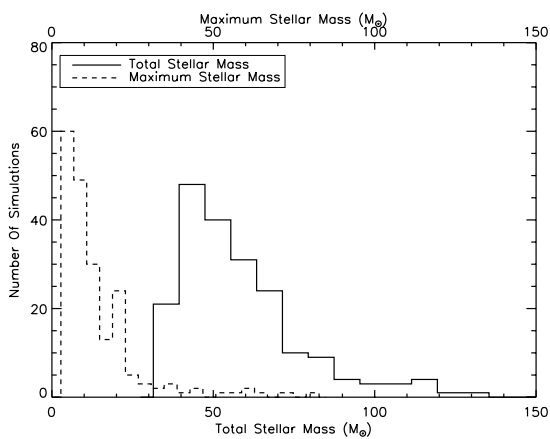}}
  \caption{Distribution of the predicted total stellar mass (full line) and
mass for the most massive star (dashed line) in a cluster for 200 SCG runs
for Mol160 (same inputs as in Fig.~\ref{dist_lumnum}).}
  \label{dist_mstars}
\end{figure}

On the other hand the distributions for the total cluster stellar
mass, and for the mass of the most massive member (see Fig.~
\ref{dist_mstars}) are rather peaked and highlight a relatively
higher predictive power of the model for these two quantities. It is
to be noted that the distributions are rather skewed, suggesting
that neither the mean nor the median are particularly suited to
characterize the peak of the distribution. Indeed we have 
computed the values that these quantities assume at the peak of their 
distributions, to have a more representative value for the masses 
and use them in the following arguments.

\begin{figure}[ht]
  \centering
  \hspace{-1cm}
  \resizebox{\hsize}{!}{\includegraphics[angle=90]{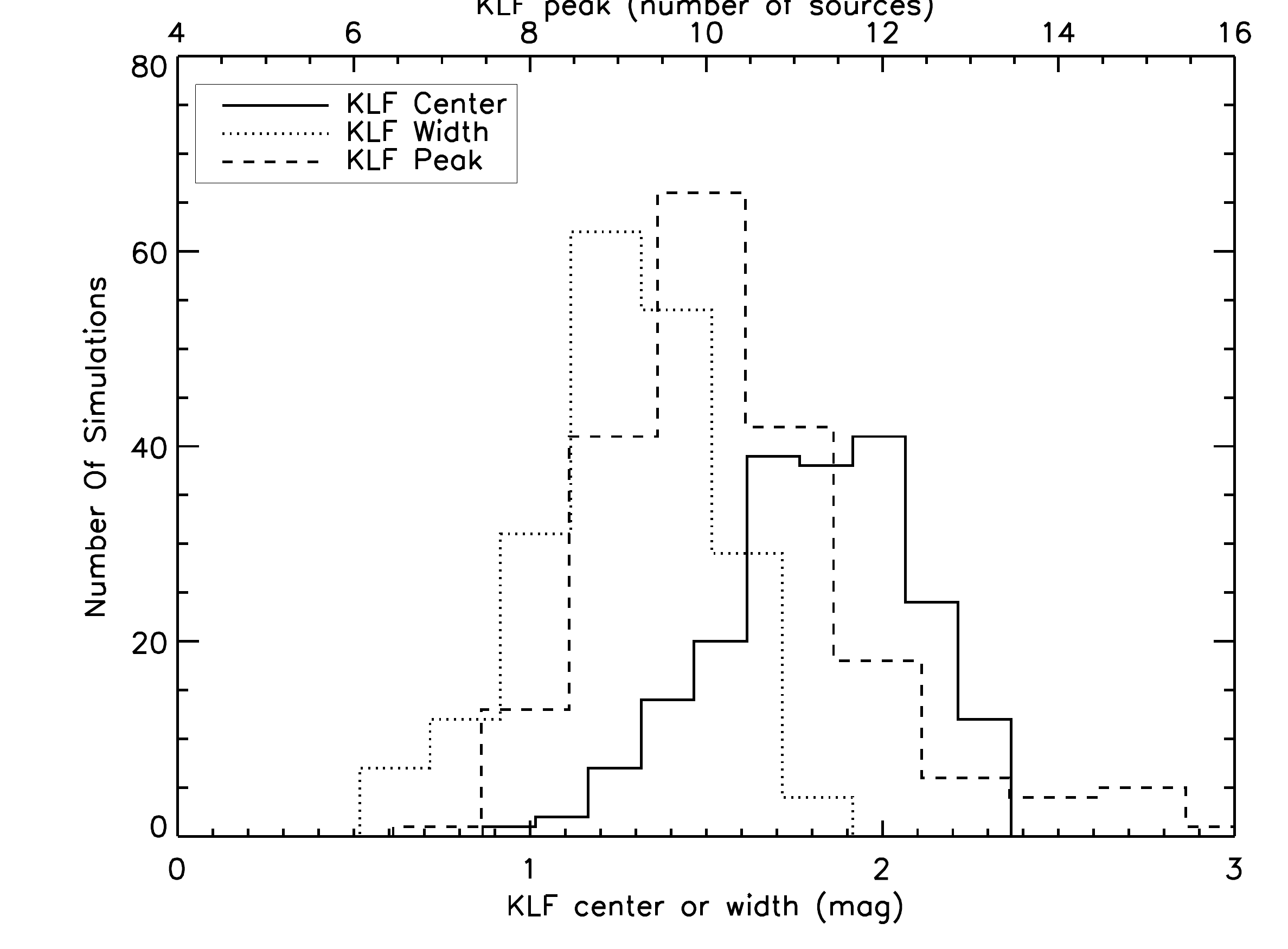}}
  \caption{Distribution of the predicted center magnitude (full line - bottom
X-axis scale), width (dotted line - bottom X-axis scale) and peak value
(dashed line - top X-axis scale) of the predicted Gaussian-fitted KLFs for
200 SCG runs for Mol160 (same inputs as in Fig.~\ref{dist_lumnum}).}
  \label{dist_klf}
\end{figure}

Concerning the reproducibility of the KLF, for each of the 200 runs the resulting KLF was
fitted with a Gaussian function and the center, peak and $\sigma$
were determined. Fig.~\ref{dist_klf} reports the distribution of
these three parameters for the 200 runs and shows that the all of
them are remarkably peaked and symmetric. The formal r.m.s. spread
for the three quantities, estimated via a Gaussian fit to the
distributions in the figure, is $\approx 0.3$~mag for the KLF center,
$\approx$ 12\% for the KLF peak (about 1.2 sources out of a mean KLF
peak of 10), and $\approx 0.25$~mag for the KLF FWHM. 

\begin{figure}[ht]
  \centering
  \hspace{-1cm}
  \resizebox{\hsize}{!}{\includegraphics[angle=90]{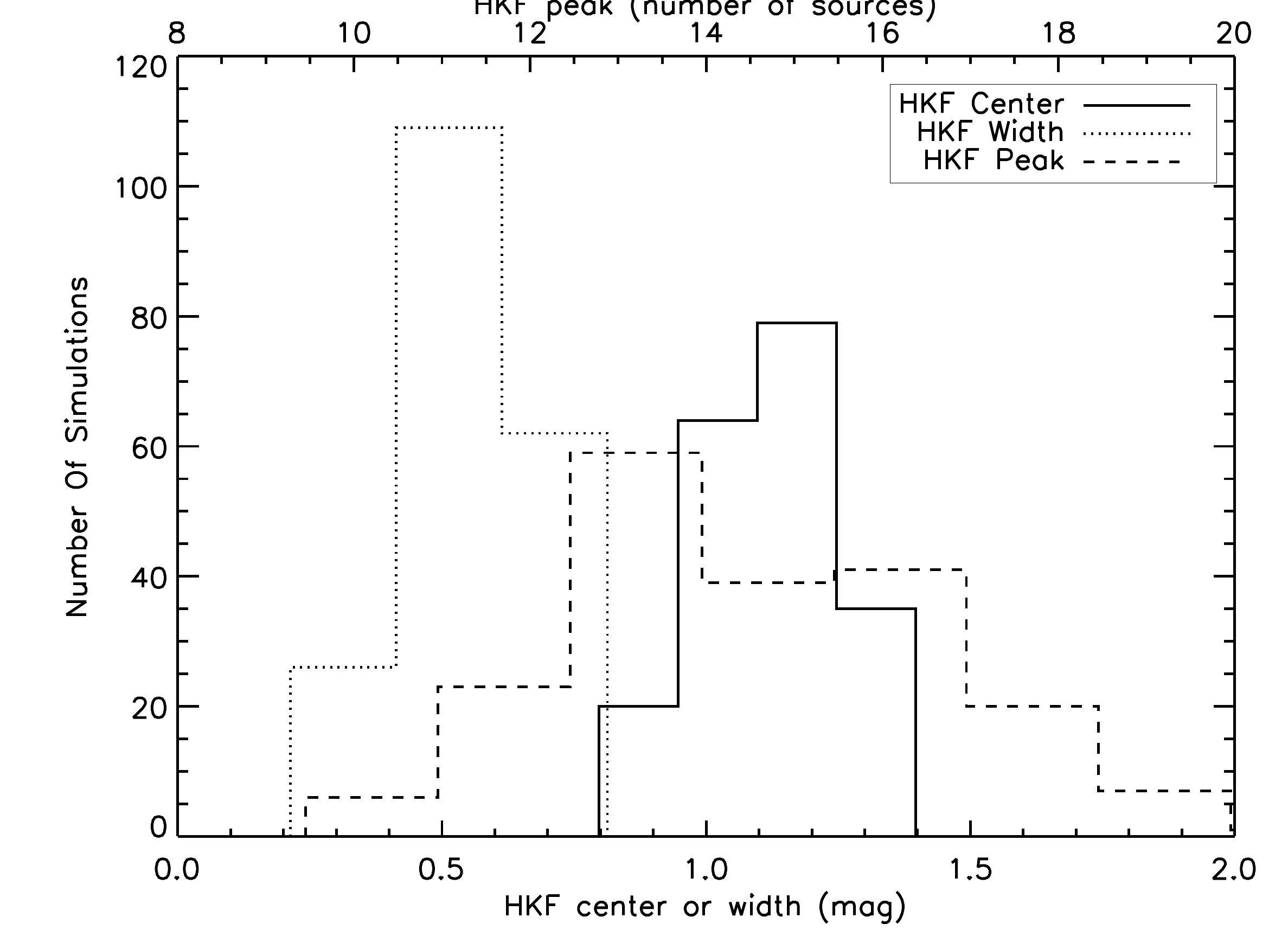}}
  \caption{Distribution of the predicted center color (full line - bottom
X-axis scale), width (dotted line - bottom X-axis scale) and peak value
(dashed line - top X-axis scale) of the predicted Gaussian-fitted HKCFs for
200 SCG runs for Mol160.}
  \label{dist_hkcf}
\end{figure}

We have made a similar analysis for HKCF (H-\ks\ color function; see
Sect.~\ref{cm_anal}). Figure~
\ref{dist_hkcf} shows the distribution of Gaussian function centers,
peaks and $\sigma$'s for HKCFs obtained for the same 200 runs used
previously for the KLFs. Gaussian fits to the three distributions in the
figure give an r.m.s. that is $\approx 0.15$~mag for the HKCF center and
$\approx 0.14$~mag for the HKCF FWHM, while the 'peak' distribution is
flatter and has an r.m.s. value of  $\approx$ 21\%
for the HKCF peak (about 3.2 sources out of a mean HKCF peak of 15). It is worthwhile to stress   that since the position which is assigned to each simulated star in the cluster is different in each of the 200 runs of the model (for any given set of input parameters), the scatter in the properties of the synthetic KLFs and HKCFs also statistically tends to account for the effects of extinction variations in the cluster's hosting clump, which may in principle be relevant in such heavily embedded systems (see Tab. \ref{clusters}).

We conclude that the model results, for a given set of input parameters, 
have a good reproducibility, except concerning the total luminosity. The model therefore has 
a quite fine predictive power concerning the median properties of a synthetic 
cluster.  Indeed, the spread of KLF center magnitudes is less than 
the bin amplitude used in building the KLFs for the simulations (and that 
will be used for the rest of the work); the median synthetic KLF 
therefore provides a good representation of the cluster luminosity distribution. 

 In conclusion, 200 simulation runs for each combination of input parameters 
(IMF and SFH) seem enough to achieve a sufficient statistical significance for 
the properties of the synthetic observables (KLFs and HKCFs). Although the distributions for the KLFs' (HKCFs') parameters seem rather
symmetrical, we will adopt the median KLF (HKCF) of the 200 runs as a more
reliable characterization for that particular parameters' set. Using the mean KLF
(HKCF) for the comparison does not significantly alter the results.

\subsubsection{Exploring the SCG parameter space: Cluster Parameters}
\label{gmdg_expl-par}

After verifying the robustness of model results for independent runs
for the same input parameters, we now want to verify the
sensitivity of the model results against changes of these
parameters. We will first concentrate on simulated cluster physical
parameters (number of cluster members, total luminosity, stellar
mass distribution), and in the next paragraph we will examine how
the KLFs and the appearance of the color-magnitude diagrams, which
are the main observables we will use in our analysis, behave in this
respect.

\paragraph{Number of stars N$_{stars}$ - } As a general rule the older the
cluster is allowed to be, irrespective of the detailed SFH adopted,
the higher  is the number of produced stars. This is easily understood
since the SCG  cluster building is stopped when the number of the \ks
-detectable stars  equals the number of observed objects; if a cluster
is old the stars will be  intrinsically fainter due to the shape of
Pre-MS tracks, and it will  statistically be less likely to
extract stars bright enough to be detectable.  N$_{stars}$ does not
significantly depend on the IMF choice as long as  t$_1\leq10^6$~yrs,
while for older systems IMF1 (Kroupa et al.~\cite{kroupa})  will
produce nearly twice as many stars as IMF3 (Salpeter~\cite{salpeter})
with IMF2 (Scalo~\cite{scalo}) in between.

\paragraph{Stellar Masses - } Likewise, the total stellar mass and the mass of
the most massive star will be higher the older the cluster is allowed to be.
If an IMF1 cluster is a very old SB or a CR with t$_1$=10$^8$~yrs and
t$_2$=10$^7$~yrs for example, M$_{\star Tot}$ and M$_{\star Max}$ will be
respectively a factor of 5 and 2 higher with respect to clusters which are
younger and/or are allowed to form stars until recent times (i.e. allowing a
CR with t$_2=10^4$~yrs). The explanation follows directly from the argument
made for the N$_{stars}$ behavior above; matching the number of \ks -detected
stars in a relatively old cluster with intrinsically fainter stars will
require that stars will have to be on average more massive objects, and this
will clearly result also in a higher total stellar mass.

Going from IMF1 to IMF3 both M$_{\star Tot}$ and M$_{\star Max}$ will
significantly increase, as expected. The trend of M$_{\star
Tot}$ with cluster age is less pronounced because with IMF2 and IMF3
it is statistically more likely to produce relatively more massive
(and hence more easily detectable in \ks) stars requiring a lower
number of star extractions and hence a lower relative total mass at the end
of the simulation. The age-trend of M$_{\star Max}$ is instead the
same (only shifted toward higher masses) because the probability of
extracting a massive star is the same for all ages and is only
function of the chosen IMF.

\paragraph{Total Stellar Luminosities and Massive Object Luminosities - } The
total stellar luminosity, like the luminosity of the most massive star
(L$_{\star Max}$), exhibits the same behaviour as
M$_{\star Max}$. This is easily understood given the steep power-law
dependence of the stellar luminosity on mass, and confirms that the total
luminosity (L$_{Tot}$) will be largely dominated by the most massive stellar
object in the cluster: L$_{Tot}$ $\propto$ L$_{\star Max}$.
Of great interest is the ratio between L$_{\star Max}$ and
L$_{Tot}$; for the large majority of clusters its value varies between 0.6 and 0.8. 
This is 
further confirmation that global properties of our clusters are dominated by
the most massive source. This ratio does not present any particular dependence
on the value of M$_{\star Max}$, or of N$_{stars}$; only for the most
populated clusters (clusters with of the order of a hundred members, such as
Mol103, where the contribution of a great number of low-mass sources 
becomes important, do we find a lower value for this ratio.

\subsubsection{Exploring the SCG parameter space: KLF variations}
\label{gmdg_klf}

\begin{figure}[ht]
  \centering
  \hspace{-1cm}
  \resizebox{\hsize}{!}{\includegraphics[angle=90]{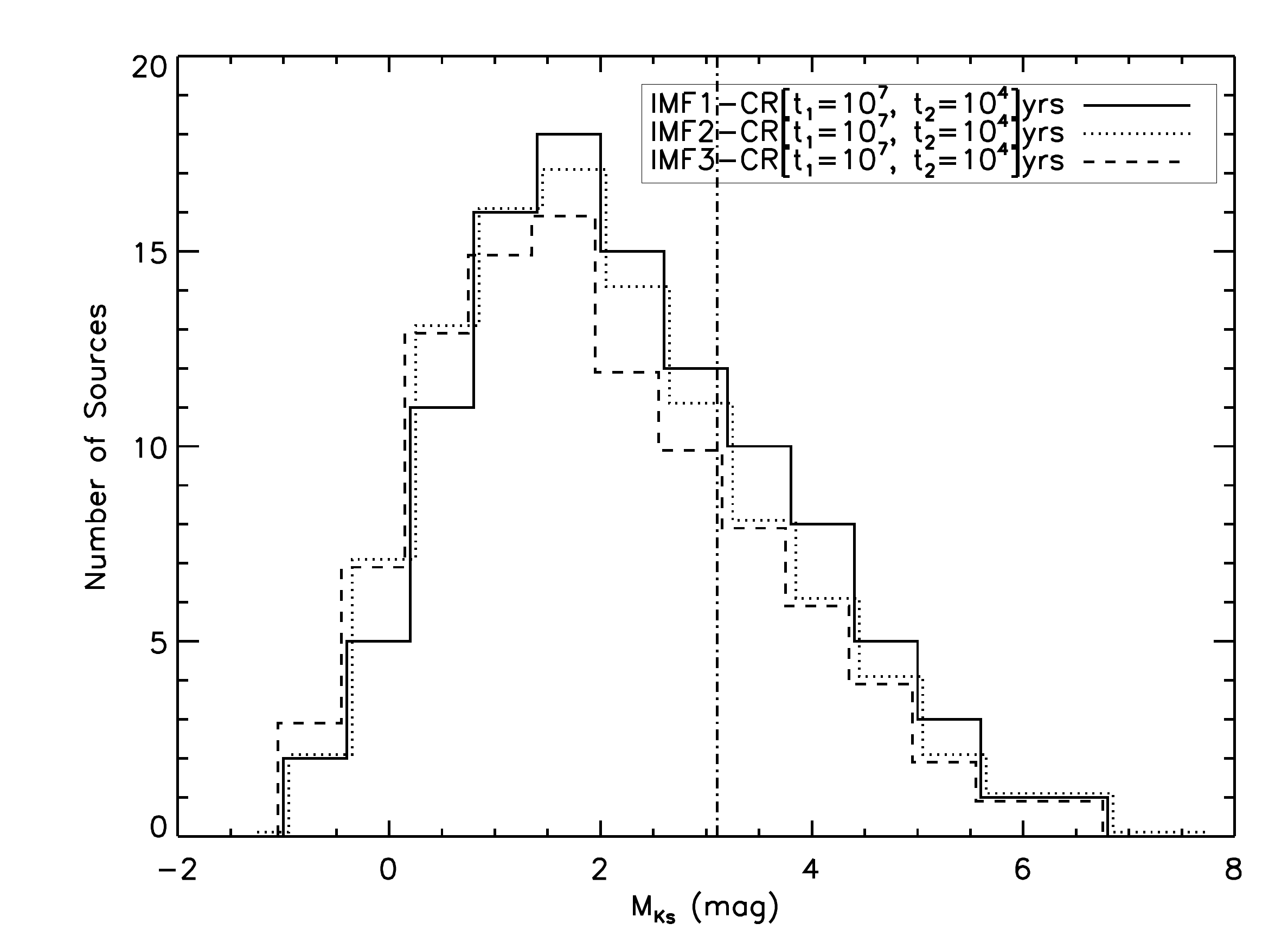}}
  \caption{KLF (using the absolute \ks\ magnitudes) for Mol3 predicted by
SCG for a CR cluster with t$_1$=10$^7$~yrs and t$_2$=10$^4$~yrs, for three
different choices of the IMF (line styles as indicated in the figure). The
dash-dotted line represents the completeness limit for this source given the
\ks\ limiting magnitude.}
  \label{gmdg_imf}
\end{figure}

We will now briefly analyze the diagnostic power of the KLF and the HKCF against changes in IMF and SFH choices. Figure~\ref{gmdg_imf} shows the KLFs predicted for source Mol3 adopting the same SFH parameters
(as indicated in the figure) and using the three different IMF choices.

The shape of the resulting KLF changes throughout the M$_K$ range; going from
Kroupa et al.'s IMF1 to Salpeter's IMF3 the distribution gets more skewed
toward lower magnitudes; this was expected since IMF1 produces more lower
mass stars than IMF3. One can certainly argue that the change is not dramatic,
but on the other hand the modification does not affect one or two bins but the
entire KLF consistently. The change is more apparent in the region between the
peak and the completeness limit  than at the bright end of the KLF, and for
this reason the ability of the model to discriminate among different IMFs will
be better for those sources, as Mol3 in the figure, where the KLF's peak is
clearly detected above the completeness limit.

\begin{figure}[ht]
  \centering
  \hspace{-1cm}
  \resizebox{\hsize}{!}{\includegraphics[angle=90]{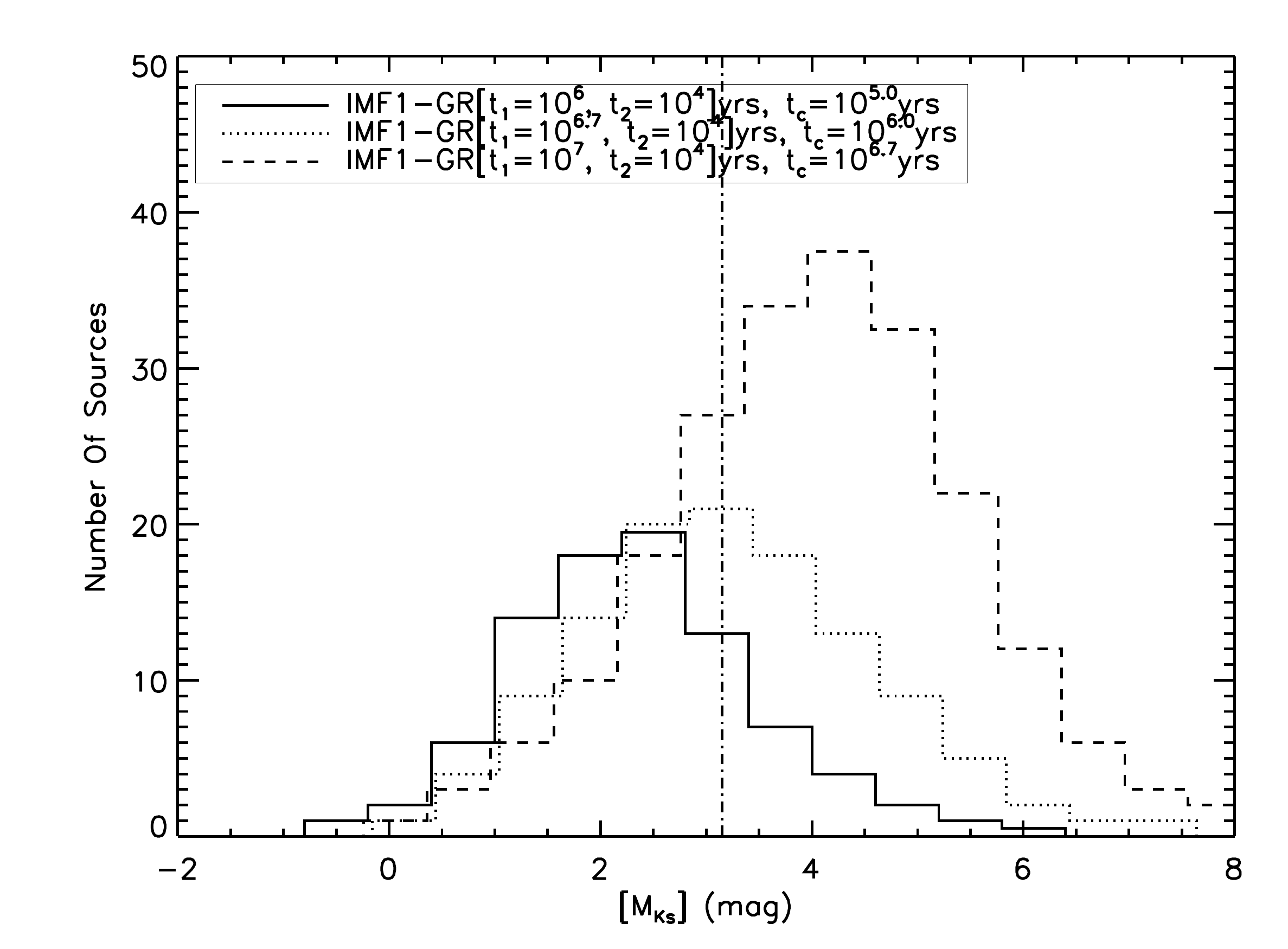}}
  \caption{KLF (using the absolute \ks\ magnitudes) for Mol3 predicted by
SCG for an IMF3 cluster with different ages as indicated in the plot. Older clusters produce a KLF peaked tower lower magnitudes. The dash-dotted line represents the completeness limit for this source given
the \ks\ limiting magnitude.}
  \label{gmdg_age}
\end{figure}

The difference in predicted KLFs is much more dramatic if different age ranges
are assumed, keeping fixed the shape of the SFH and the IMF, as it is apparent in
Fig.~\ref{gmdg_age}. The peak of the KLF shifts considerably toward higher
magnitudes as the median stellar ages (t$_c$) increase. A similar trend is observed comparing SB models with different
ages, although SB models always produce KLFs which are considerably narrower
than CR or GR models. Larger cluster ages would shift the peak of the KLF beyond the completeness limit; in other words, our analysis is not sensitive to ages for the majority of stars in excess of  $\sim5\,10^6\div 10^7$ yrs; besides, such old clusters would be hard to justify given the fact that they are still heavily embedded in dense clumps.

\begin{figure}[ht]
  \centering
  \hspace{-1cm}
  \resizebox{\hsize}{!}{\includegraphics[angle=90]{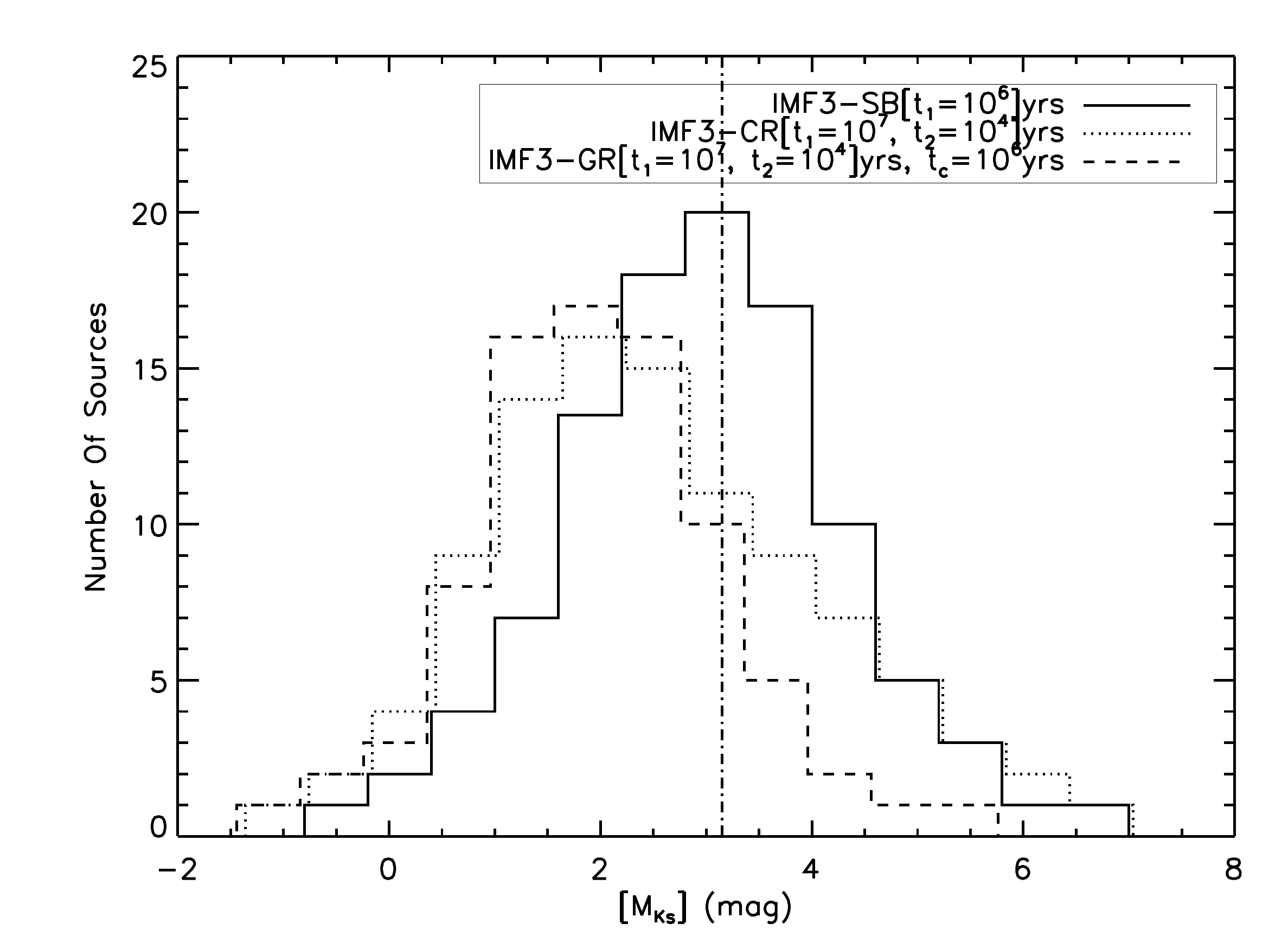}}
  \caption{KLF (using the absolute \ks\ magnitudes) for Mol3 predicted by
SCG for an IMF3 SB cluster with t$_1$=10$^6$~yrs (full line), CR with
t$_1$=10$^7$~yrs and t$_2$=10$^4$~yrs (dotted line),  and GR with same t$_1$
and t$_2$ and t$_c$=10$^6$~yrs (dashed line). The dash-dotted line represents
the completeness limit for this source given the \ks\ limiting magnitude.}
  \label{gmdg_sfh}
\end{figure}

Finally, we briefly show how the KLFs change for different choices of the
SFHs. Figure \ref{gmdg_sfh} shows the KLFs obtained for a SB with
t$_1$=10$^6$~yrs, compared with a CR with t$_1$=10$^7$~yrs and
t$_2$=10$^4$~yrs and a GR with the same start and end star formation period,
and with a peak times for star formation rate of t$_c$=10$^6$~yrs. The KLFs
are clearly different, with a peak magnitude which is quite sensitive to the
formation rate typology and the peak time for star production. 

As concerns the [H-K$_s$] color functions, they are found insensitive to the choice of IMF. Similarly to the case of the KLFs, the main differences between synthetic HKCFs are
more evident for different age ranges especially in the number of detectable stars.

\subsection{Comparing Observed and Synthetic KLFs and HKCFs}
\label{scgresults}

The detailed comparison of the model KLFs and HKCFs functions to those observed was
carried out only for those sources where the number of detected
stars was sufficient (I$_c \geq$15) to allow a statistically significant comparison 
(see Col. 3 of Table~\ref{clusters}), and where submm information was available to
allow meaningful estimates of extinction (this excludes Mol15 and Mol99). 
The number of clusters fulfilling these criteria were 16 out of 23 detected clusters. 
The comparison of the observed KLFs and HKCFs (KLF$_{Obs}$, HKCF$_{Obs}$), with the
synthetic ones produced by SCG (KLF$_{Syn}$ HKCF$_{Syn}$), for the full set
of input parameters (IMF, SFH and age parameters) was carried out
automatically; to ease the process, the observed and
synthetic functions were computed on the same M$_K$ and H-K grid.

The comparison procedure between synthetic and observed KLFs is described below, 
but it is the same for HKCFs. The KLFs are first compared bin by bin (the
comparison being limited to those bins brighter than the completeness limit) 
identifying with $i$ each bin of the observed KLF, starting
from i=1 for the lower-M$_K$ non-zero bin to N for the bin where the
completeness limit for that source is reached (the number N will clearly be
different for each cluster). 
In the case of HKCFs, we exclude objects with H and K magnitudes greater than the 
observed limiting magnitudes for these bands. 
A matching flag $m_i$ is set to 1 for those bins where the number of
sources coincide within the 1$\sigma$ Poissonian error bar of the observed
KLF, i.e.:

\begin{equation}
      |N_{\star Syn_i} - N_{\star Obs_i}| \leq \sqrt{N_{\star Obs_i}}
\label{eq_match}
\end{equation}

The total number of bins where a match is found is divided by the total
number of bins useful for the comparison to get a KLF compatibility figure
(in \%) as

\begin{equation}
      C = 100\times {{\sum_{i=1} ^N m_i}\over{N}}
\label{eq_c}
\end{equation}

The higher is C, the better is the overall match between KLF$_{Obs}$
and KLF$_{Syn}$.


However, the same value of C may result from bins concentrated
in the low-M$_K$ end of the KLF, where there are few sources, or in the region
around the peak and in the proximity of the completeness limit, where instead
there are more sources and hence higher statistical significance. A further 
figure of merit is then introduced,

\begin{equation}
      W = \sum_{i=1} ^N  {{m_i \cdot\ N_{\star Obs_i} }\over{N_{\star Obs_{tot}}}},
\label{eq_w}
\end{equation}

where N$_{\star Obs_{tot}}$ is the total number of sources present in all the
bins used for comparison. This parameter weights each matching bin by its
relative richness, favoring the bins closer to the completeness limit and the
KLF peak over those in the bright tail of the KLF, and favoring for HKCF the
bin closer to the peak of the distribution. This choice is due to the fact 
that the KLF (HKCF) peak region is the the most sensitive to changes in the 
SCG input parameters.

In this automatic procedure we select those models for which the parameters C and W 
(at the same time for KLFs and HKCFs) are maximum. For Mol8B, Mol45 and Mol84 
the observed KLF has a very irregular and multiple-peaked shape which cannot be
matched by any model, and are therefore discarded from further considerations. We 
are then left with 14 clusters for which a series of models can be found with at least 
75\% of the bins matching the observations. We find that the best values of C and W 
are never found for one single set of parameters, but rather we identify ranges of 
parameter values which produce the best match; in other words there is a level of 
degeneracy that the models cannot remove, and this varies from source to source. In 4 
clusters (Mol109, 110, 136 and 151) this degeneracy is essentially complete and the model is not 
able to make any prediction; in one case (Mol148) the comparison selects models with very old stellar ages but with total stellar luminosities by far in excess of the measured bolometric luminosity obtained integrating the integrated observed for this region from the mid-IR to the millimeter (see Tab. \S\ref{clusters}). In the 9 remaining cases some degeneracy persists especially in the
IMF, confirming (\S\ref{gmdg_klf}) that our models are weakly selective on the IMFs, 
but there are clear indications concerning the SFH and ages. 

\begin{table}[t]
\setlength{\tabcolsep}{0.075in}
\begin{flushleft}
\caption{Results for SCG runs on detected clusters}
\label{gmdg_restab}  
\begin{tabular}{lccccccc}\hline\hline
Source & IMF  & t$_{10\%}$ & t$_{med}$ & t$_{90\%}$ & N$_{\star}$ & M$_{\star Max}$ & M$_{\star Tot}$   \\
Mol & & \multicolumn{3}{c}{yrs} & & \msun & \msun  \\ \hline
3 & 1     &  $10^{4.98}$ & $10^{5.40}$ & $10^{5.81}$ & 91   & 3.7  & 61     \\
8A & 2-3     &  $10^{4.71}$ & $10^{5.38}$ & $10^{5.83}$ & 31  & 3.8 & 30    \\
11 & 1-2-3 &  $10^{6.06}$ & $10^{6.41}$ & $10^{6.82}$ & 48   & 3.8  & 47      \\
28 & 3     &  $10^{6.03}$ & $10^{6.12}$ & $10^{6.24}$ & 77   & 9.9  & 105     \\
50 & 1     &  $10^{6.16}$ & $10^{6.48}$ & $10^{6.66}$ & 53   & 3.5  & 36      \\
103 & 1-2   &  $10^{6.57}$ & $10^{6.70}$ & $10^{6.83}$ & 115  & 4.0  & 80     \\
139 & 3     &  $10^{5.34}$ & $10^{5.96}$ & $10^{6.46}$ & 25   & 2.9  & 16      \\
143 & 2     &  $10^{6.57}$ & $10^{6.70}$ & $10^{6.82}$ & 27   & 3.1  & 21      \\
160 & 1-2-3 &  $10^{6.57}$ & $10^{6.7}$ & $10^{6.83}$ & 89   & 4.3  & 63     \\
\hline
\end{tabular}        
\end{flushleft}      
\end{table}         
                     
Table ~\ref{gmdg_restab} reports a summary of the results.The IMF of matching 
classes of synthetic cluster models is shown in Col.~2.
Cols.~3, 4 and 5 contain 
the times for the formation of 10\%, 50\% and 90\% of the total number of cluster 
members; these values are obtained as the median of the values that these times have in all models that match the observations. Col. 6 is 
the number of cluster members N$_{\star}$; Col.~7 shows the mass of the most massive 
object M$_{\star Max}$; Cols.~8 reports the total stellar mass of the cluster 
M$_{\star Tot}$. We stress again that 
the analysis selects classes of models rather than single models; the values reported in Table \ref{gmdg_restab} are the median of the values that the parameters assume within the class of matching models. The table shows 
that for some fields multiple IMFs are compatible with the data and, in general, 
SFHs with constant (CR) or Gaussian (GR) star formation rate provide acceptable 
solutions for certain age ranges (as reported in the Table). Simulations with SFHs 
with a single burst are, in general, not accepted. Our modeling is insensitive to bulk stellar ages in a cluster in excess of $5\,10^6 \div 10^7$ years (\S\ref{gmdg_klf}).

\section{Discussion}
\label{disc}

\subsection{Cluster Ages and Star Formation Histories}
\label{history}

Perhaps the most important result of this work is that in all clusters where the comparison of observed KLFs with the ones predicted by the SCG model is possible (see previous paragraph), the observations are consistent with a star formation which goes on
over time intervals that in most cases have a spread between about few $10^{5}$ and 
few $10^{6}$ yrs , and with a median cluster age of a few $10^{6}$ yrs. In most cases we cannot discriminate clearly between a
constant or variable SFR but we are confident that we can exclude that on average the stars in our clusters
are coeval and originating from a single burst of formation. Detailed studies
toward the Orion Nebula Cluster show that stars have been forming for at least
10 t$_{dyn}$, or 20 t$_{ff}$ (Palla \& Stahler~\cite{PS99}; Hoogerwerf et al.~
\cite{hoog}, Hillenbrand ~\cite{Hill97}), and our results would seem to generalize this on a larger sample
of intermediate and high-mass star forming regions. 

In  principle it can be argued that our analysis is incomplete since we did not take in consideration longer wavelength data which could pinpoint heavily extincted objects barely visible, or not visible at all, in the near infrared. This however, does not appreciably modify our conclusions about the age spread within the clusters. Indeed, Vig et al. (\cite{Vig07}) applied a different  analysis to the specific region Mol075, a field not included in our final analysis (Table \ref{gmdg_restab}) because the background-subtracted KLF is populated by too few objects for a statistically significant model comparison. Vig et al. also considered \textit{Spitzer} IRAC and MIPS data, looking for the brighter and redder objects in the area covered by submillimeter emission. In this way they could identify the younger and more massive objects in the field with an estimated age of the order of 10$^6$ years or less. This approach, however, is not sensitive to low mass and relatively older pre-MS objects, for which our method is ideally designed. While for this particular field, for the reasons explained above, we cannot perform a direct comparison to our approach, it is clear that the inclusion of longer wavelength data in the analysis might have identified a different, younger, population of objects, rather amplifying  the observed age spread deduced for the clusters.

Models of cluster formation via competitive accretion seemingly succeed in
delivering an IMF close to the observed ones thanks to the spread of the
accretion rates consequent to the competitive accretion mechanism, but the
prediction that all stars are formed in about $5 \times 10^{5}$~yrs (Bonnel et
al.~\cite{bonnellVnBa}) for typical conditions in young clusters, corresponding
to a dynamical time or so, seems to be in disagreement with our results. We
instead favour scenarios (Tan \& McKee~\cite{tanMc}) where stars keep forming
over several free-fall times thus providing the required age spread. The finding that the most massive object in the fields considered in this work
are still being formed or have just finished a phase of intense
accretion (Molinari et al.~\cite{Moli08a}) is a further indication that star
formation \textbf{seems} to be a long-duration process in the life of a
molecular clump.

\begin{figure*}[ht]
   \centering
   \resizebox{\hsize}{!}{\includegraphics[angle=90]{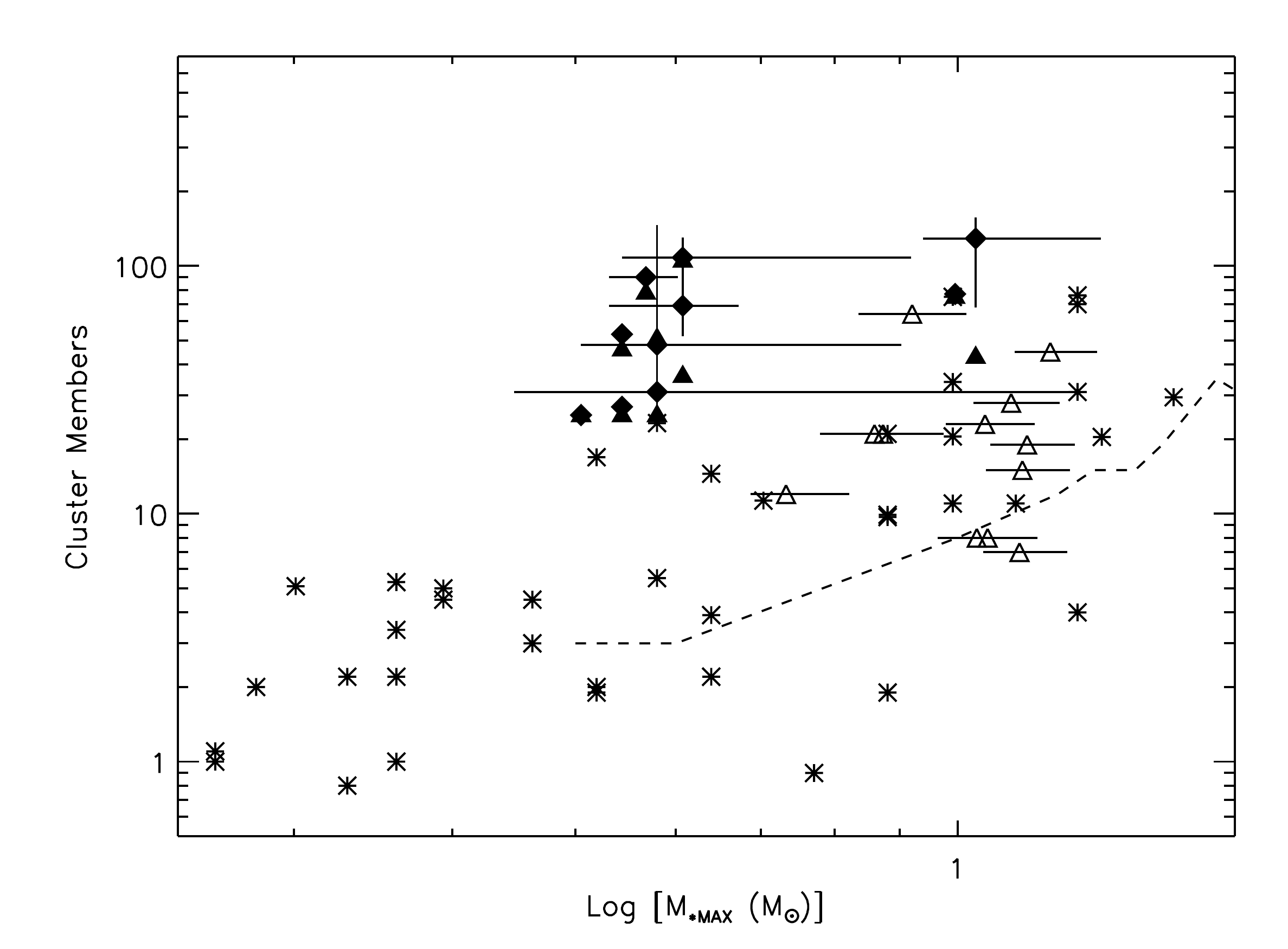}}
  \caption{Number of cluster members as a function of Mass of highest mass
star. Asterisks are for Testi et al.'s (\cite{testi1}, \cite{testi2}) Herbig
Ae/Be sample. Other point are our source sample, where the full diamonds are the
N$_{\star}$ obtained from our clusters simulation analysis; the full
lines passing through the full diamonds represent the total spread of the 
parameters from all SCG models that match the observed KLF and HKCF. Full 
triangles are the same clusters where this time we use the observationally 
derived I$_c$ instead of the model-provided N$_{\star}$. Empty triangles are 
those clusters which were not analyzed with SCG, or which exhibit complete model 
degeneracy; in these cases I$_c$ was used for the cluster membership (from 
Table~\ref{clusters}), while the maximum stellar mass has been derived assuming 
that half of the bolometric luminosity is generated by a single ZAMS star 
(the horizontal lines through the empty triangles represent the mass spread 
assuming that a fraction between 30\% and 100\% of L$_{bol}$ come from a single star)}.  
  \label{herbig}
\end{figure*}

How do our clusters compare to relatively more evolved system, like those observed for a sample of Herbig
Ae/Be by Testi et al. (\cite{testi1}, \cite{testi2}) ? Figure~\ref{herbig} shows the relationship between the mass of the most massive 
source and the total number of cluster stars as provided by the SCG simulations 
for our modeled clusters. The filled symbols represent the clusters which we could model (Tab. \ref{gmdg_restab}); the empty symbols represent the clusters which instead could not be modeled for a variety of reasons (see \S\ref{scgresults}), while Testi et al.'s clusters are reported as 
asterisks (see figure caption for detailed explanation of the symbols). The figure suggests that  the clusters presented in this study are richer than those
surrounding Herbig Ae/Be stars for any given value of the most massive 
star in each cluster. The trend persists if we use similar indicators 
(e.g. I$_c$, the full triangles in the figure).
Furthermore, we note that while
the limiting magnitudes of our observations and those of Testi et
al. (\cite{testi2}) are similar, higher A$_V$ values toward our sources and the
typically greater distance from the Sun would justify the non detection of 
the fainter cluster members predicted by the SCG models.
It is thus likely that the values of I$_c$ derived from our observations
tend to underestimate the cluster memberships.

\begin{figure}[ht]
   \centering
   \resizebox{\hsize}{!}{\includegraphics[angle=90]{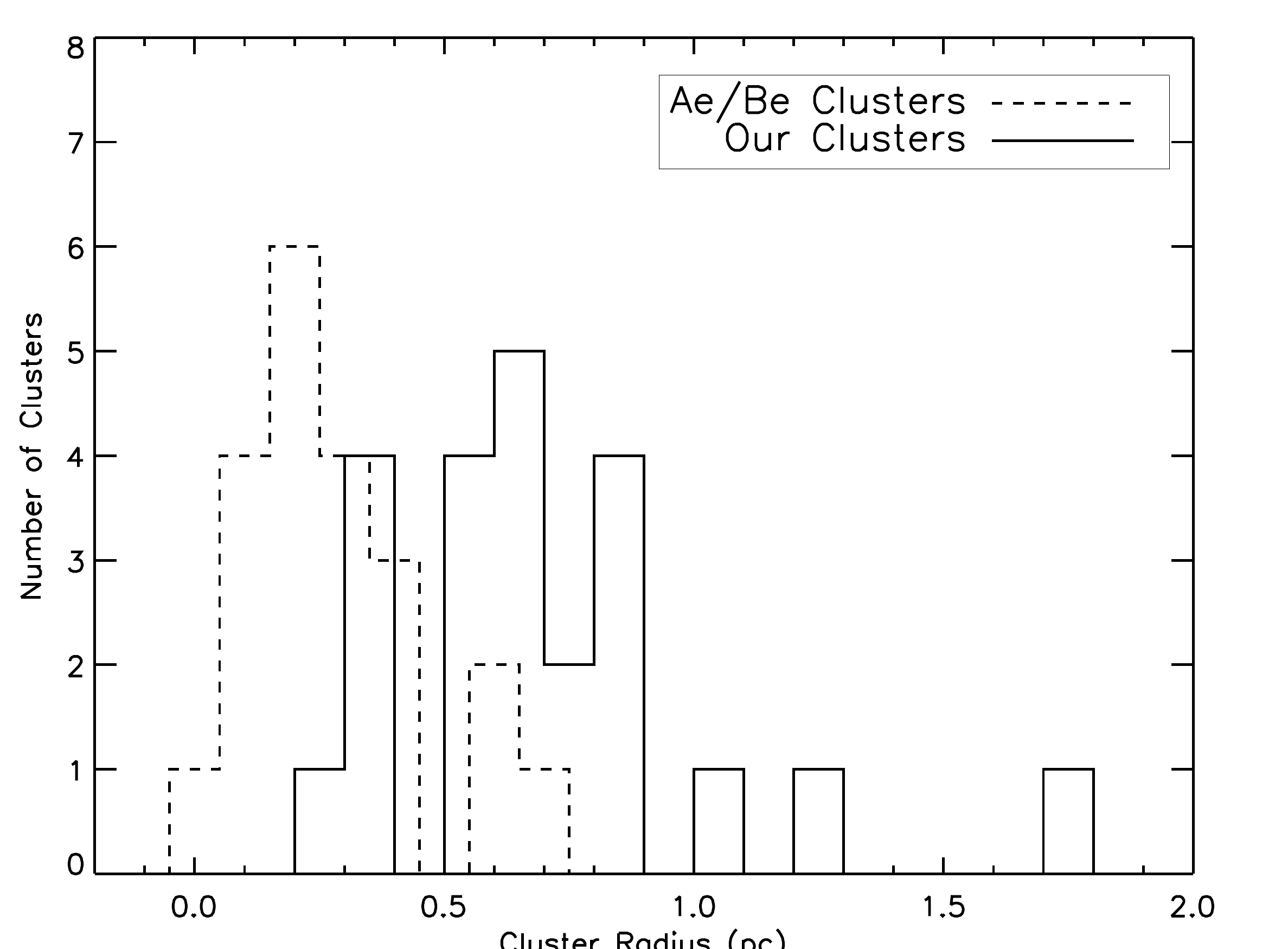}}
  \caption{Distribution of radii for our clusters (full line) 
and those associated with Herbig Ae/Be stars (from Testi et al. (\cite{testi4}), 
dashed line).}
  \label{rclu2}
\end{figure}

This evidence is clear for values of the most massive star in the cluster below $\sim 10$\msun, where the 9 clusters for which we could compare observations with SCG predictions lie (the diamonds). In the clusters for which this could not be done (the empty triangles), an estimate of the mass for the highest-mass star was made by assuming that a fraction between 30 and 100\%\ of the bolometric luminosity is due to a single ZAMS star. In this case the trend for richer clusters compared to Herbig stars (i.e., the asterisks) would tend to become marginal. These estimates, however, place the latter clusters systematically to the right in the plot, compared to the 9 modeled clusters; indeed, if we were to estimate in the same way a maximum stellar mass also for the 9 modeled clusters, we would obtain values in excess (between a factor two and three) to those provided by the detailed SCG modeling. In other words,  the evidence that our clusters are richer than those around Herbig stars is marginal at worst (i.e., using the most conservative approach of estimating the mass of the highest-mass star).

The plausibility of this interpretation is strengthened by the recent results of Baumgardt \& Kroupa (\cite{BK07}) who carried out extensive numerical 
simulations of the evolution of stellar clusters as a function of, among other 
parameters, the cluster gas content. They show that for a wide 
range of initial conditions and star formation efficiencies, the dispersal 
of the gas with age causes the cluster to expand overall and disperse a fraction of 
the stars originally belonging to the cluster. Besides, as the cluster expands its 
decreasing stellar density would make the low-density outer regions of the 
cluster ever more difficult to detect against the field stars (especially 
in the Galactic plane, where all these objects lie), mimicking a smaller 
cluster from an observational viewpoint. Fig.~\ref{rclu2} shows the distribution of radii 
for our clusters (full line) and for those surrounding Ae/Be stars; radii 
have been derived with the same analysis in the two samples and the histogram 
clearly shows that our clusters are indeed larger in size, confirming the 
prediction of Baumgardt \& Kroupa (\cite{BK07}). 

\begin{figure}[t]
   \centering
   \resizebox{\hsize}{!}{\includegraphics[angle=90]{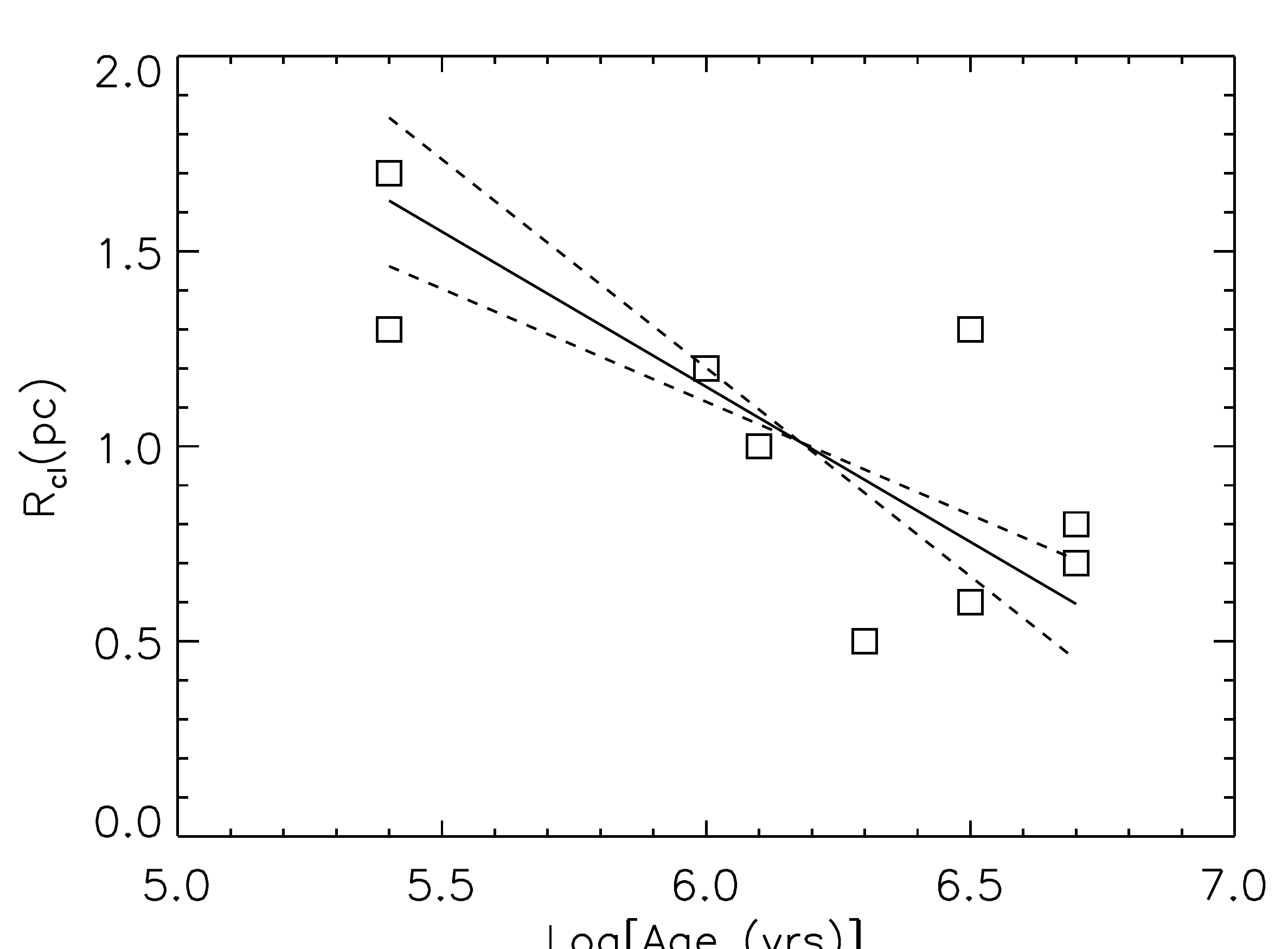}}
  \caption{Cluster radii (from 
Tab.~\ref{clusters}) as a function of the cluster median ages (from 
Tab.~\ref{gmdg_restab}). Dashed lines 
represent the linear fits obtained fitting in turn one plotted variable as a function 
of the other; the full line is the bisector of the two dashed lines and represents 
the fit which minimizes the quadratic geometric (i.e. not along the X or Y axis 
alone) distance of the data from the fit. Following Isobe et al. (\cite{Iso90}), 
this is the adequate approach when the nature of the data scatter around the 
linear fit is not known (and it is not due to classical measurement uncertainties); the slope of the full line is $-0.8\pm 0.2$ and the 1$\sigma$ spread is within the area enclosed by the two dashed lines.}
  \label{age-rclu}
\end{figure}

This age effect on clusters size is also revealed within our clusters sample. Fig. \ref{age-rclu} presents the relationship between the clusters radii as derived from the 
observations and their ages as derived from the modeling. The reported correlation has a Spearman rank correlation coefficient of $\sim -0.6$ indicating a good correlation, with a significance of 92\% (between 2 and 3$\sigma$). Ongoing gas dispersal is certainly plausible in our clusters, given the common detection in these systems of molecular outflows  (Zhang et al. \cite{Zhang01}, \cite{Zhang05}) which are highly effective in 
transferring material away from the forming objects and possibly out of the 
star forming region; parsec-scale flows are indeed commonly observed also from low-mass 
YSOs.

The final stage of gas dispersal, eventually leading to optically revealed clusters like those around Ae/Be stars, might be triggered by the powerful winds and radiation fields from newborn 
O and B stars; indications are (Molinari et al.~\cite{Moli08a}, \cite{Moli08b}) 
that the most massive objects forming in the densest regions of the clumps 
hosting our clusters may have not yet reached the ZAMS, or are just starting to 
develop their H{\sc ii} regions. It is quite likely that this event will 
mark the moment of maximum efficiency of gas dispersal and further 
evolution of our clusters toward the Ae/Be's ones. 

\subsection{Physical \textit{vs.} Statistical Models for Cluster Formation}

Figure~\ref{herbig} can also be used as a diagnostic to discriminate between
different classes of models for the origin of clusters. Testi et al.
(\cite{testi4}) called \textit{physical} the class of models which imply a
physical relationship between the most massive star that forms and other
environmental properties like the cluster richness or the mass of the gaseous
clump where the stars originate from; examples are the "turbulent core" (McKee
\& Tan \cite{MKT03}), the "coalescence" (Bonnell et al., \cite{bonnell}), or
the competitive accretion models (Bonnell \& Bate, \cite{bonnellBa}). In a
second class of models, called \textit{statistical}, the relationship
between the most massive star in a cluster and its richness arises from 
the higher probability of finding the rare massive
stars in rich clusters rather than in isolation (Bonnell \& Clarke
\cite{bonnellCl}). If clusters are populated by randomly
picking stars from the field stars' initial mass function, and considering
a cluster membership-size distribution in the form of an appropriate power law,
then the observations of Testi et al.~(\cite{testi3}) can be naturally
explained. Nevertheless, this model predicts that a
significant fraction of high-mass stars are still associated with relatively
poorly populated clusters, in other words that massive stars can be found
both in high-N clusters and, to a lesser extent, in low-N clusters.

The dashed line in Fig.~\ref{herbig}  is the upper boundary of the region which should contain 25\% of the cluster realizations obtained by randomly extracting stars from the IMF (Testi et al. \cite{testi4}).
Indeed, if we consider our measurements of I$_c$ for our entire sample 
(i.e. the full and empty triangles), there is fraction of about 15\% of the clusters which is found marginally below the dashed line. However, it has to be noted that our modeling was possible only for
clusters above a membership threshold derived with I$_c$, it is thus a somewhat 
biased subsample toward rich clusters. In the extreme assumption that the fields with no detected cluster are cases of systems composed by a single heavily extincted star, and thus would fall below the dashed line in Fig. \ref{herbig}, then this fraction would approach the 25\% level. This, however, is an extreme case because, as we already discussed, the high value of the extinction derived from submillimeter maps may be the 
reason for not detecting clusters in at least a number of observed fields.


\begin{figure}[h]
  \centering
  \resizebox{\hsize}{!}{\includegraphics[angle=90]{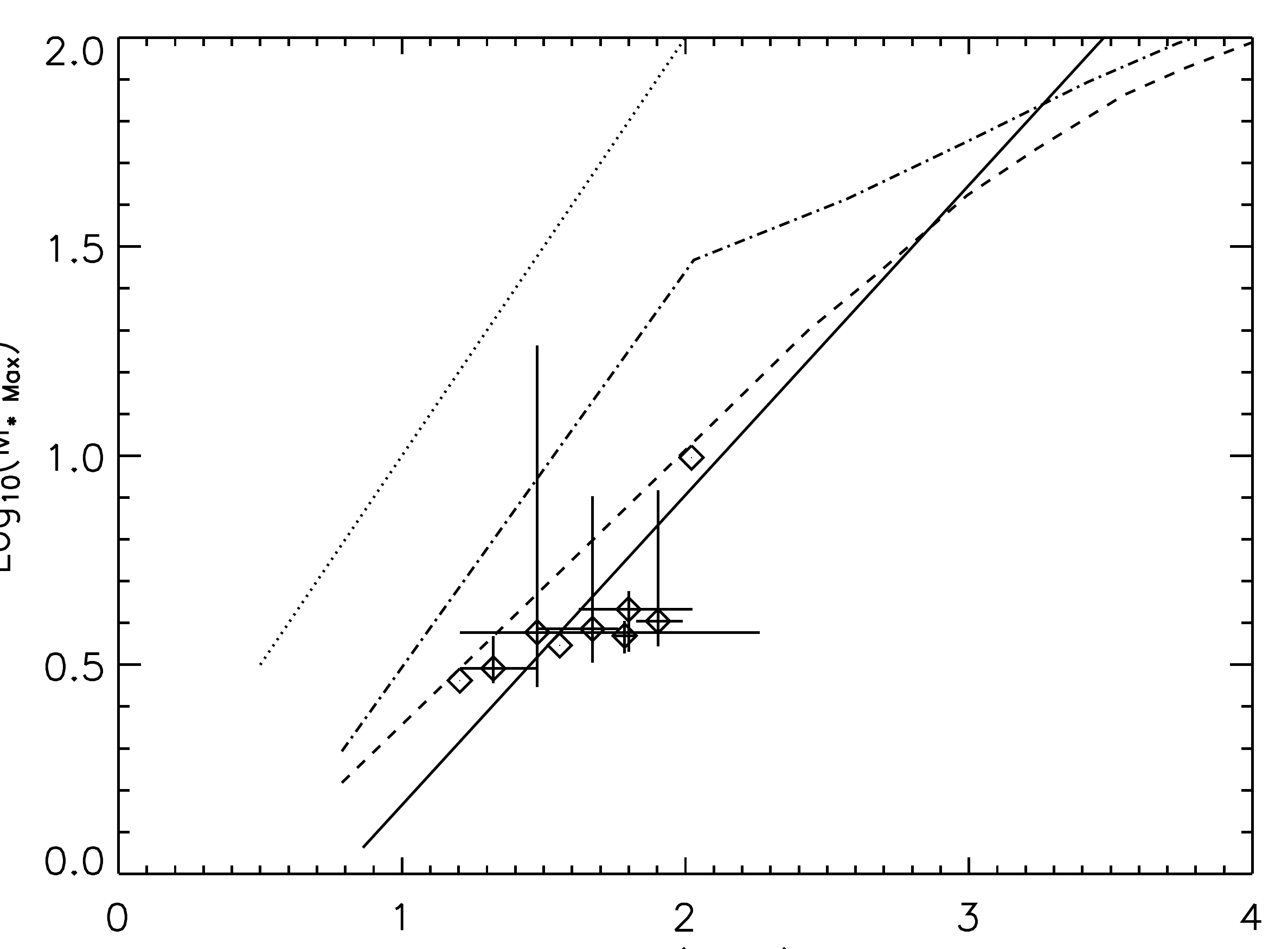}}
  \caption{Mass for the highest mass star as a function of the total stellar 
mass for the 9 modeled clusters (see Table \ref{gmdg_restab}); the bars 
associated to each cluster (the diamond symbols) represent the total span 
of the parameter values for the classes of models selected by our analysis 
(the diamond marks the median values). The dash-dotted, and dashed lines 
represents the relationship obtained for numerical simulations of clusters 
drawn from pure \textit{random sampling} of the IMF, and using a  so-called 
\textit{sorted sampling}, following Weidner \& Kroupa (\cite{WK06}). The 
full line is a semi-analytical approximation of this relationship obtained by Weidner \& Kroupa 
(\cite{WK04}). The dotted line is the limit where a cluster is made of just 
one star.}
  \label{mmax-mstars}
\end{figure}

As a further diagnostic between physical and statistical cluster 
models, Weidner \& Kroupa (\cite{WK06}) recently argued that a non-trivial 
correlation exists between the highest-mass star in a cluster, M$_{\star Max}$, 
and the total stellar mass of the cluster itself, M$_{\star Tot}$ 
(Fig.~\ref{mmax-mstars}). 
Numerical simulations show that the relationship obtained by pure random 
sampling of the IMF with an imposed physical limit of 150~\msun\ for 
the maximum stellar mass (the dashed-dot line 
in Fig.~\ref{mmax-mstars}) is clearly not representing our data.

A substantially different result (the dashed line) is obtained if 
cluster members are picked in ascending order and constrained to total 
cluster masses distributed according to a cluster total mass function 
(Weidner \& Kroupa~\cite{WK06}). Basically, this second option represents 
the fact that drawing 10 cluster of 100 stars will not deliver the same 
M$_{\star Max}$=f(M$_{\star Tot}$) as drawing 1 cluster of 1000 stars. This 
trend closely resembles a semi-analytical approximation of 
M$_{\star Max}$=f(M$_{\star Tot}$) obtained by Weidner \& Kroupa (\cite{WK04}), 
again assuming that total cluster stellar masses are distributed according to a power-law 
mass function. Weidner \& Kroupa (\cite{WK06}) suggest that this 
\textit{sorted sampling} way of populating a cluster can be physically 
understood in terms of a pre-stellar clump where initial low-amplitude 
perturbations start low-mass star formation; as further perturbations 
with larger amplitude grow, higher mass stars will start to form until 
the feedback from the latter will start disrupting the cloud. This scenario 
of an organised star formation where low-mass stars are the first ones to 
form, is the same we favor (see \S\ref{history}) considering the age 
spread we find in our clusters in which, based on independent considerations 
(Molinari et al.~\cite{Moli08a}), the most massive star may have not been 
formed at all. By the way, this latter possibility does not change the substance of 
the agreement between our data and the \textit{physical} cluster models 
in Fig.~\ref{mmax-mstars}, since the late addition of the highest mass star 
currently not yet visible in the near-IR would shift the points toward the 
top-right of the plot.

We verified \textit{a posteriori} that the range of M$_{\star Max}$ and M$_{\star Tot}$ 
parameters values explored by our simulations is much wider than the area spanned by the bars attached to the single points, and also includes regions compatible with the random sampling cluster model. We then conclude that the result of Fig. \ref{mmax-mstars} is not likely to be produced by 
a biased sampling of the clusters' physical parameters explored in our models.

The predictions of the \textit{sorted sampling} above descibed, with which our data points best agree, are also in good agreement with the results from 
simulations of clusters forming in a competitive accretion scenario 
(Bonnell, Vine \& Bate~\cite{bonnellVnBa}). This model, however, seems 
excluded by the observed ages and age spreads in our clusters which are 
in clear disagreement with the predicted cluster formation timescales 
of 2$\div$3 free-fall times.

\subsection{Influence of Binarity on the Interpretation of Age Spread}

\begin{figure}
  \centering
  \resizebox{\hsize}{!}{\includegraphics[angle=90]{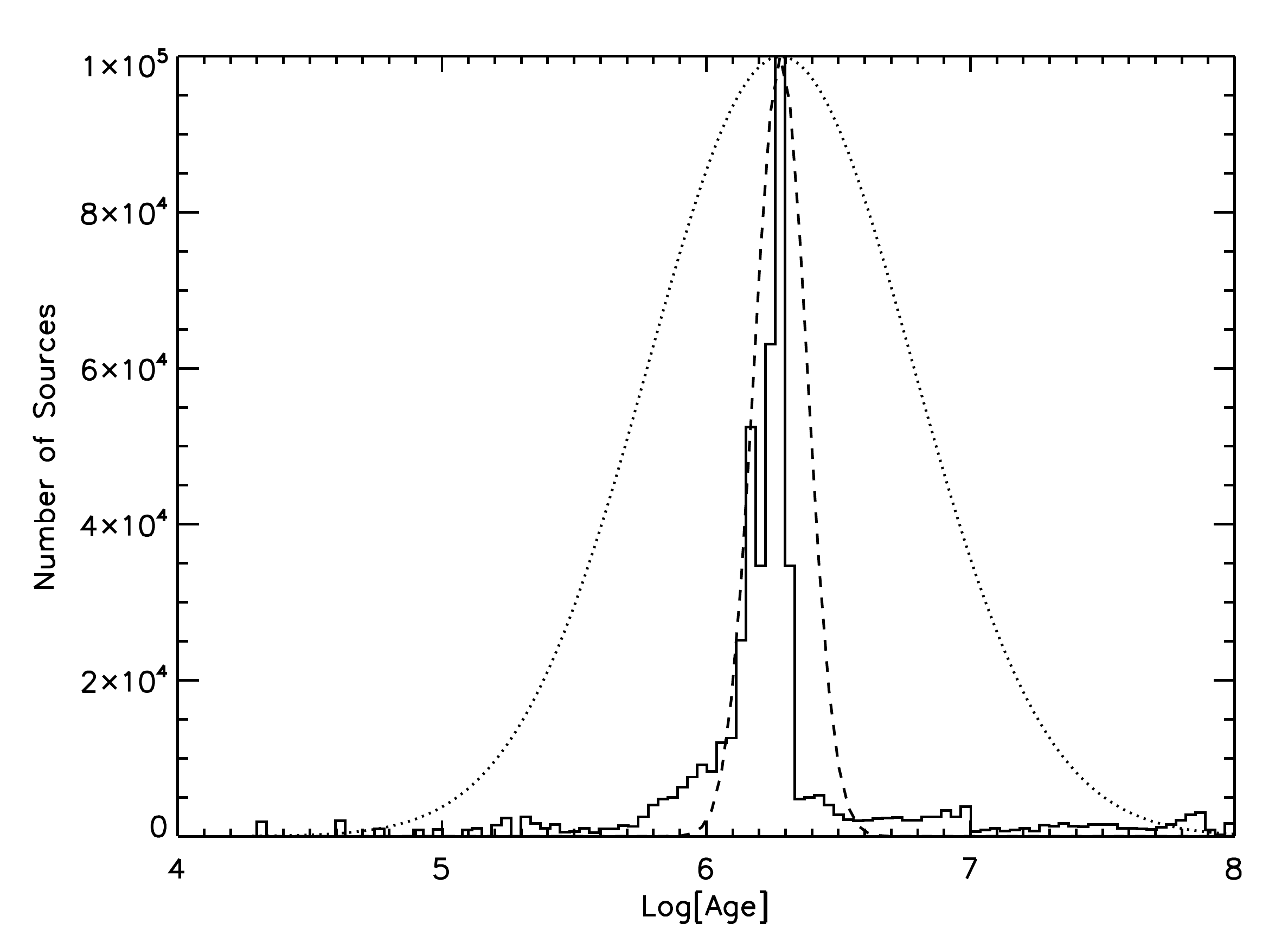}}
  \caption{Full line represents the age spread resulting from the simulations of Weidner, Kroupa and Maschberger 
(\cite{Weid}) for a cluster with an input age of $2 \times 10^{6}$~yrs and  
100\% binarity fraction. The two gaussians are the age weight functions used in our 
simulations of Gaussian Star Formation Histories, with $\sigma$=0.1 (dashed 
line) and 0.5 (dotted line) respectively. In case of constant SFH models, the 
adopted age spread is comparable to the $\sigma$=0.5 distribution in the figure.}
  \label{agespread}
\end{figure}

Weidner, Kroupa \& Maschberger (\cite{Weid}) carried out extensive 
numerical simulations  to determine how the presence of unresolved binary/multiple 
stars can affect the observational properties of a young cluster in a massive 
star forming region.  Assuming a 100\% of binarity in a cluster of coeval sources, 
they find that unresolved binaries may lead an observer to conclude that instead 
a significant age spread is present in the cluster; the full line in 
Fig.~\ref{agespread} shows the \textit{a posteriori} age determination assuming 
that all binaries are unresolved. We see that the measured age spread for the 
large majority of stars simulated in Weidner et al.'s simulation is comparable 
to a Log(age) Gaussian distribution with $\sigma$=0.1, which is one of the 
possible choices of Star Formation Histories in our cluster models. However, 
the comparison of our observed KLFs with the SCG models (Table~\ref{gmdg_restab}) 
suggests age spreads much bigger than this, and more comparable to a Log(age) 
distribution with $\sigma$=0.5 (the dotted line in Fig. \ref{agespread}). 
We then conclude that unresolved binaries in our clusters cannot account 
for the observed age spread.

\section{Conclusions}

The main results of this work are the following:

\begin{itemize}

 \item We have imaged in the J, H, and \ks\ NIR bands 26 intermediate and
high-mass star forming regions selected from a larger sample of sources and
spanning a range in luminosities and presumed youth. We have identified
the presence of 23 young stellar clusters in 22 fields.

\item Revealed clusters have richness indicator values
between ten and several tens of objects and have median radius of 0.7~pc. 
Compared to clusters around Herbig Ae/Be stars, our clusters seem 
richer and larger for any given mass for the most massive star in each 
cluster. Color-color diagrams show that these clusters are young: many sources
show colors typical of young pre-MS objects with an intrinsic IR excess arising
from warm circumstellar dust. This is confirmed by the analysis of
color-magnitude diagrams where a significant fraction of stars in each cluster
are found in correspondence to the Pre-MS evolutionary tracks even after conservative 
de-reddening is applied.

 \item We cannot perform a direct inversion of stellar luminosities (and
colors) into masses
and ages; we use a Synthetic cluster Generator (SCG) model to create statistically-significant 
cluster simulations for different initial parameters (IMF,
SFH, source ages and their distribution), and compare the synthetic KLFs and
HKCFs with the observed (field star-subtracted) ones. For the fraction of clusters for which this comparison selects a well-confined region of the parameter space, we conclude that
star formation in these regions \textit{cannot} be represented with a single burst, 
but is a process that is spread out in time. Clusters have mean ages of 
a few $10^{6}$~yrs; the ages of most of the clusters members are spread, 
within each cluster, between a few $10^{5}$~yrs to a few
$10^{6}$~yrs. Together with the independent evidence that the most 
massive stars in these systems are very young, or not even on the ZAMS yet, 
this result is difficult to reconcile with any model predicting cluster 
formation in a crossing time.

\item The cluster radii seem to be inversely proportional to their age, 
as also confirmed by the comparison of cluster parameters with those typical 
of Ae/Be systems, which are smaller and less rich. As suggested by numerical 
simulations in the literature, dispersal of intra-cluster gas (by, e.g., 
molecular outflows or radiation fields from massive stars) may lead to the 
loss of a fraction of cluster stellar population, thus indeed leading to smaller and less rich clusters. 
Our results seem in line with this prediction.

\item The relation between the mass of the most massive
star in a cluster and the cluster's richness indicator suggests that a  
\textit{physical} rather than \textit{statistical} nature of the cluster 
origin is more likely. 

\end{itemize}


\begin{acknowledgements}
We thank the anonymous referee whose comments greatly helped improving the 
quality of the data presentation and the overall readability of the paper. 
We also thank P. Kroupa for his comments and suggestions.
\end{acknowledgements}

\clearpage

\Online

\begin{appendix}
\section{K$_s$ images for all observed fields}
\label{app_images}

This appendix contains the K$_s$ images for all observed fields; these are also available at http://galatea.ifsi-roma.inaf.it/faustini/K-images/

\begin{figure*}[h]
  \centering
  \resizebox{16cm}{!}{\includegraphics[angle=-90]{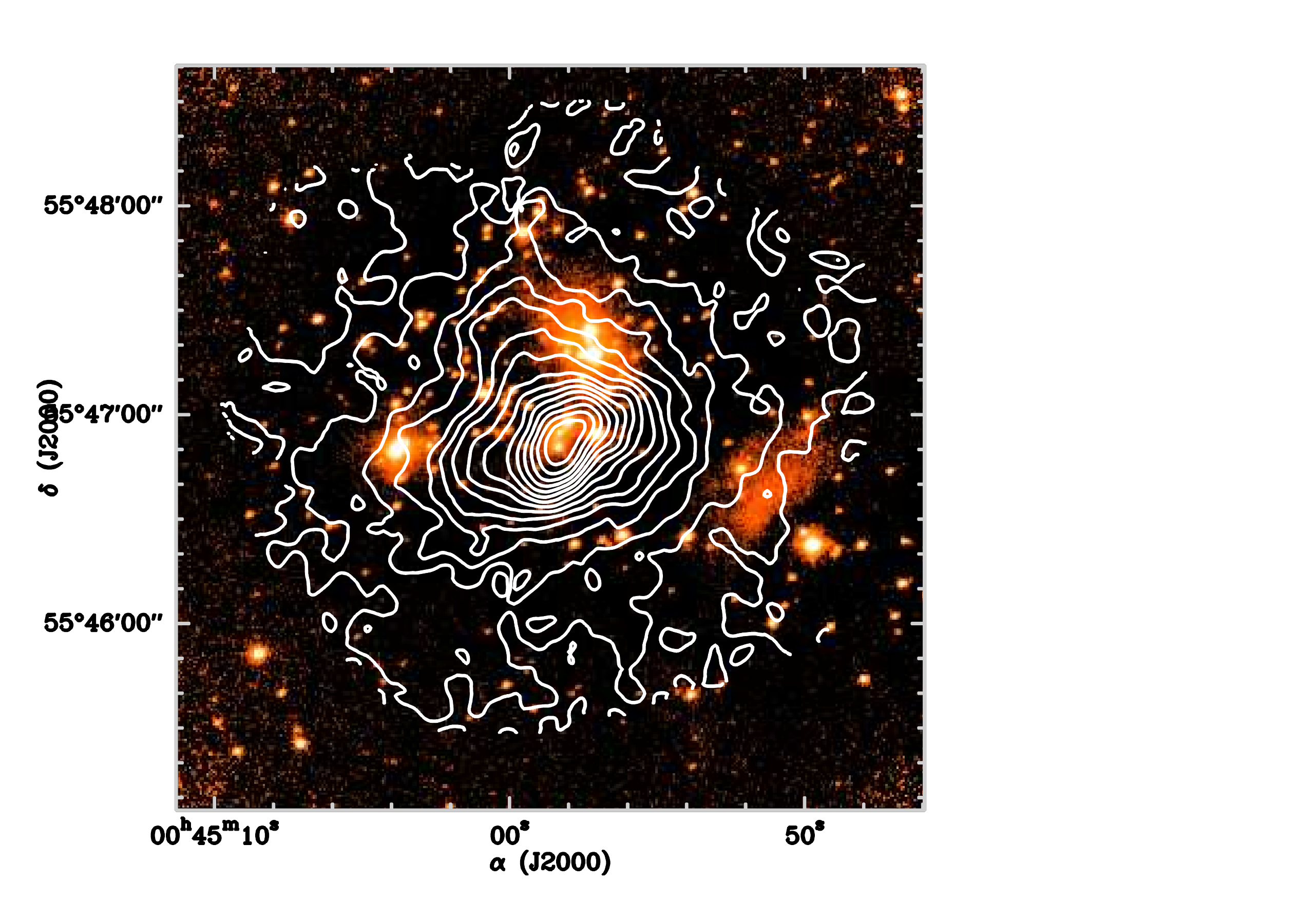}} 
  \caption{K$_s$ image of field Mol003 with superimposed SCUBA 850\um\ continuum in white contours (Molinari et al. \cite{Moli08a}).}
  \label{map_mol003}
\end{figure*}

\begin{figure*}[h]
  \centering
\vspace{-2cm}
  \resizebox{16cm}{!}{\includegraphics[angle=-90]{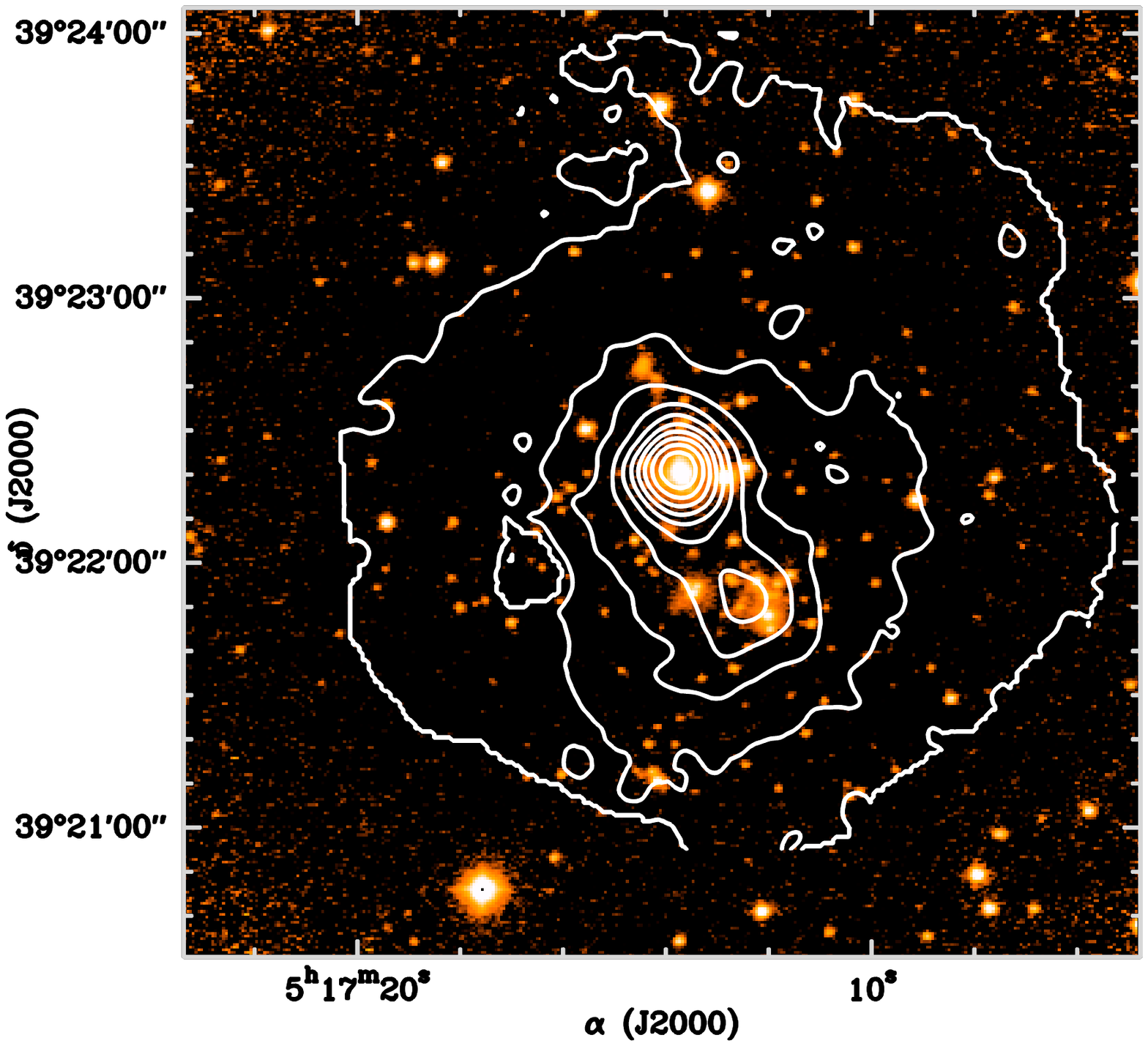}}
  \caption{K$_s$ image of field Mol008 with superimposed SCUBA 850\um\ continuum in white contours (Molinari et al. \cite{Moli08a}).}
  \label{map_mol008}
\end{figure*}

\clearpage

\begin{figure*}[h]
  \centering
  \resizebox{16cm}{!}{\includegraphics[angle=-90]{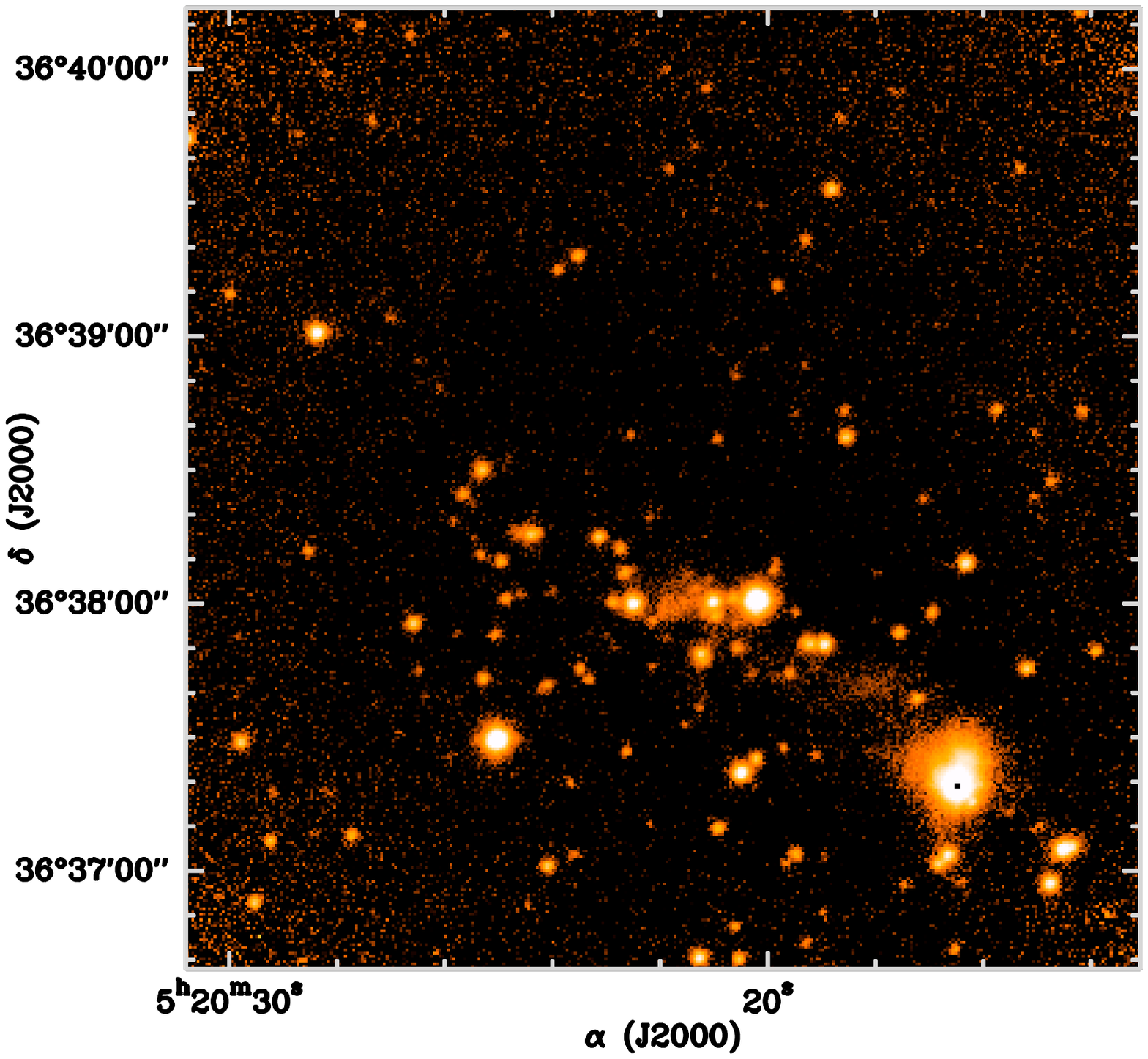}} 
  \caption{K$_s$ image of field Mol009; no submillimeter image is available for this field.}
  \label{map_mol009}
\end{figure*}

\begin{figure*}[h]
  \centering
\vspace{-2cm}
  \resizebox{16cm}{!}{\includegraphics[angle=-90]{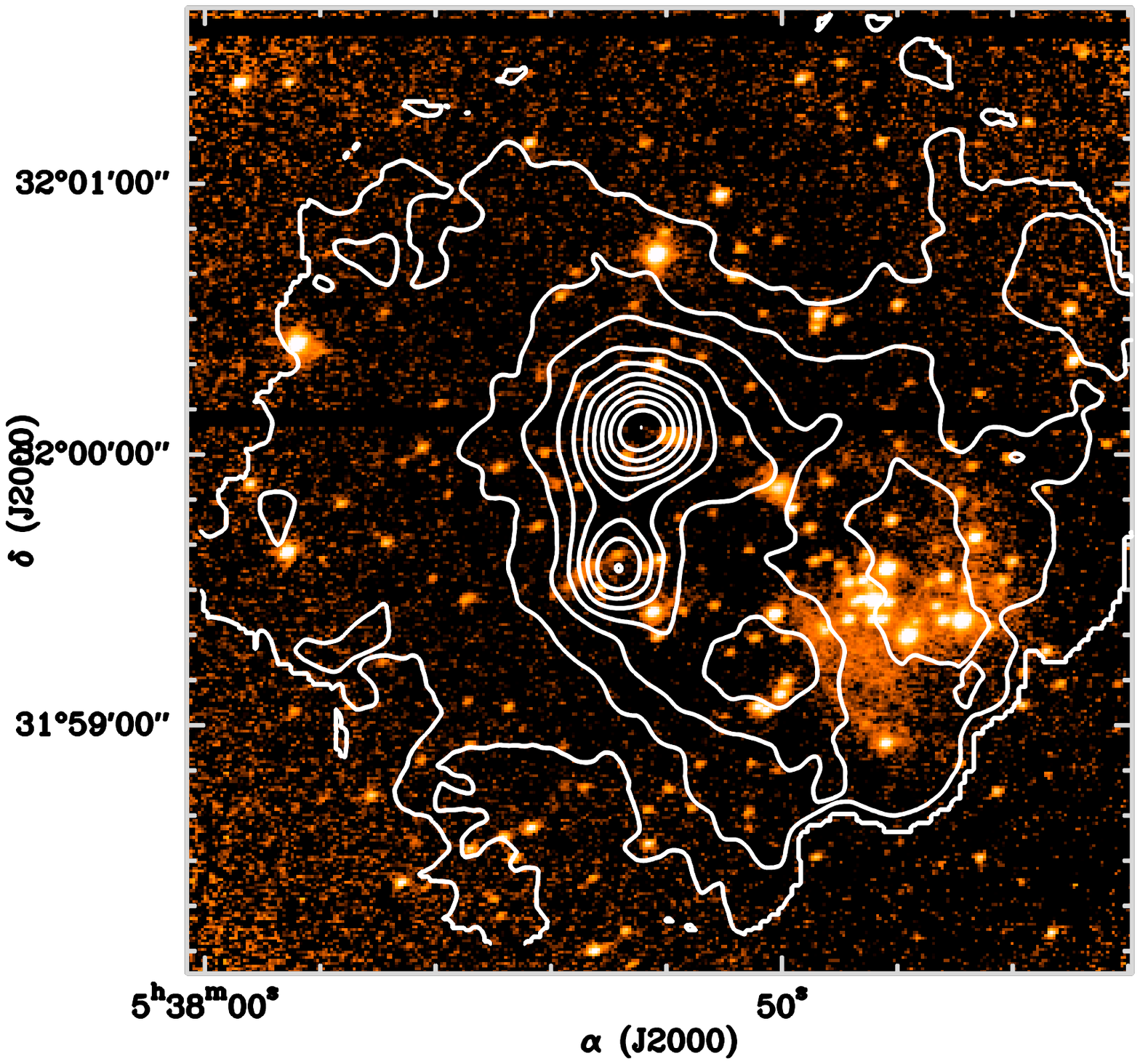}}
  \caption{K$_s$ image of field Mol011 with superimposed SCUBA 850\um\ continuum in white contours (Molinari et al. \cite{Moli08a}).}
  \label{map_mol011}
\end{figure*}

\clearpage

\begin{figure*}[h]
  \centering
  \resizebox{16cm}{!}{\includegraphics[angle=-90]{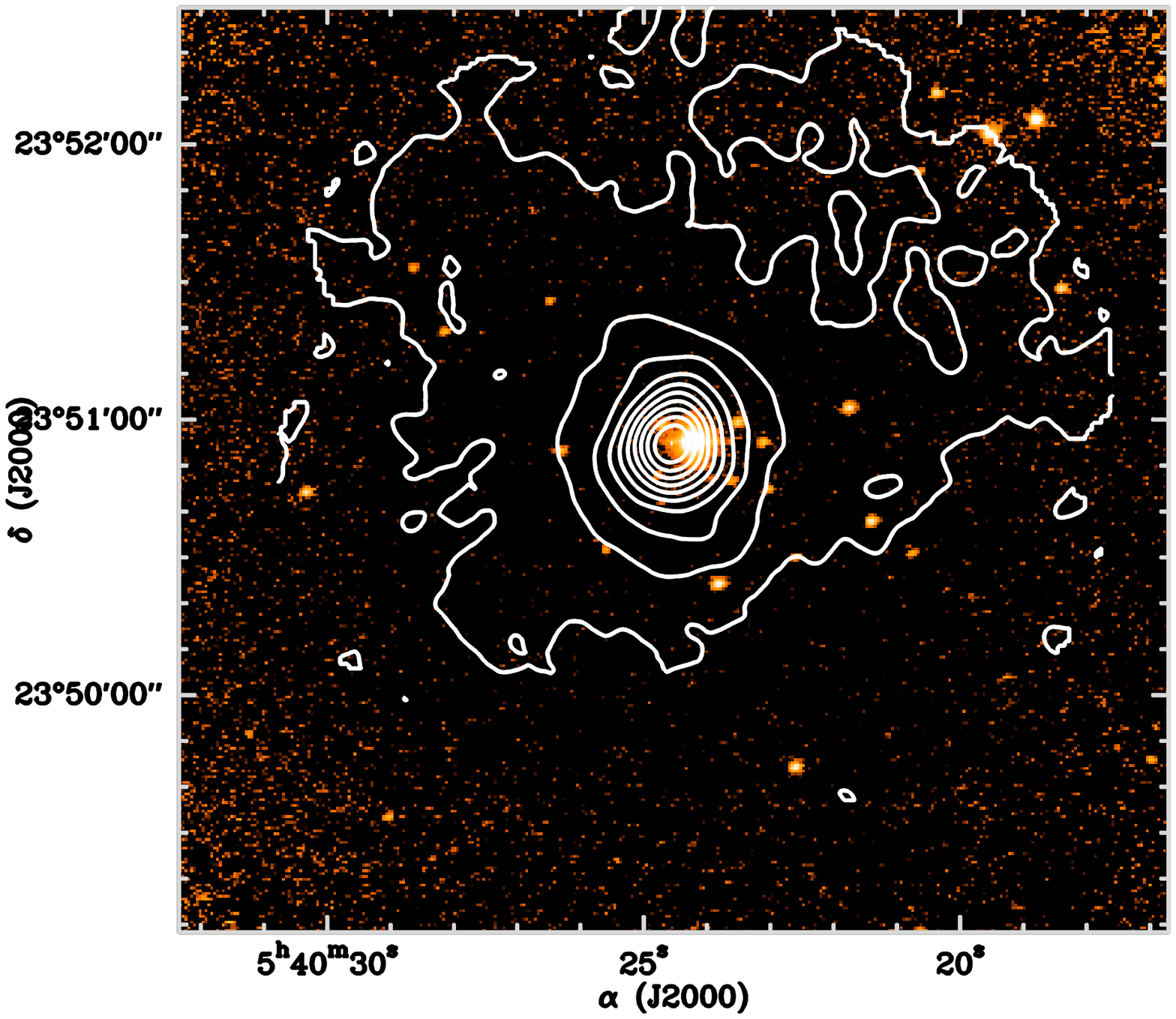}}
  \caption{K$_s$ image of field Mol012 with superimposed SCUBA 850\um\ continuum in white contours (Molinari et al. \cite{Moli08a}).}
  \label{map_mol012}
\end{figure*}

\begin{figure*}[h]
  \centering
\vspace{-2cm}
  \resizebox{16cm}{!}{\includegraphics[angle=-90]{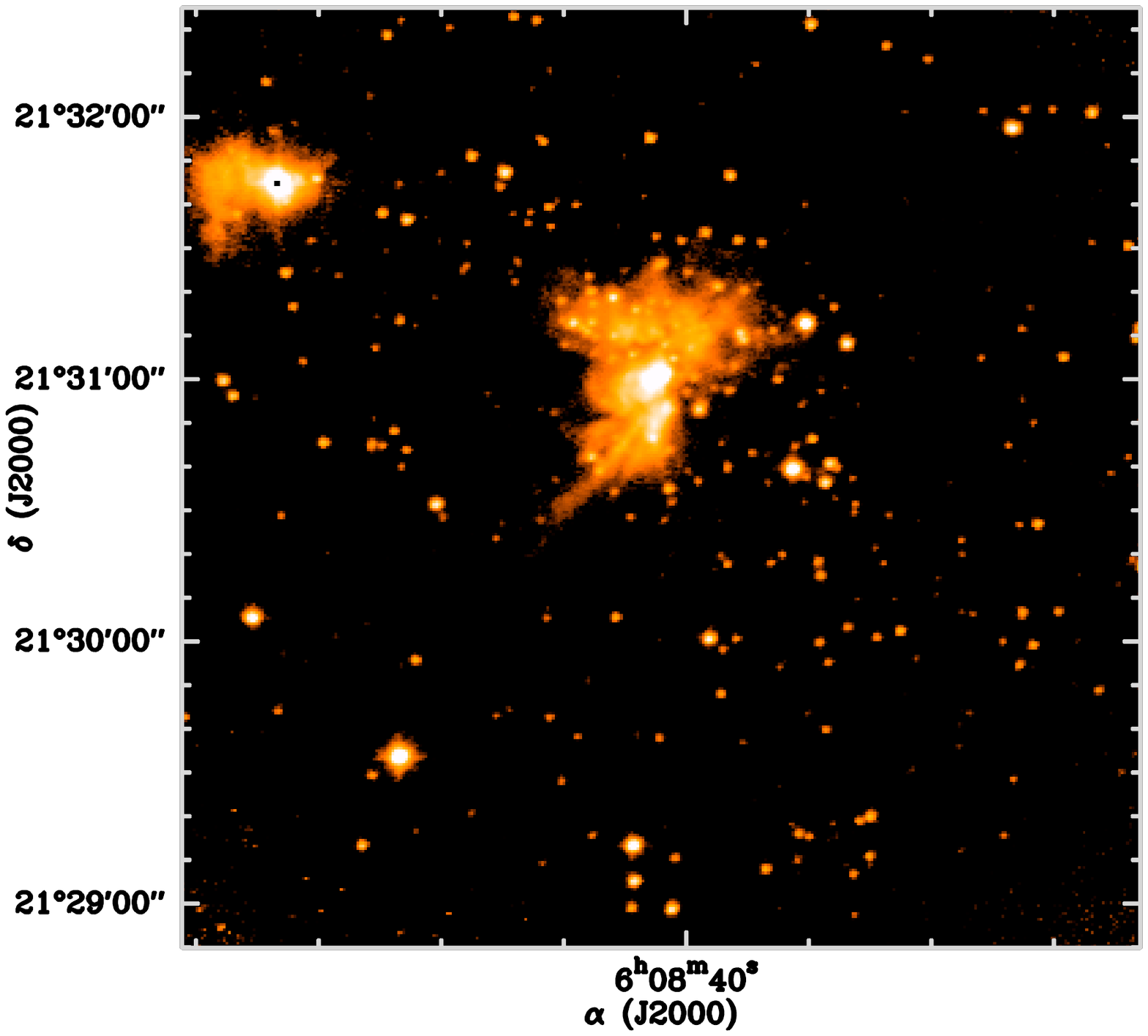}} 
  \caption{K$_s$ image of field Mol015; no submillimeter image is available for this field.}
  \label{map_mol015}
\end{figure*}

\clearpage

\begin{figure*}[h]
  \centering
  \resizebox{16cm}{!}{\includegraphics[angle=-90]{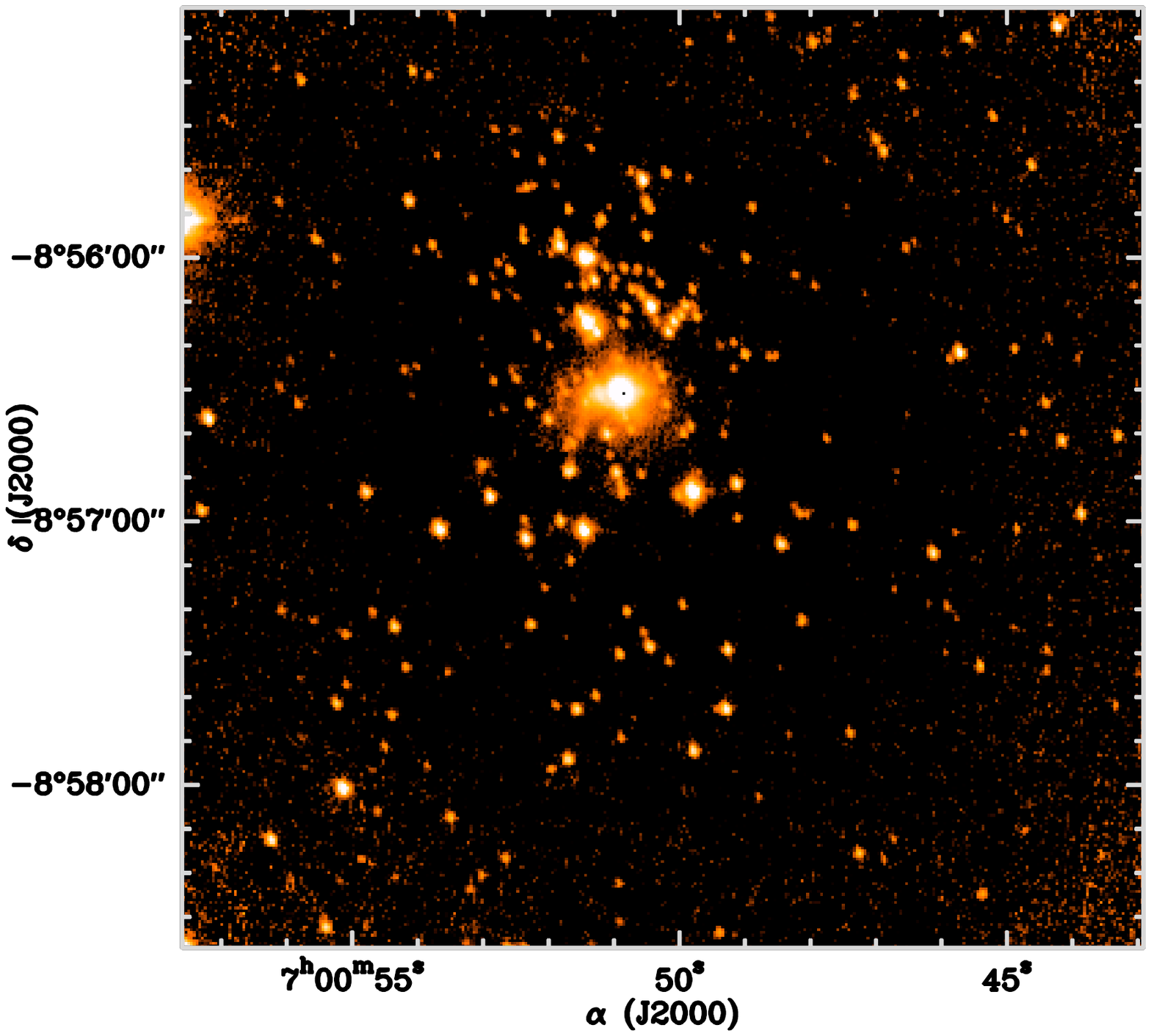}} 
  \caption{K$_s$ image of field Mol028; no submillimeter image is available for this field.}
  \label{map_mol028}
\end{figure*}

\begin{figure*}[h]
  \centering
\vspace{-2cm}
  \resizebox{16cm}{!}{\includegraphics[angle=-90]{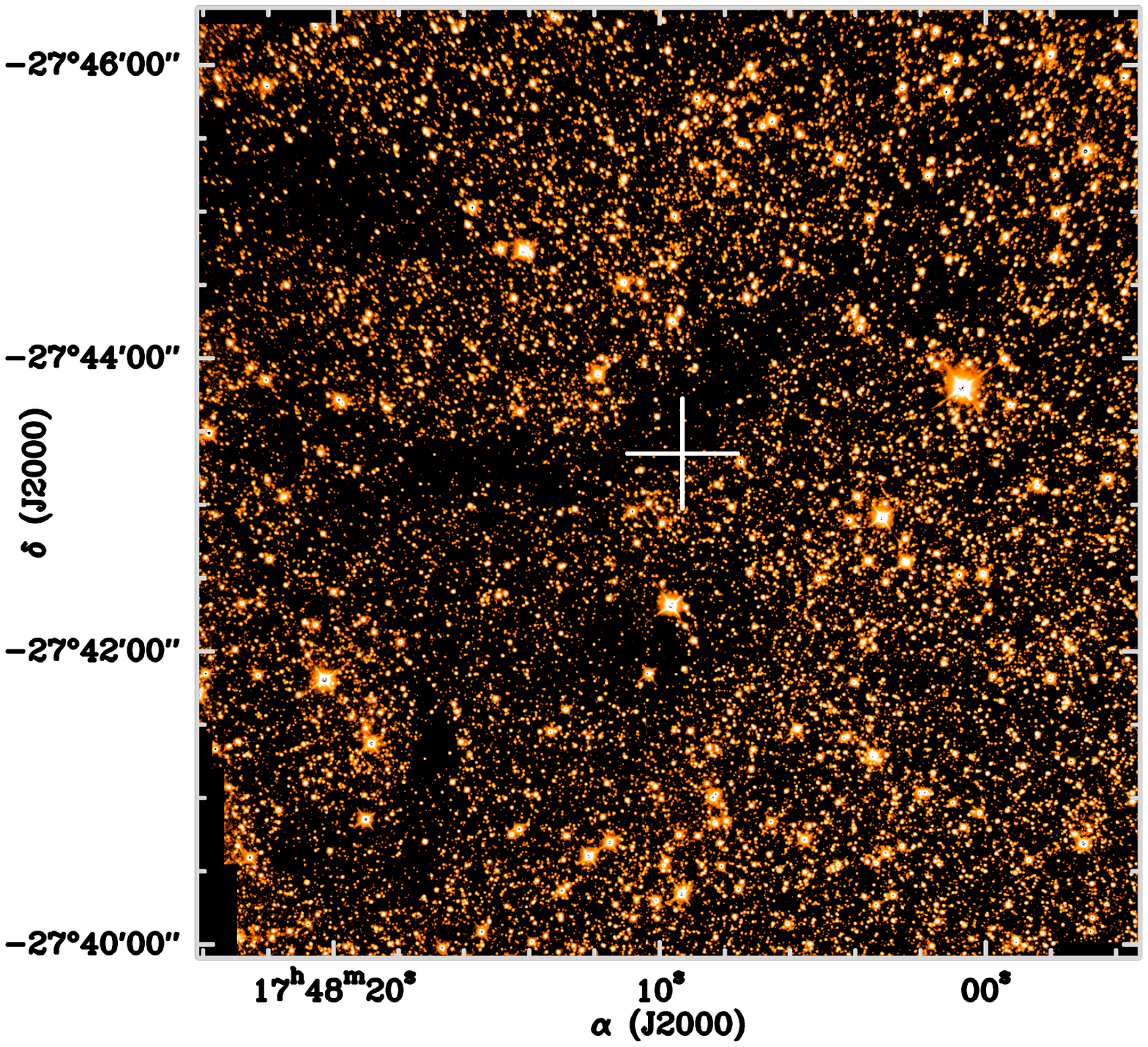}} 
  \caption{K$_s$ image of field Mol030; no submillimeter image is available for this field. The white cross marks the position of the IRAS source.}
  \label{map_mol030}
\end{figure*}

\clearpage

\begin{figure*}[h]
  \centering
  \resizebox{16cm}{!}{\includegraphics[angle=-90]{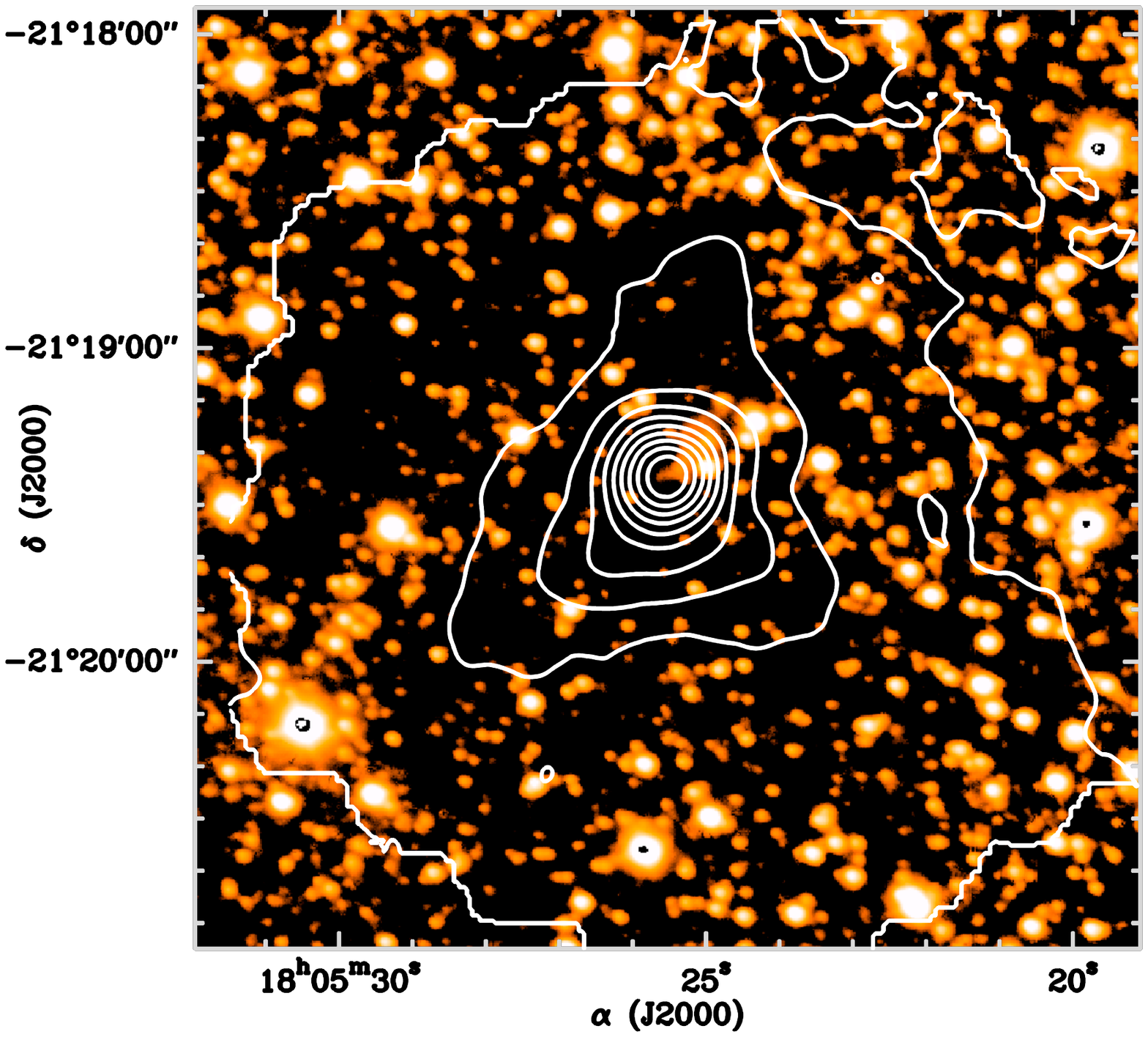}}
  \caption{K$_s$ image of field Mol038 with superimposed SCUBA 850\um\ continuum in white contours (Molinari et al. \cite{Moli08a}).}
  \label{map_mol038}
\end{figure*}

\begin{figure*}[h]
  \centering
\vspace{-2cm}
  \resizebox{16cm}{!}{\includegraphics[angle=-90]{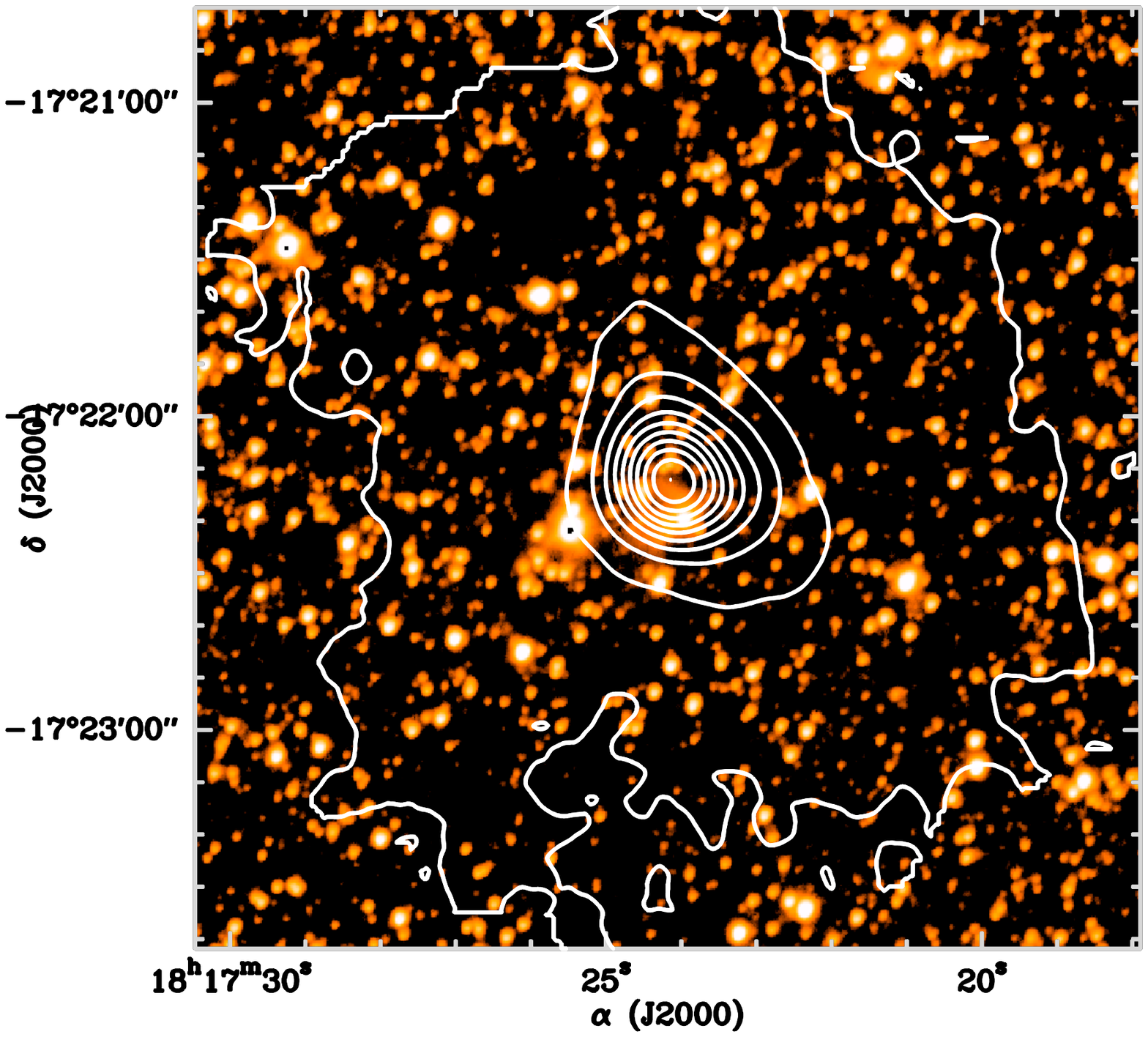}}
  \caption{K$_s$ image of field Mol045 with superimposed SCUBA 850\um\ continuum in white contours (Molinari et al. \cite{Moli08a}).}
  \label{map_mol045}
\end{figure*}

\clearpage

\begin{figure*}[h]
  \centering
  \resizebox{16cm}{!}{\includegraphics[angle=-90]{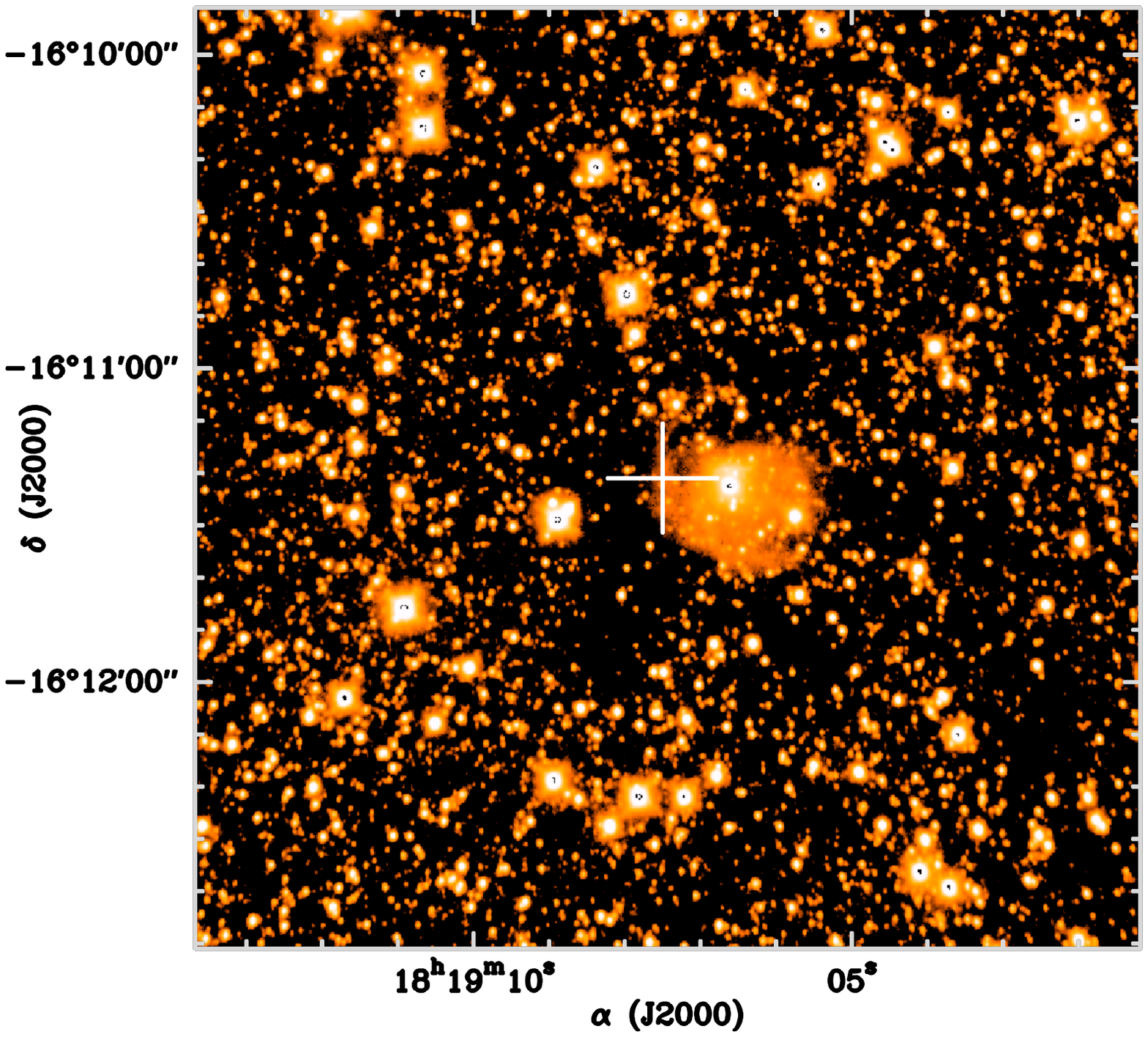}} 
  \caption{K$_s$ image of field Mol050; no submillimeter image is available for this field. The white cross marks the position of the IRAS source.}
  \label{map_mol050}
\end{figure*}

\begin{figure*}[h]
  \centering
\vspace{-2cm}
  \resizebox{16cm}{!}{\includegraphics[angle=-90]{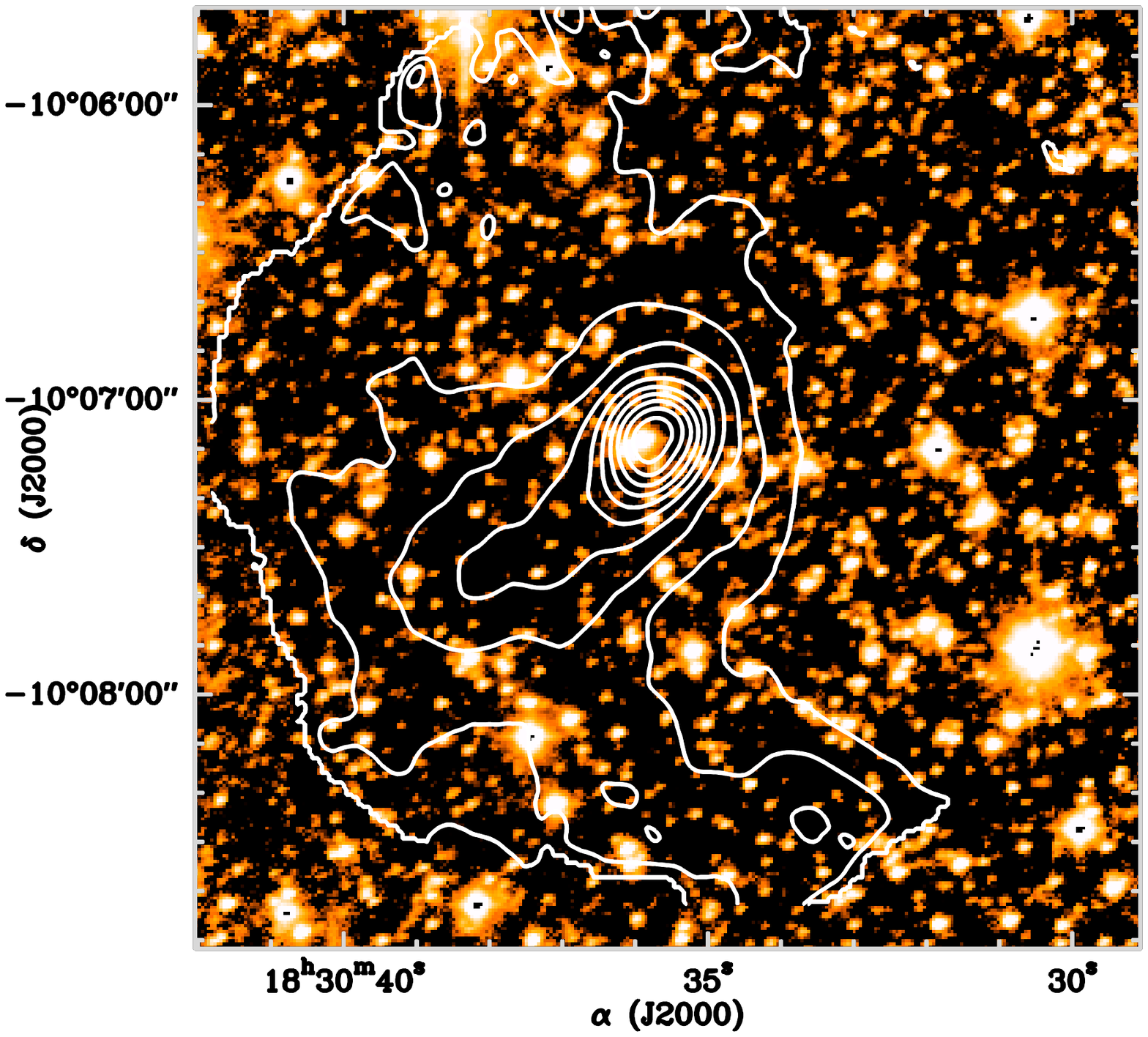}}
  \caption{K$_s$ image of field Mol059 with superimposed SCUBA 850\um\ continuum in white contours (Molinari et al. \cite{Moli08a}).}
  \label{map_mol059}
\end{figure*}

\clearpage

\begin{figure*}[h]
  \centering
  \resizebox{16cm}{!}{\includegraphics[angle=-90]{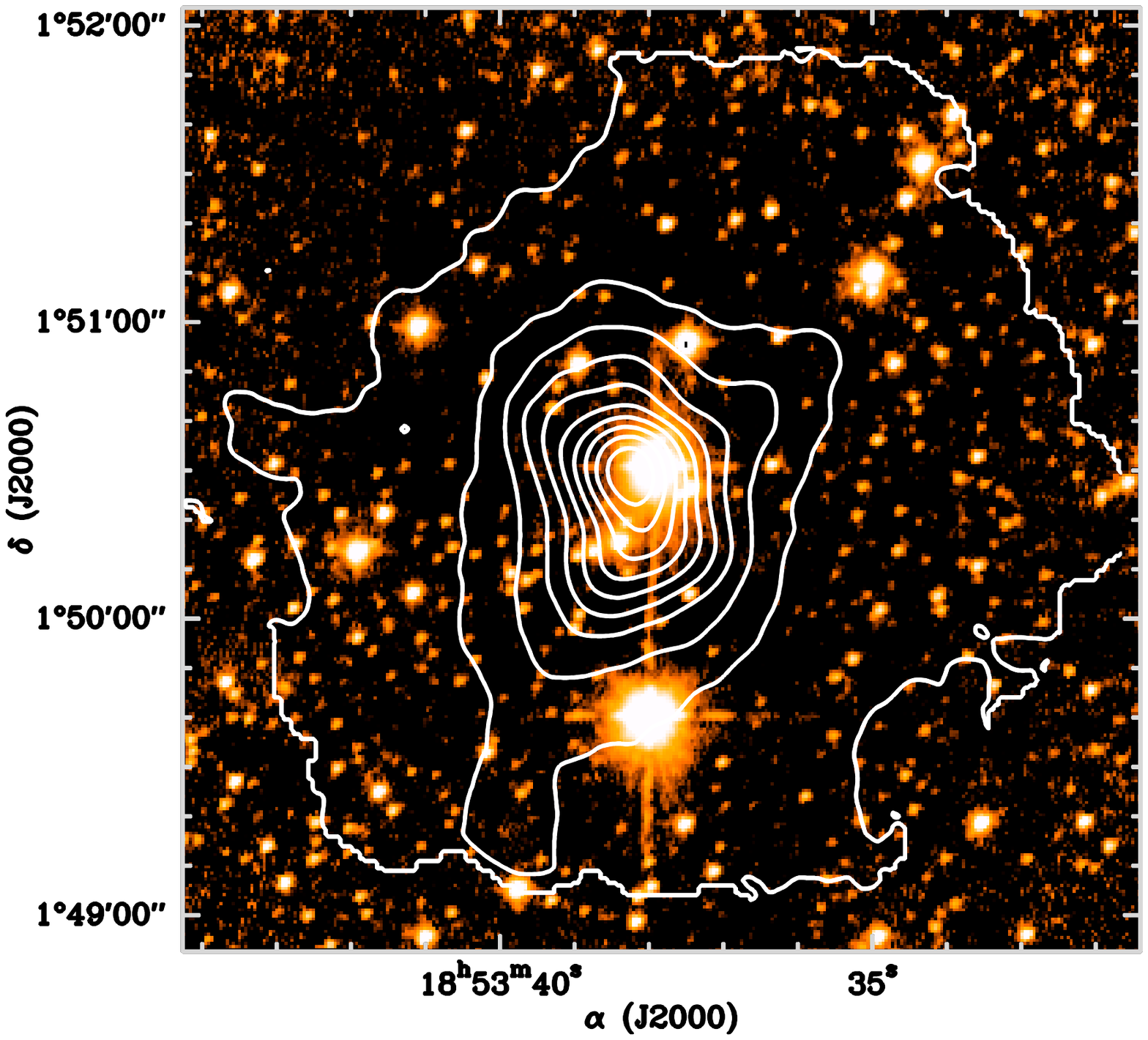}}
  \caption{K$_s$ image of field Mol075 with superimposed SCUBA 850\um\ continuum in white contours (Molinari et al. \cite{Moli08a}).}
  \label{map_mol075}
\end{figure*}

\begin{figure*}[h]
  \centering
\vspace{-2cm}
  \resizebox{16cm}{!}{\includegraphics[angle=-90]{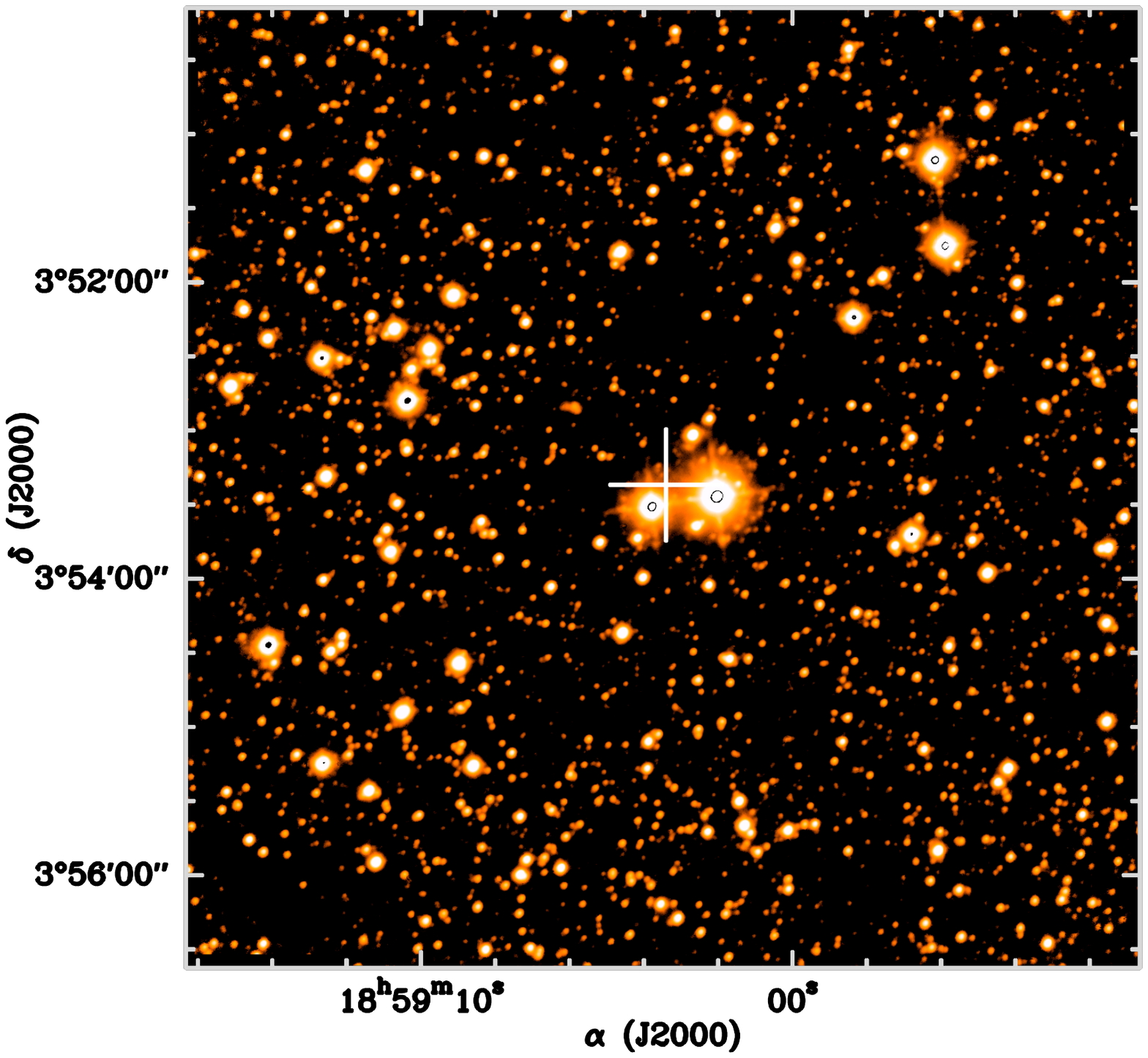}} 
  \caption{K$_s$ image of field Mol082; no submillimeter image is available for this field. The white cross marks the position of the IRAS source.}
  \label{map_mol082}
\end{figure*}

\clearpage

\begin{figure*}[h]
  \centering
  \resizebox{16cm}{!}{\includegraphics[angle=-90]{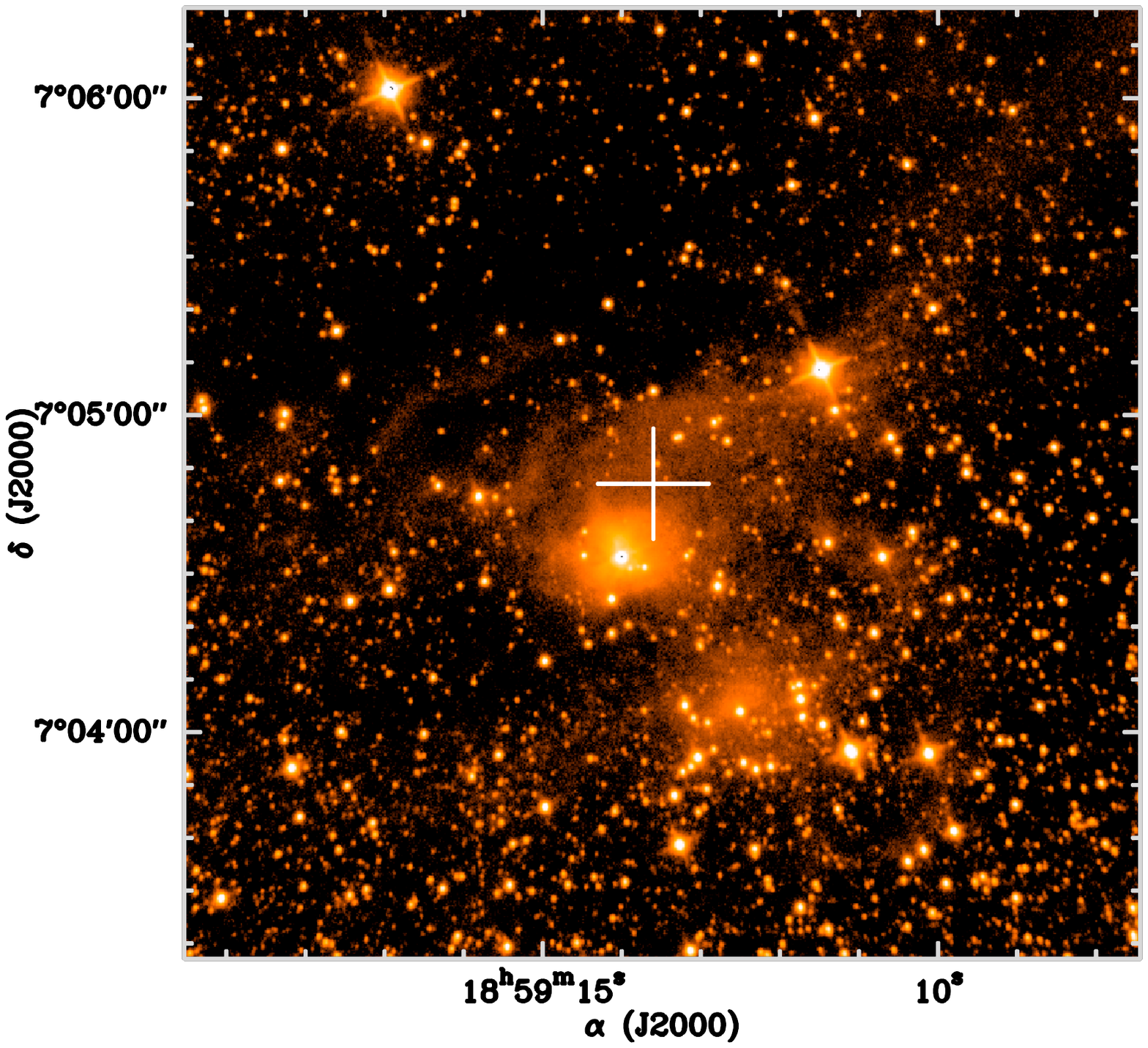}} 
  \caption{K$_s$ image of field Mol084; no submillimeter image is available for this field. The white cross marks the position of the IRAS source.}
  \label{map_mol084}
\end{figure*}

\begin{figure*}[h]
  \centering
\vspace{-2cm}
  \resizebox{16cm}{!}{\includegraphics[angle=-90]{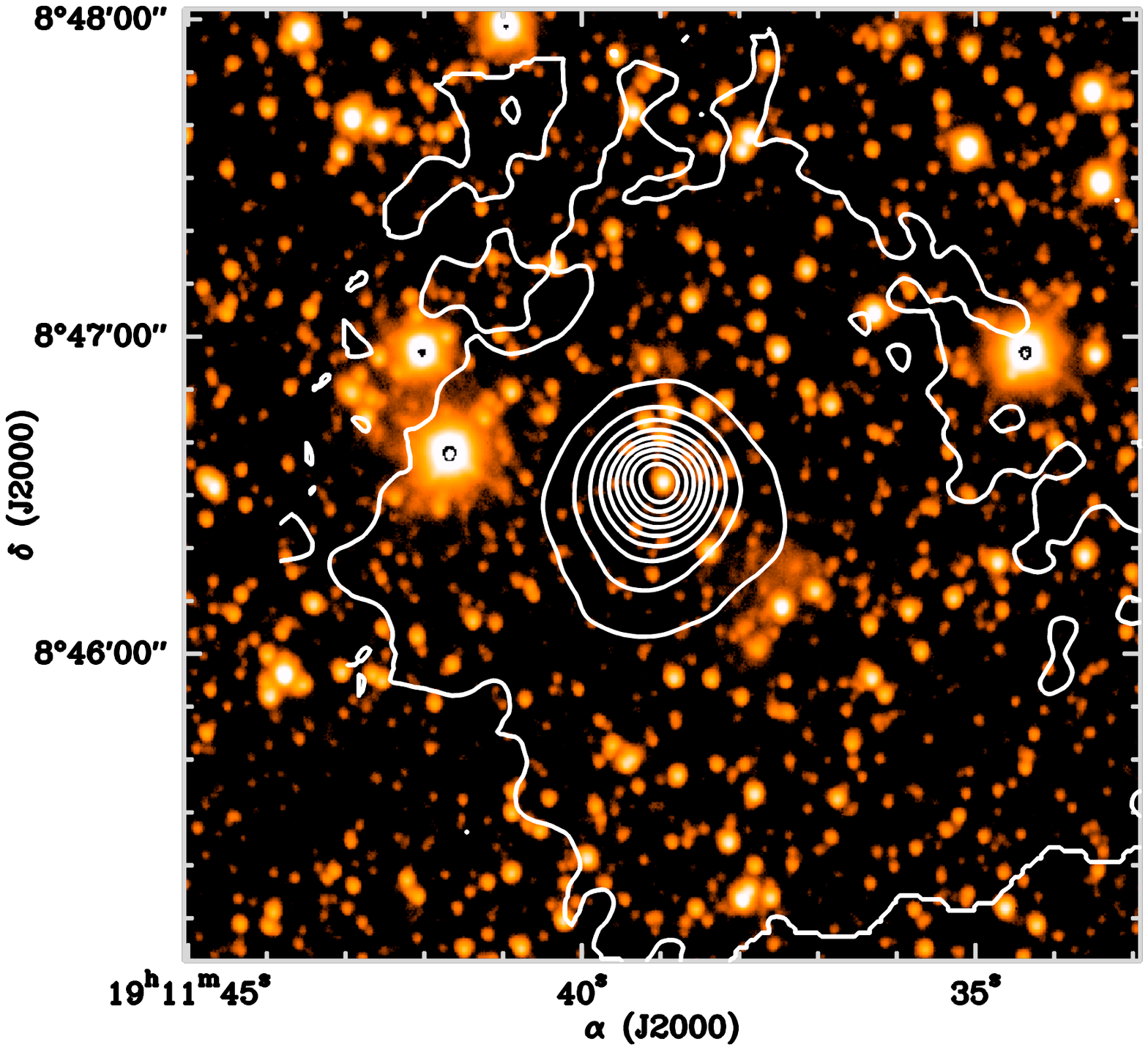}}
  \caption{K$_s$ image of field Mol098 with superimposed SCUBA 850\um\ continuum in white contours (Molinari et al. \cite{Moli08a}).}
  \label{map_mol098}
\end{figure*}

\clearpage

\begin{figure*}[h]
  \centering
  \resizebox{16cm}{!}{\includegraphics[angle=-90]{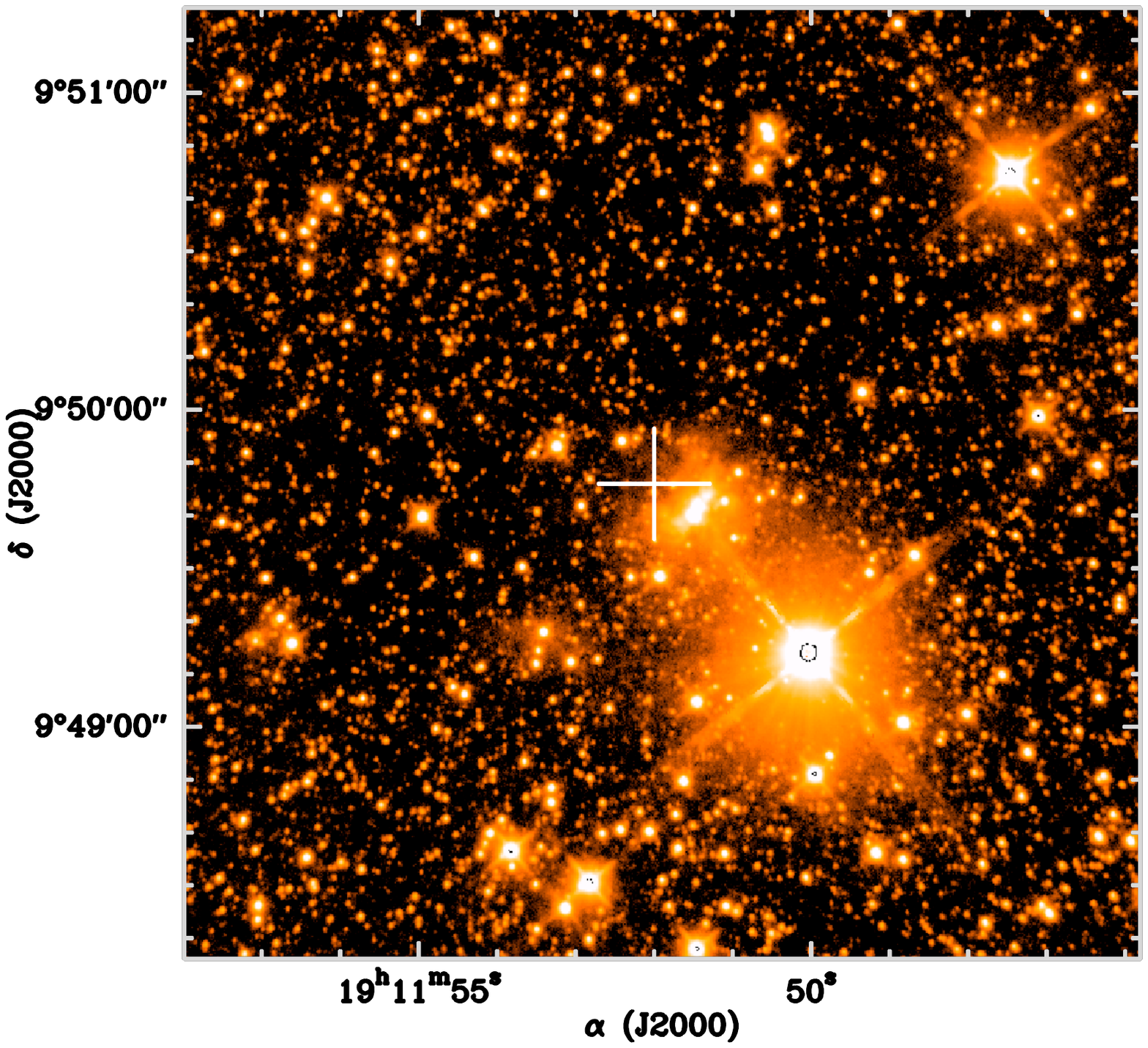}} 
  \caption{K$_s$ image of field Mol099; no submillimeter image is available for this field. The white cross marks the position of the IRAS source.}
  \label{map_mol099}
\end{figure*}

\begin{figure*}[h]
  \centering
\vspace{-2cm}
  \resizebox{16cm}{!}{\includegraphics[angle=-90]{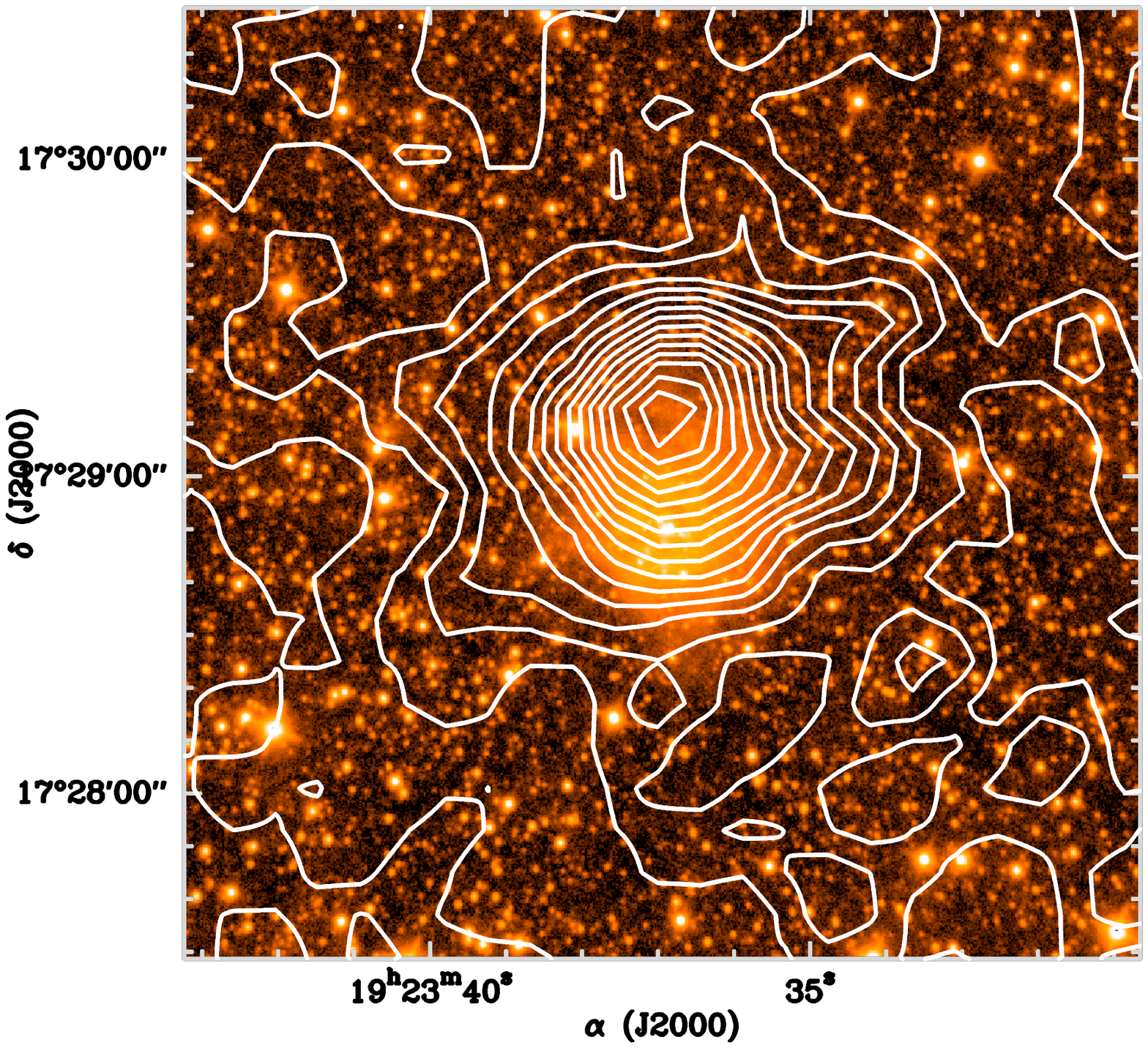}}
  \caption{K$_s$ image of field Mol103 with superimposed SIMBA 1.2mm continuum in white contours ( Beltr\'an et al. \cite{beltran06}).}
  \label{map_mol103}
\end{figure*}

\clearpage

\begin{figure*}[h]
  \centering
  \resizebox{16cm}{!}{\includegraphics[angle=-90]{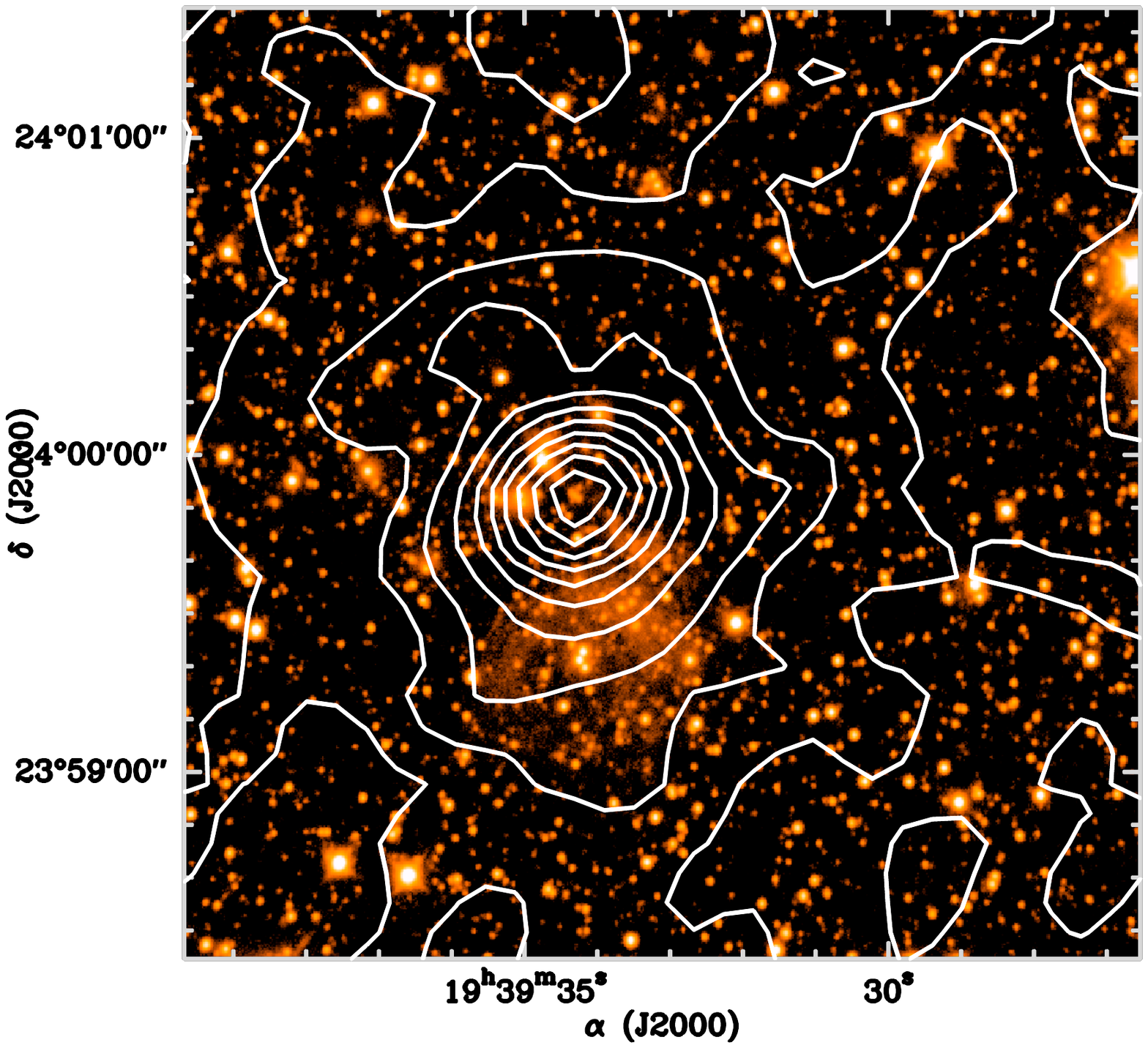}}
  \caption{K$_s$ image of field Mol109 with superimposed SIMBA 1.2mm continuum in white contours ( Beltr\'an et al. \cite{beltran06}).}
  \label{map_mol109}
\end{figure*}

\begin{figure*}[h]
  \centering
\vspace{-2cm}
  \resizebox{16cm}{!}{\includegraphics[angle=-90]{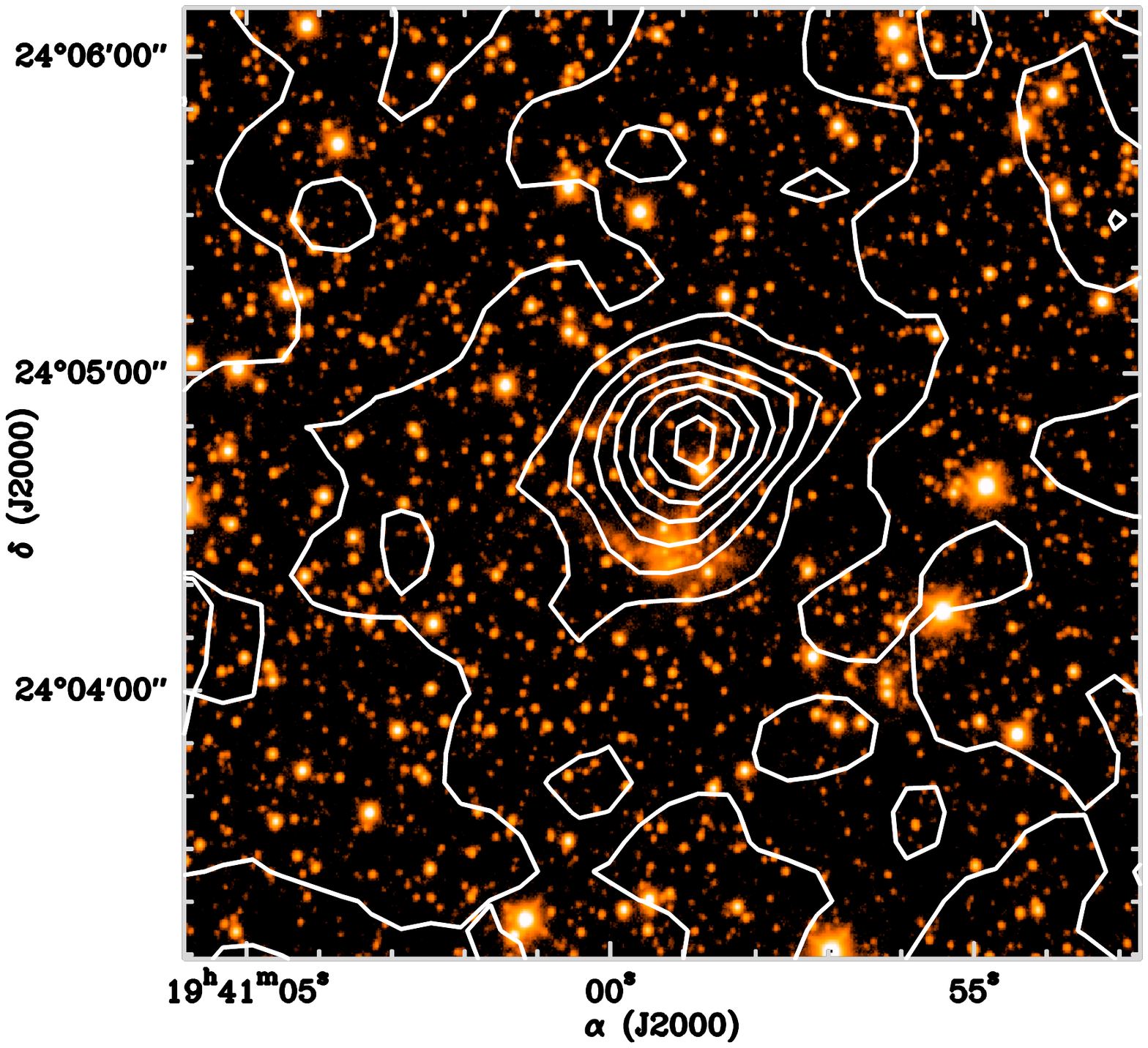}}
  \caption{K$_s$ image of field Mol110 with superimposed SIMBA 1.2mm continuum in white contours ( Beltr\'an et al. \cite{beltran06}).}
  \label{map_mol110}
\end{figure*}

\clearpage

\begin{figure*}[h]
  \centering
  \resizebox{16cm}{!}{\includegraphics[angle=-90]{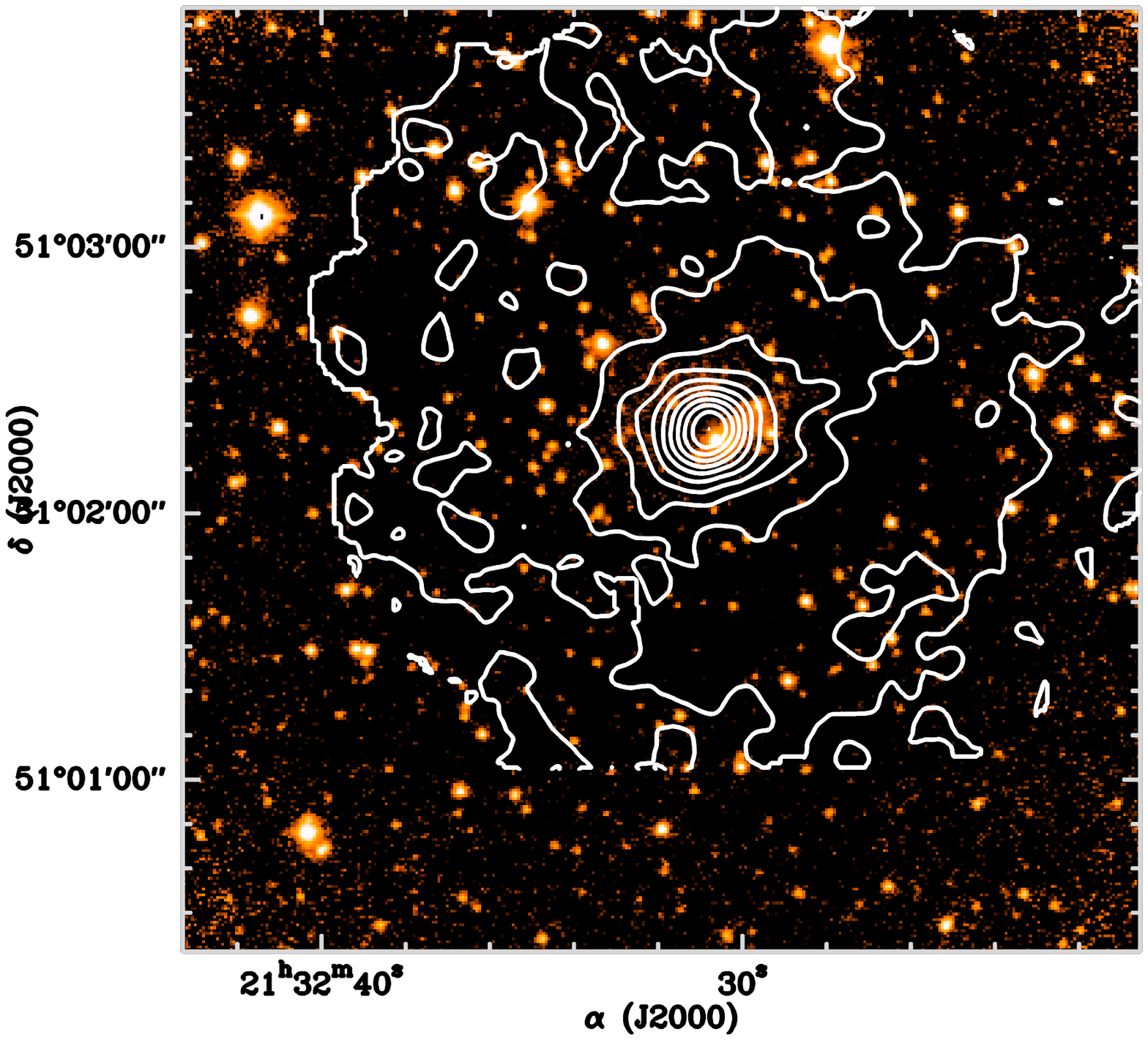}}
  \caption{K$_s$ image of field Mol136 with superimposed SCUBA 850\um\ continuum in white contours (Molinari et al. \cite{Moli08a}).}
  \label{map_mol136}
\end{figure*}

\begin{figure*}[h]
  \centering
\vspace{-2cm}
  \resizebox{16cm}{!}{\includegraphics[angle=-90]{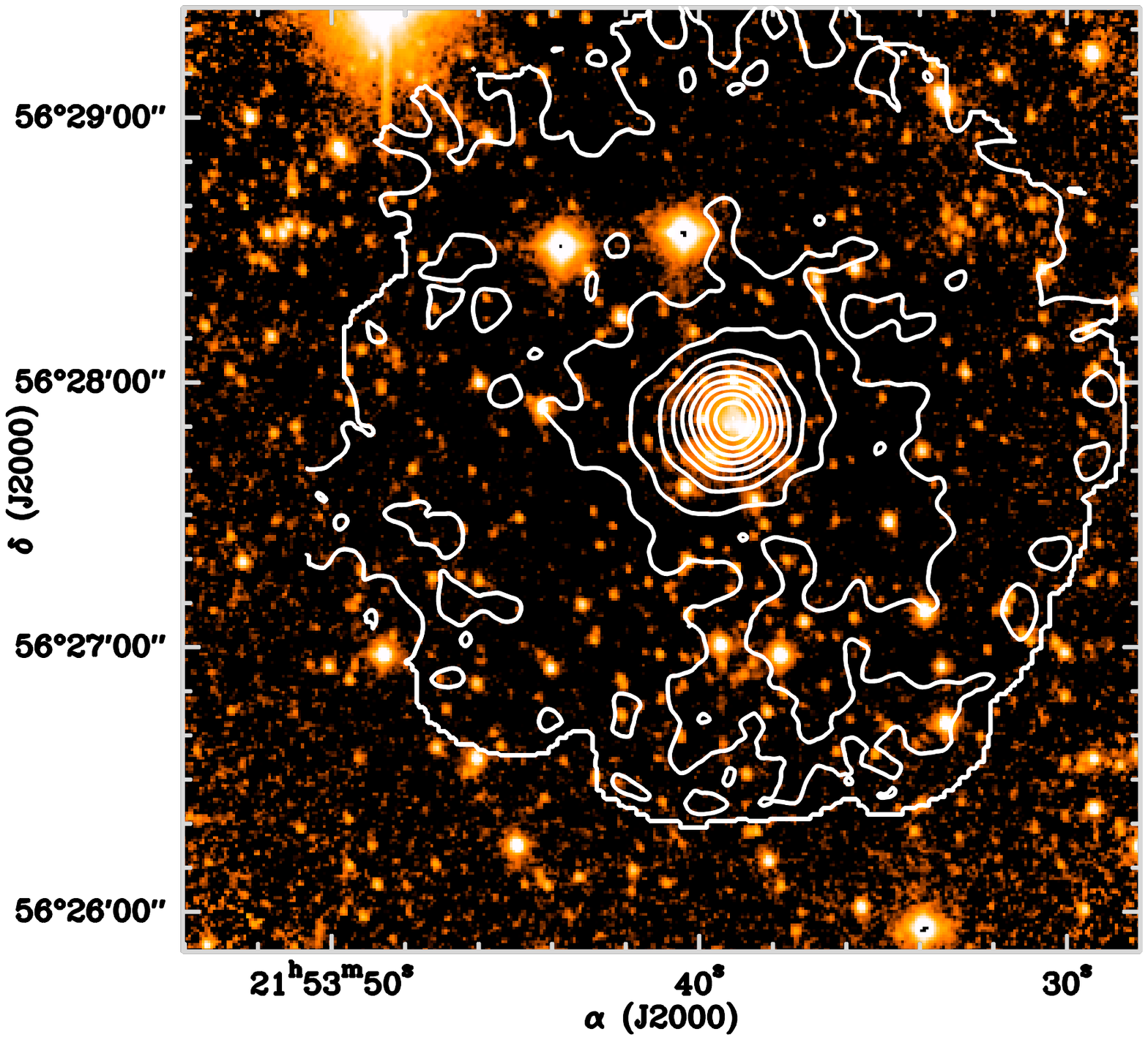}}
  \caption{K$_s$ image of field Mol139 with superimposed SCUBA 850\um\ continuum in white contours (Molinari et al. \cite{Moli08a}).}
  \label{map_mol139}
\end{figure*}

\clearpage

\begin{figure*}[h]
  \centering
  \resizebox{16cm}{!}{\includegraphics[angle=-90]{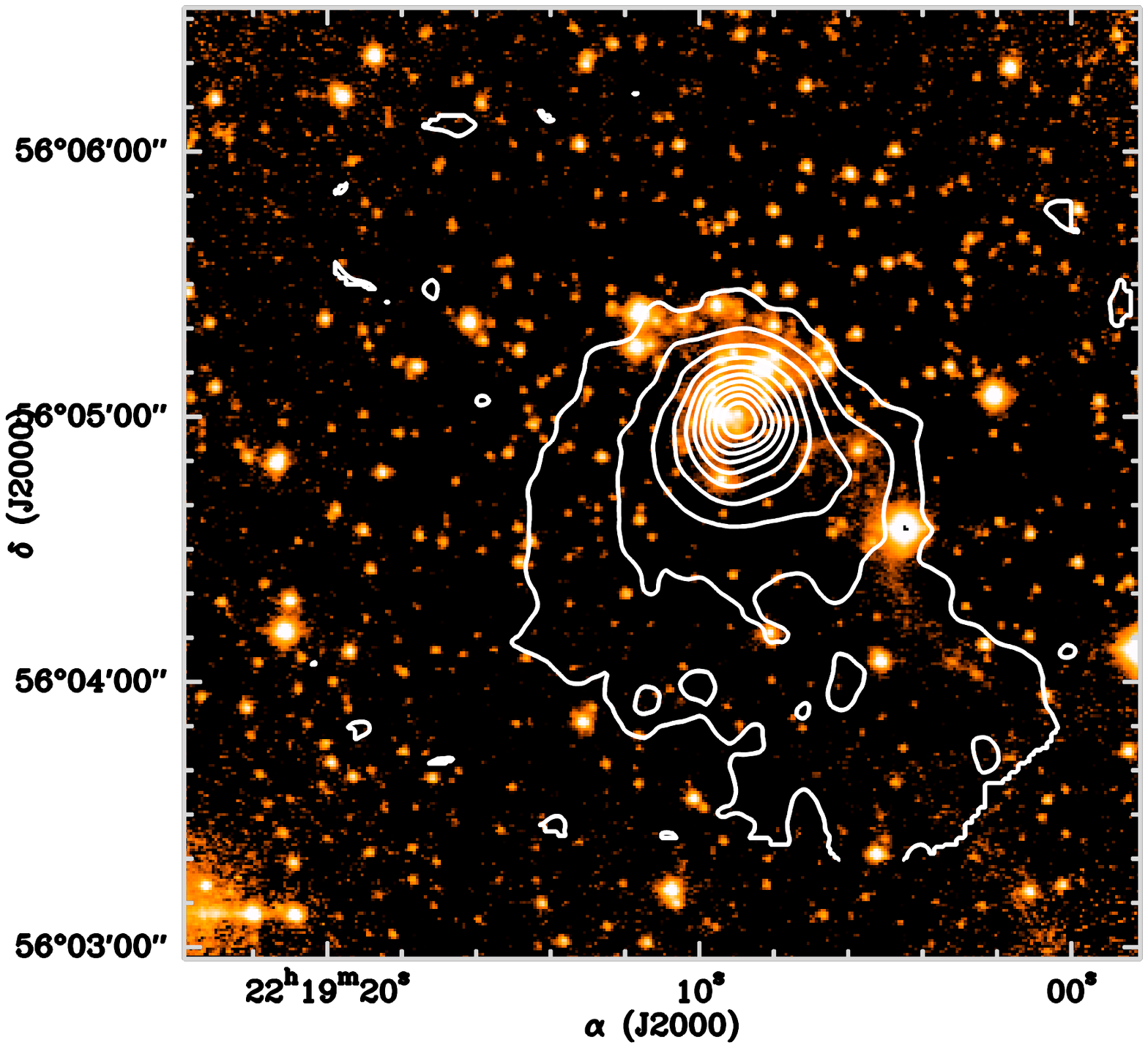}}
  \caption{K$_s$ image of field Mol143 with superimposed SCUBA 850\um\ continuum in white contours (Molinari et al. \cite{Moli08a}).}
  \label{map_mol143}
\end{figure*}

\begin{figure*}[h]
  \centering
\vspace{-2cm}
  \resizebox{16cm}{!}{\includegraphics[angle=-90]{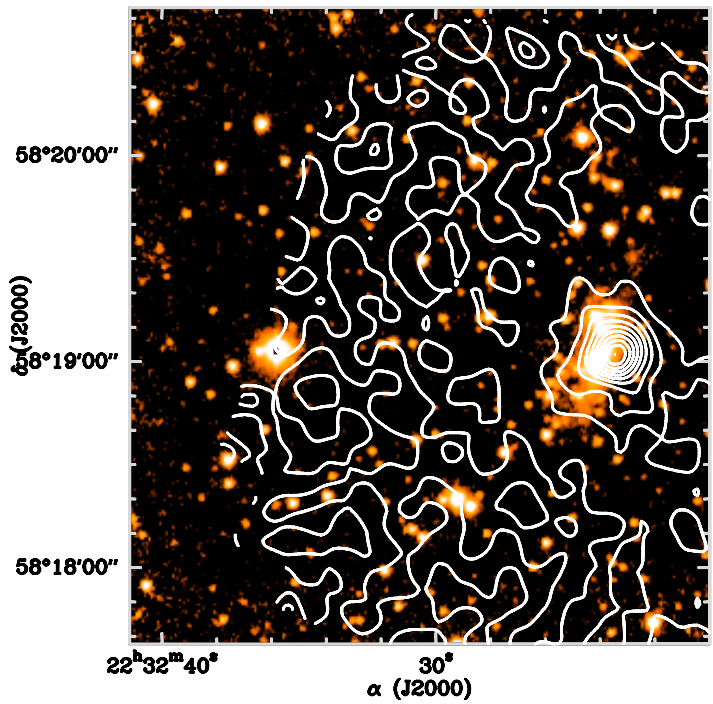}}
  \caption{K$_s$ image of field Mol148 with superimposed MAMBO 1.1mm continuum in white contours (Molinari et al. \cite{Moli08a}).}
  \label{map_mol148}
\end{figure*}

\clearpage

\begin{figure*}[h]
  \centering
  \resizebox{16cm}{!}{\includegraphics[angle=-90]{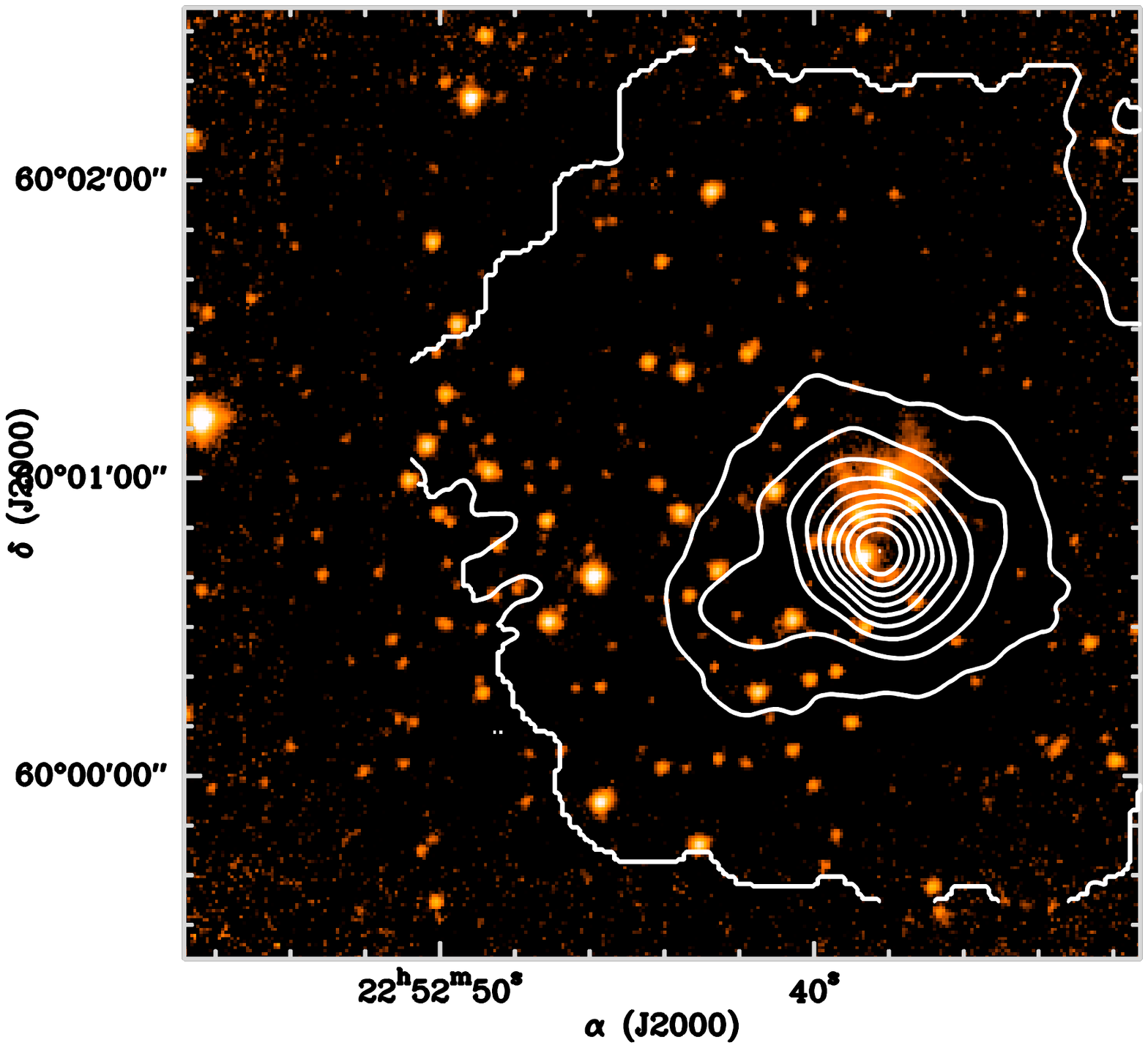}}
  \caption{K$_s$ image of field Mol151 with superimposed SCUBA 850\um\ continuum in white contours (Molinari et al. \cite{Moli08a}).}
  \label{map_mol151}
\end{figure*}

\begin{figure*}[h]
  \centering
\vspace{-2cm}
  \resizebox{16cm}{!}{\includegraphics[angle=-90]{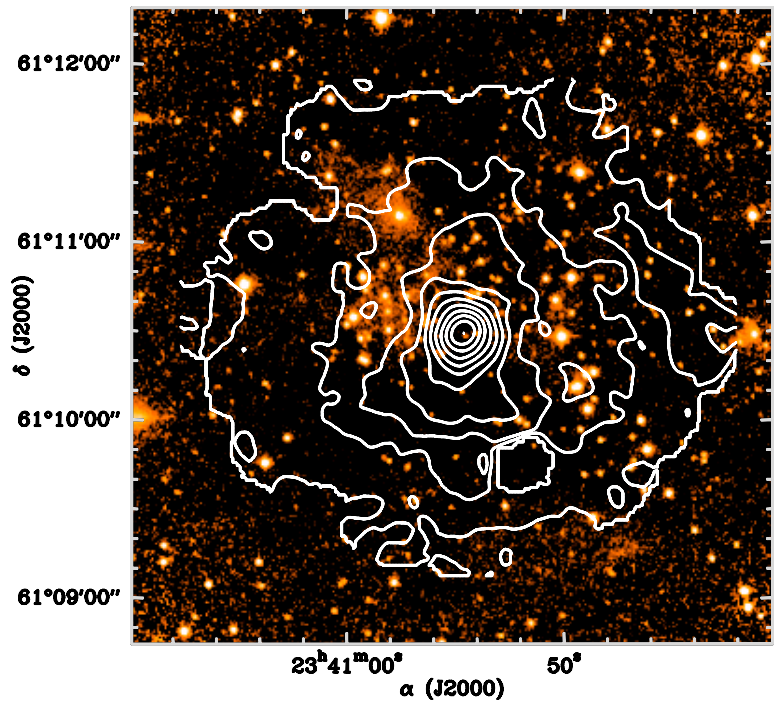}}
  \caption{K$_s$ image of field Mol160 with superimposed SCUBA 850\um\ continuum in white contours (Molinari et al. \cite{Moli08a}).}
  \label{map_mol160}
\end{figure*}
\end{appendix}

\clearpage

\begin{appendix}
\label{klf_plots}

\section{K$_s$ luminosity functions}
This appendix presents the set of background-subtracted K$_s$ luminosity functions for all detected clusters; the green vertical line represents the 90\% completeness limit as estimated from artificial star experiements (see \S\ref{phot}). Material can also be retrieved at http://galatea.ifsi-roma.inaf.it/faustini/KLF/

\begin{figure*}[h]
\centering
\resizebox{\hsize}{!}
{
\includegraphics[angle=-0,origin=c,width=6cm]{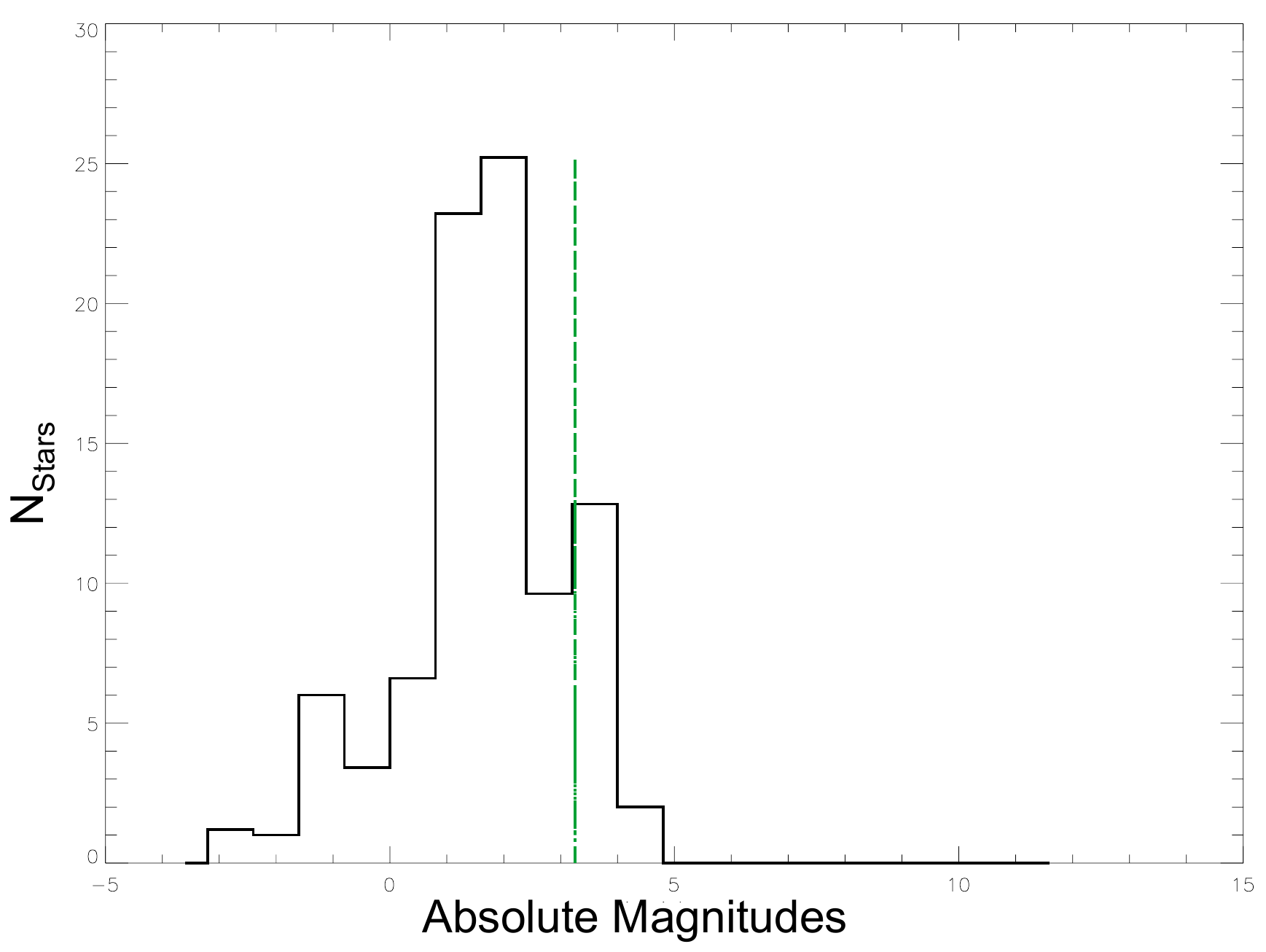} 
\includegraphics[angle=-0,origin=c,width=6cm]{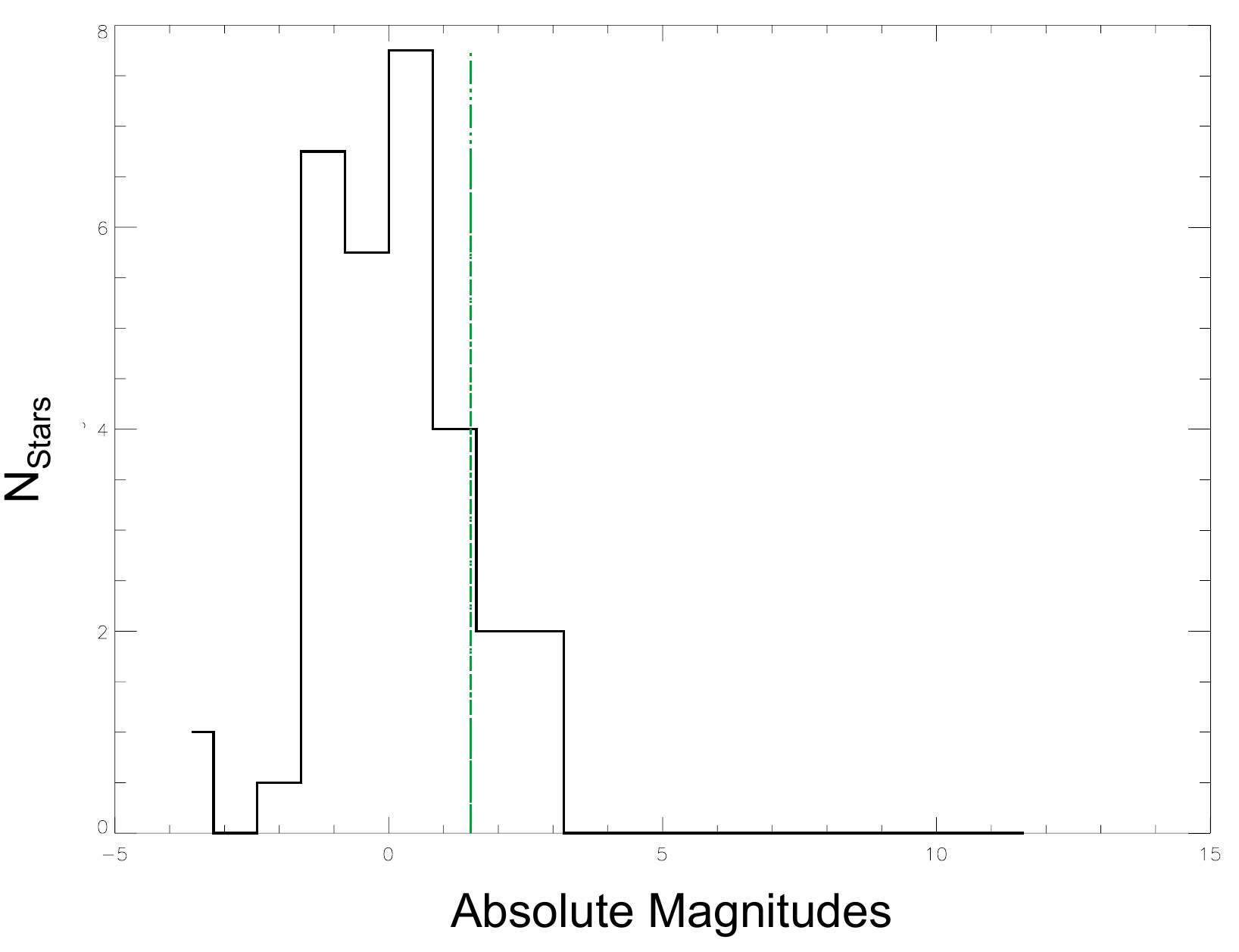} 
\includegraphics[angle=-0,origin=c,width=6cm]{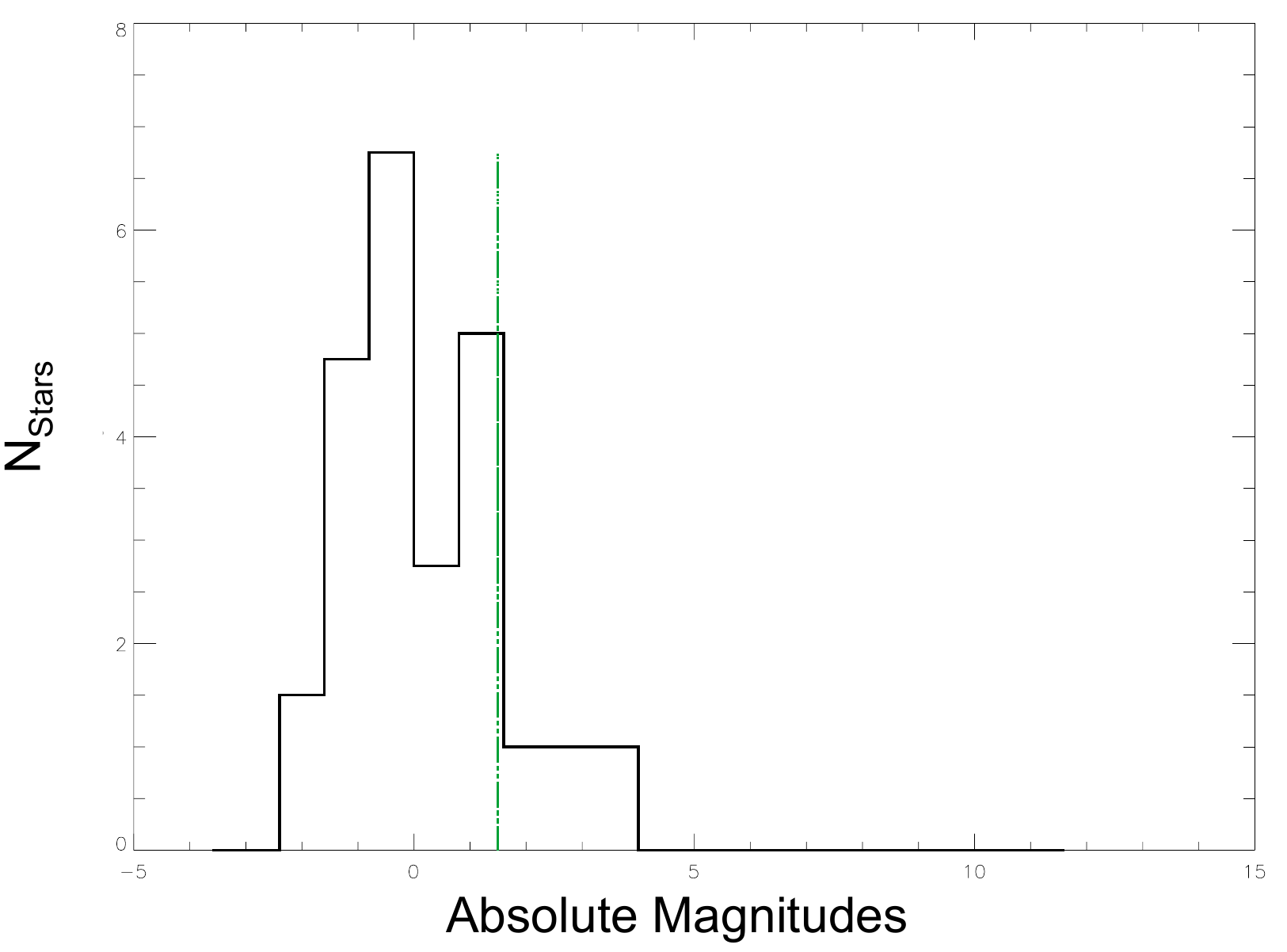}
}
\caption{Mol003 (left), Mol008A (center), Mol008B (right)}
\end{figure*}

\begin{figure*}[h]
\centering
\resizebox{\hsize}{!}
{
\includegraphics[angle=-0,origin=c,width=6cm]{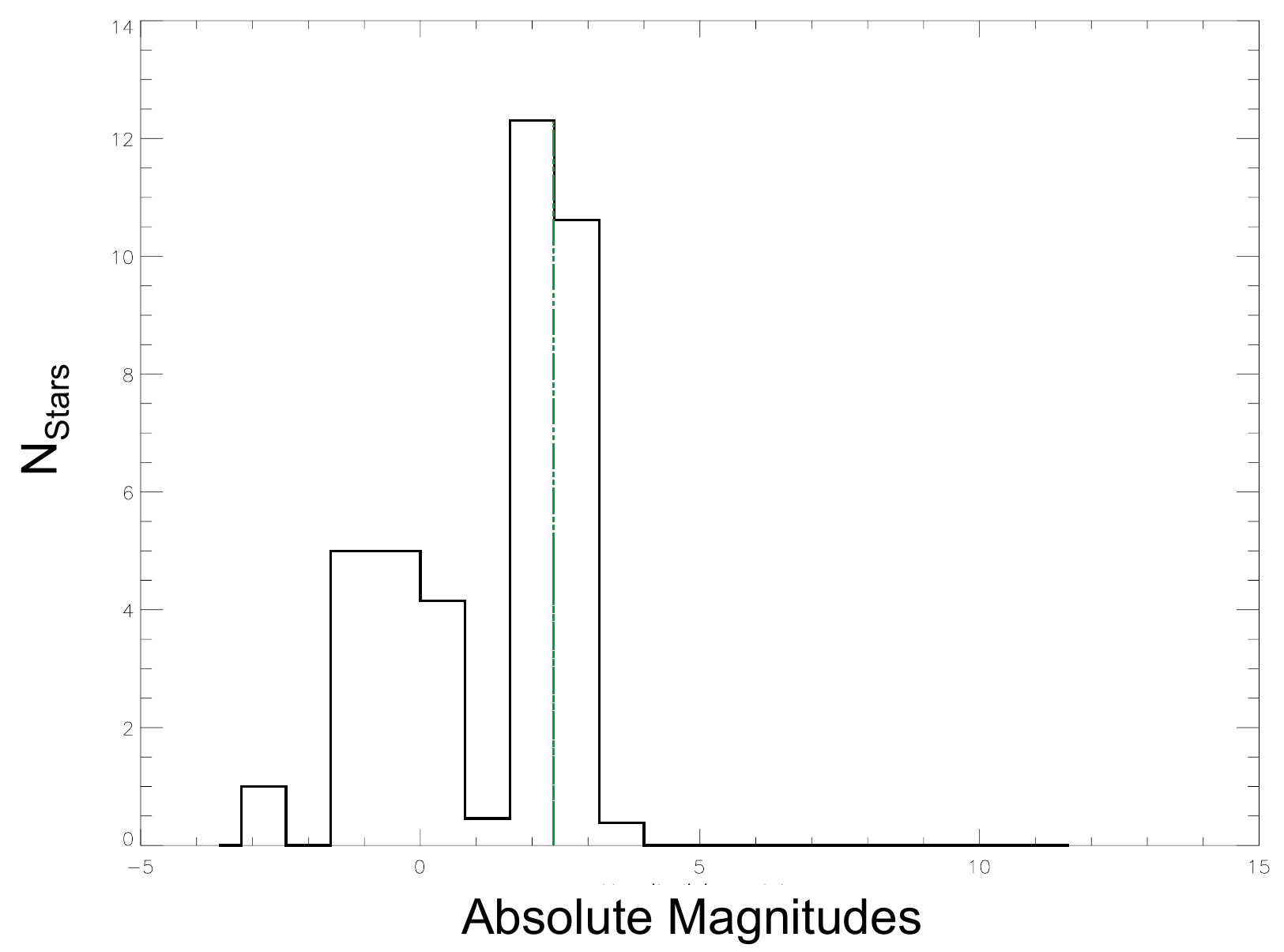} 
\includegraphics[angle=-0,origin=c,width=6cm]{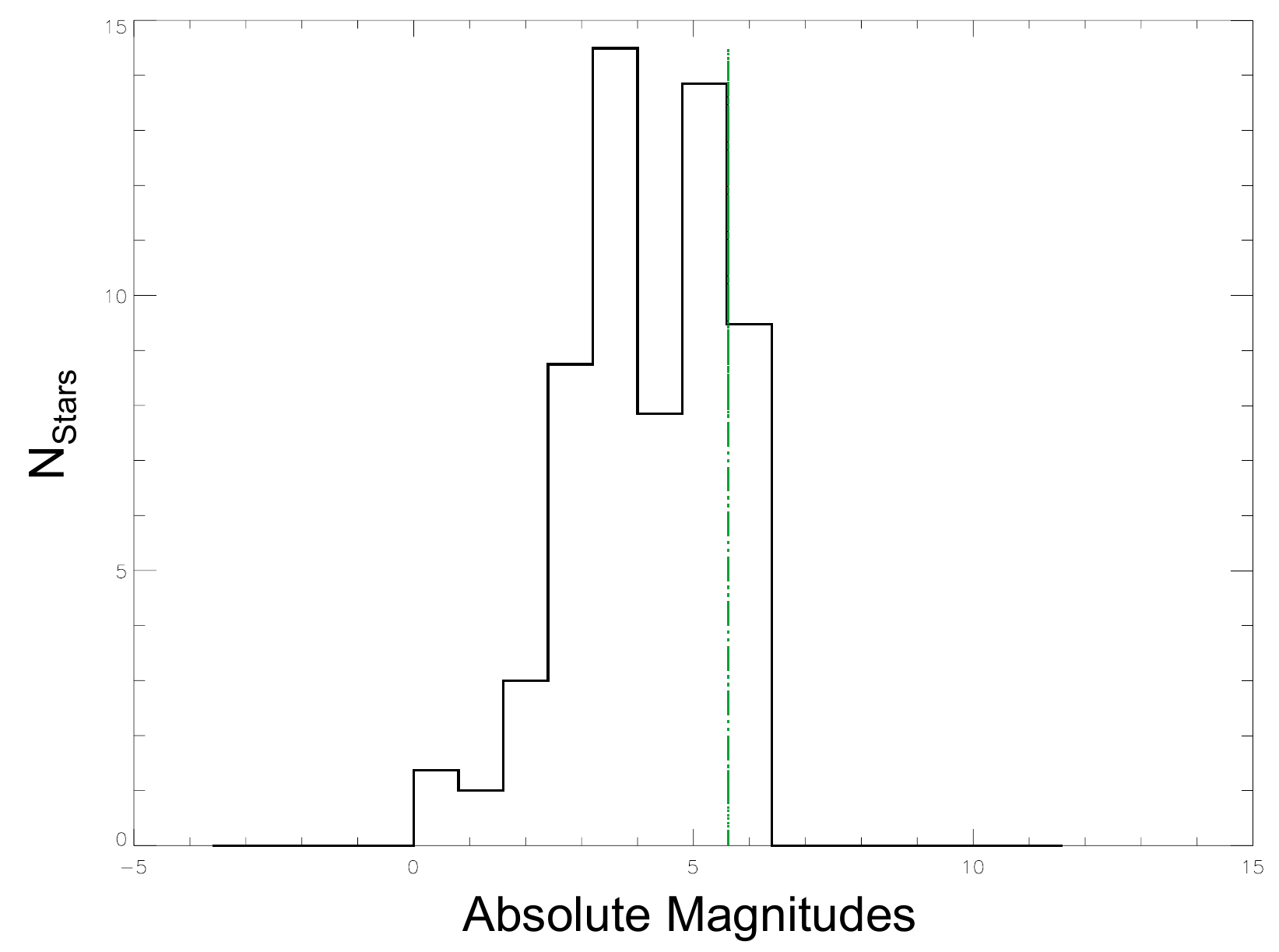} 
\includegraphics[angle=-0,origin=c,width=6cm]{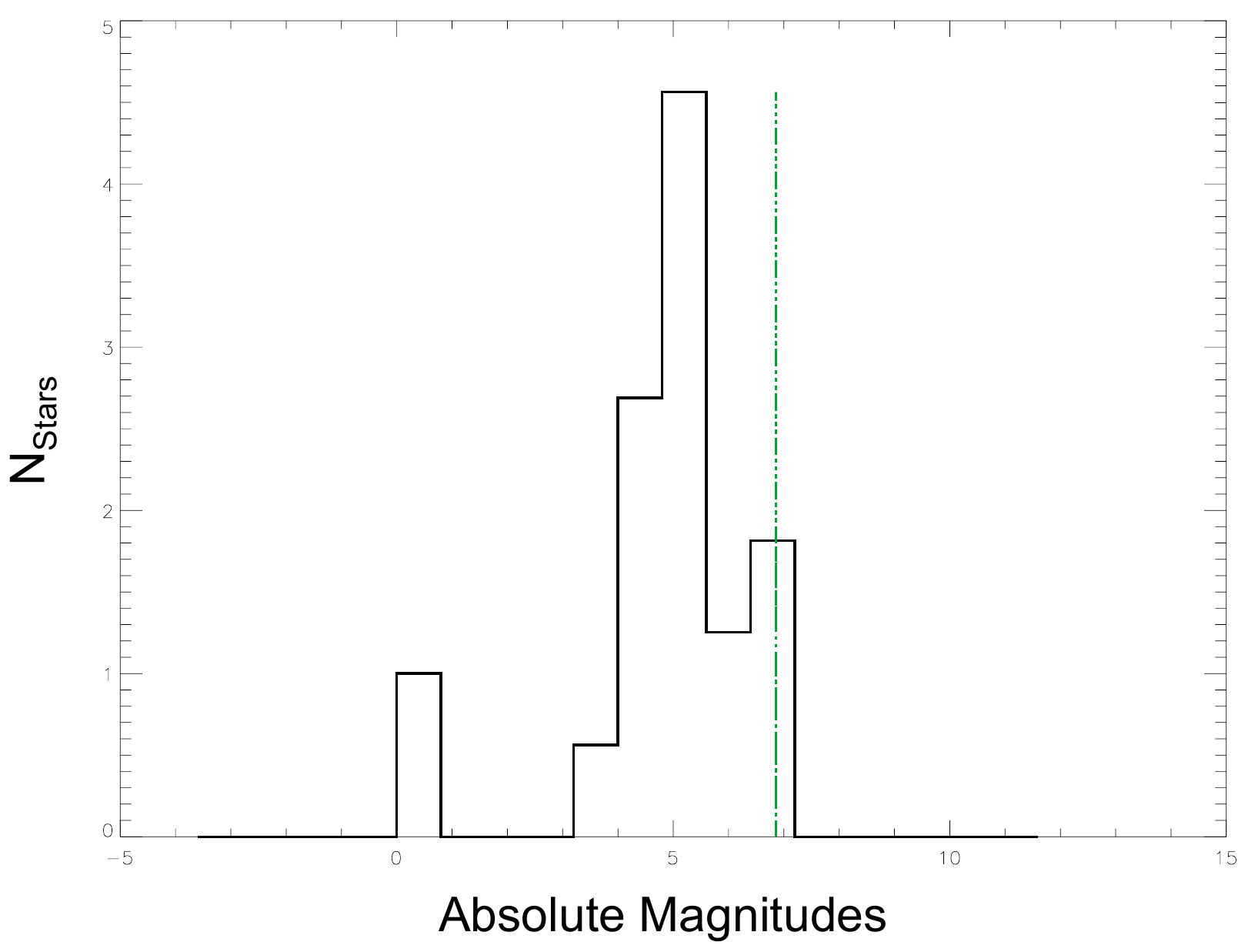}
}
\caption{Mol009 (left), Mol011 (center), Mol012 (right)}
\end{figure*}

\begin{figure*}[h]
\centering
\resizebox{\hsize}{!}
{
\includegraphics[angle=-0,origin=c,width=6cm]{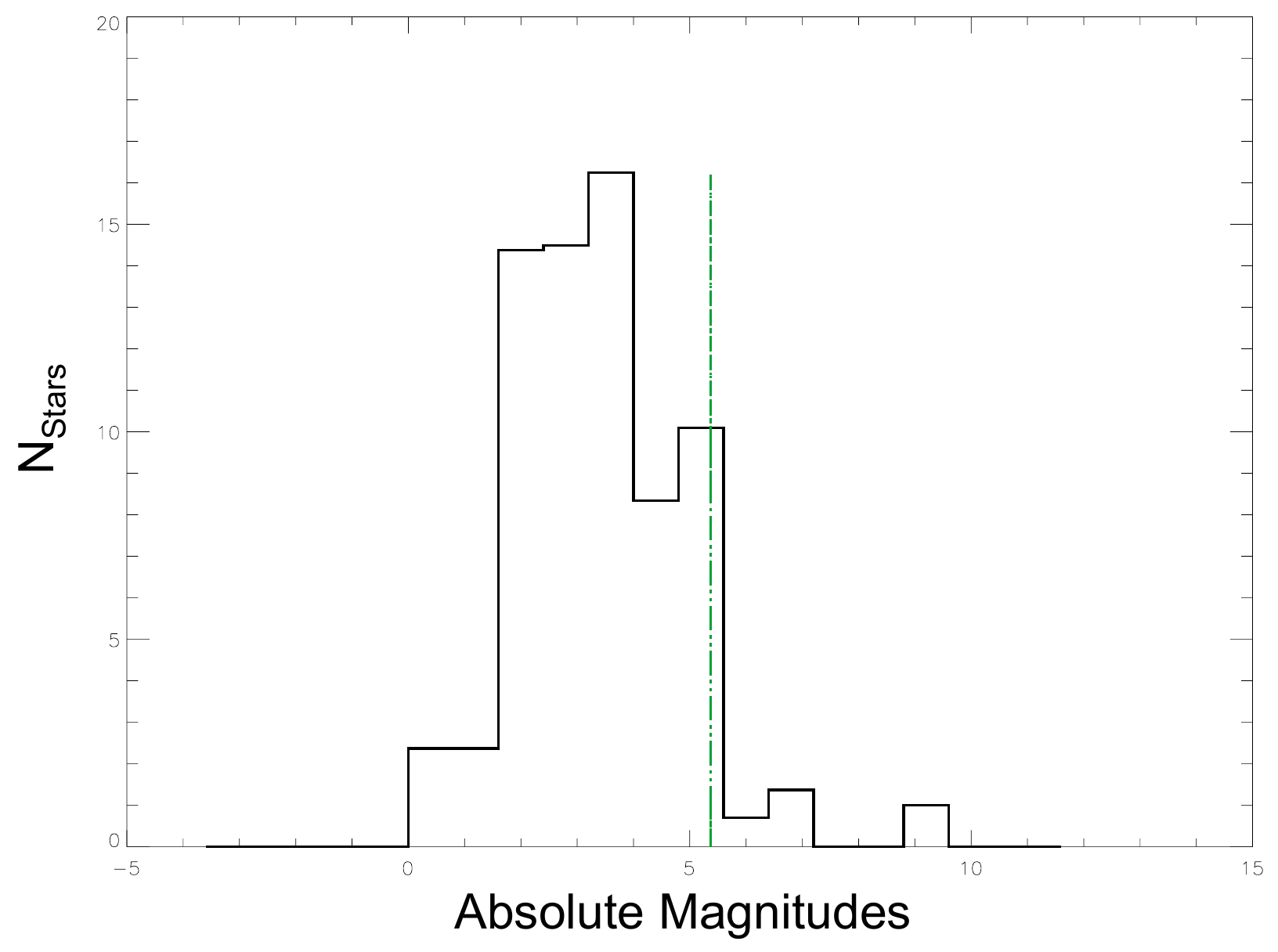} 
\includegraphics[angle=-0,origin=c,width=6cm]{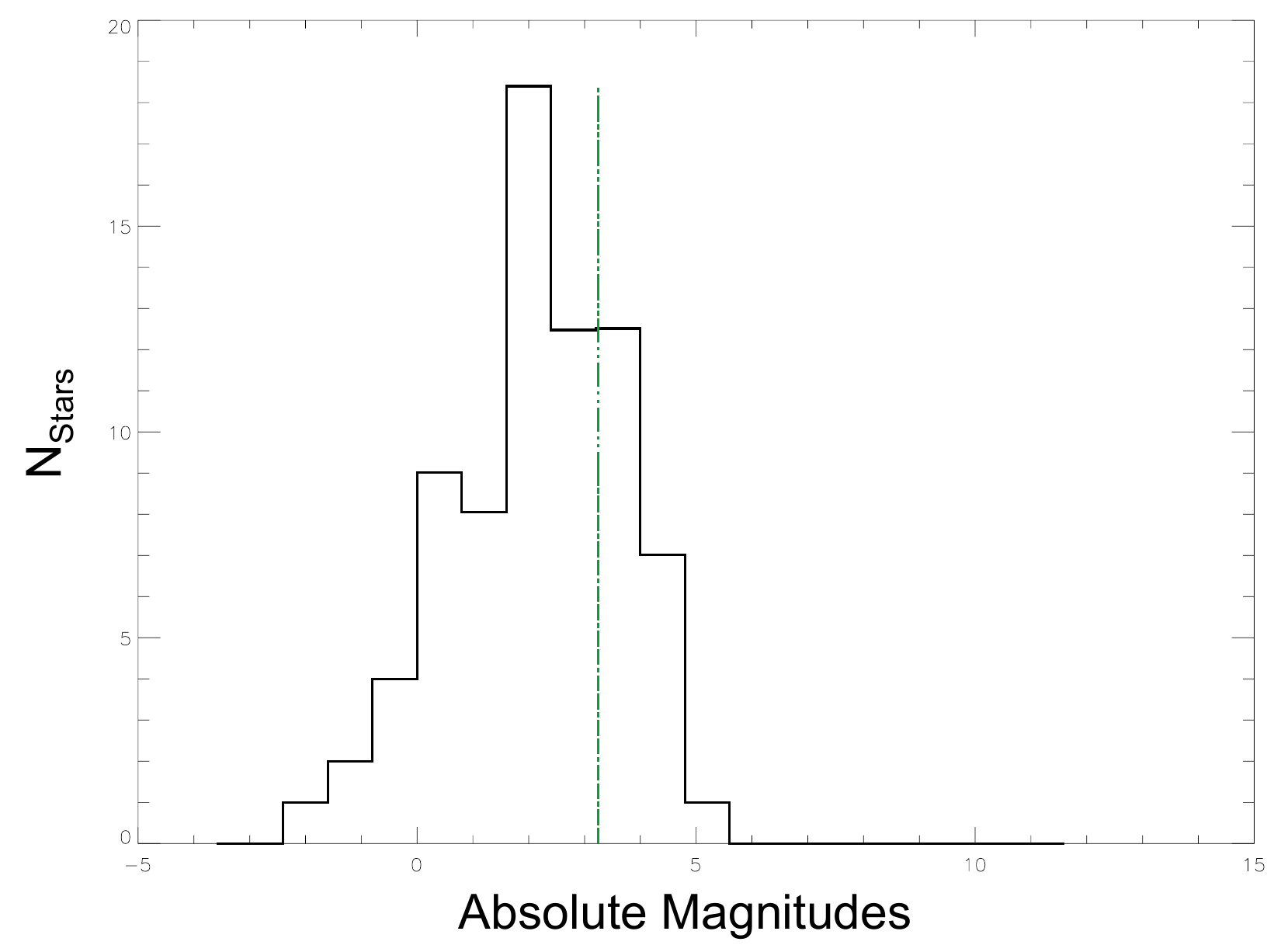} 
\includegraphics[angle=-0,origin=c,width=6cm]{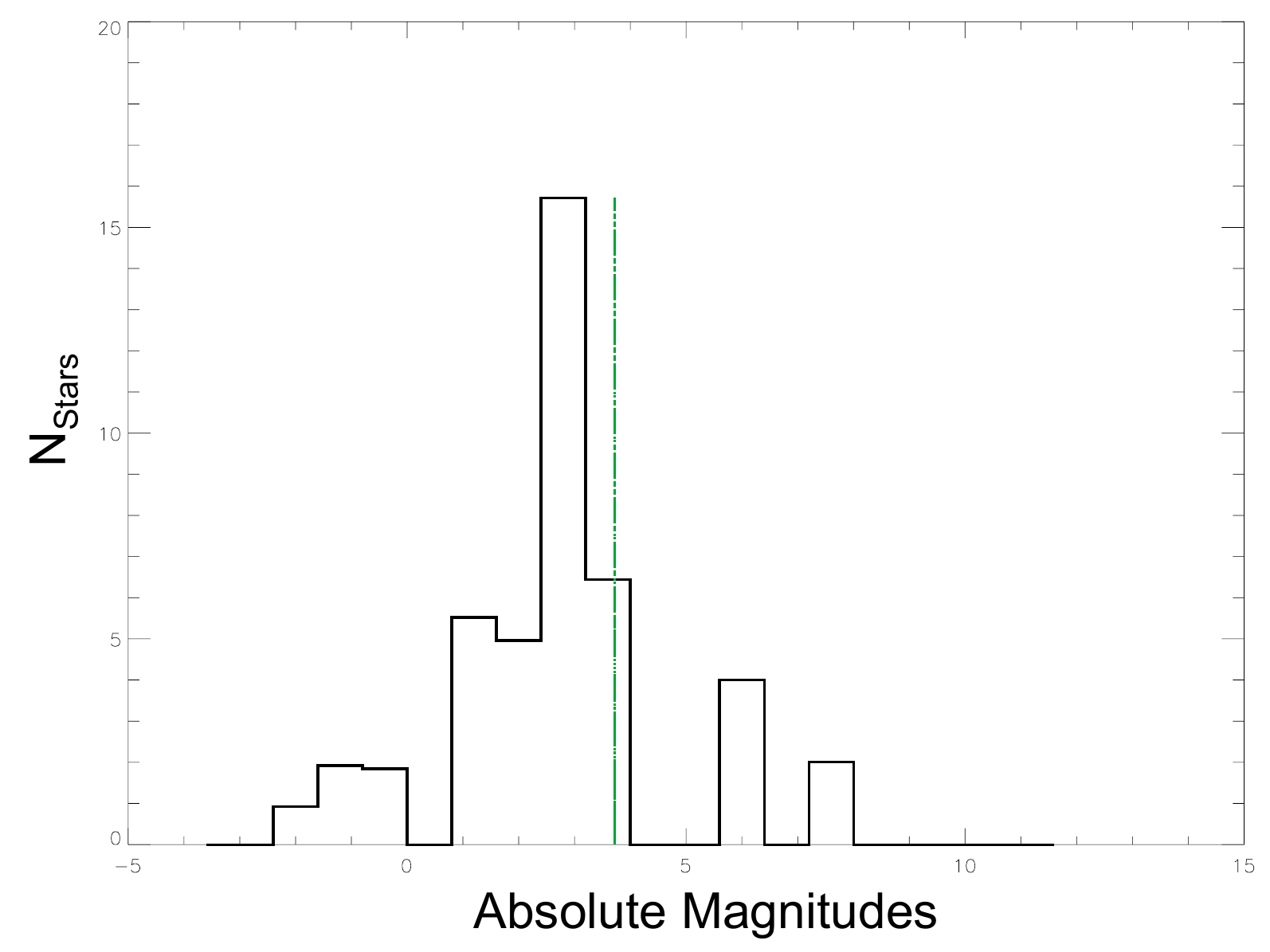}
}
\caption{Mol015 (left), Mol028 (center), Mol045 (right)}
\end{figure*}

\begin{figure*}[h]
\centering
\resizebox{\hsize}{!}
{
\includegraphics[angle=-0,origin=c,width=6cm]{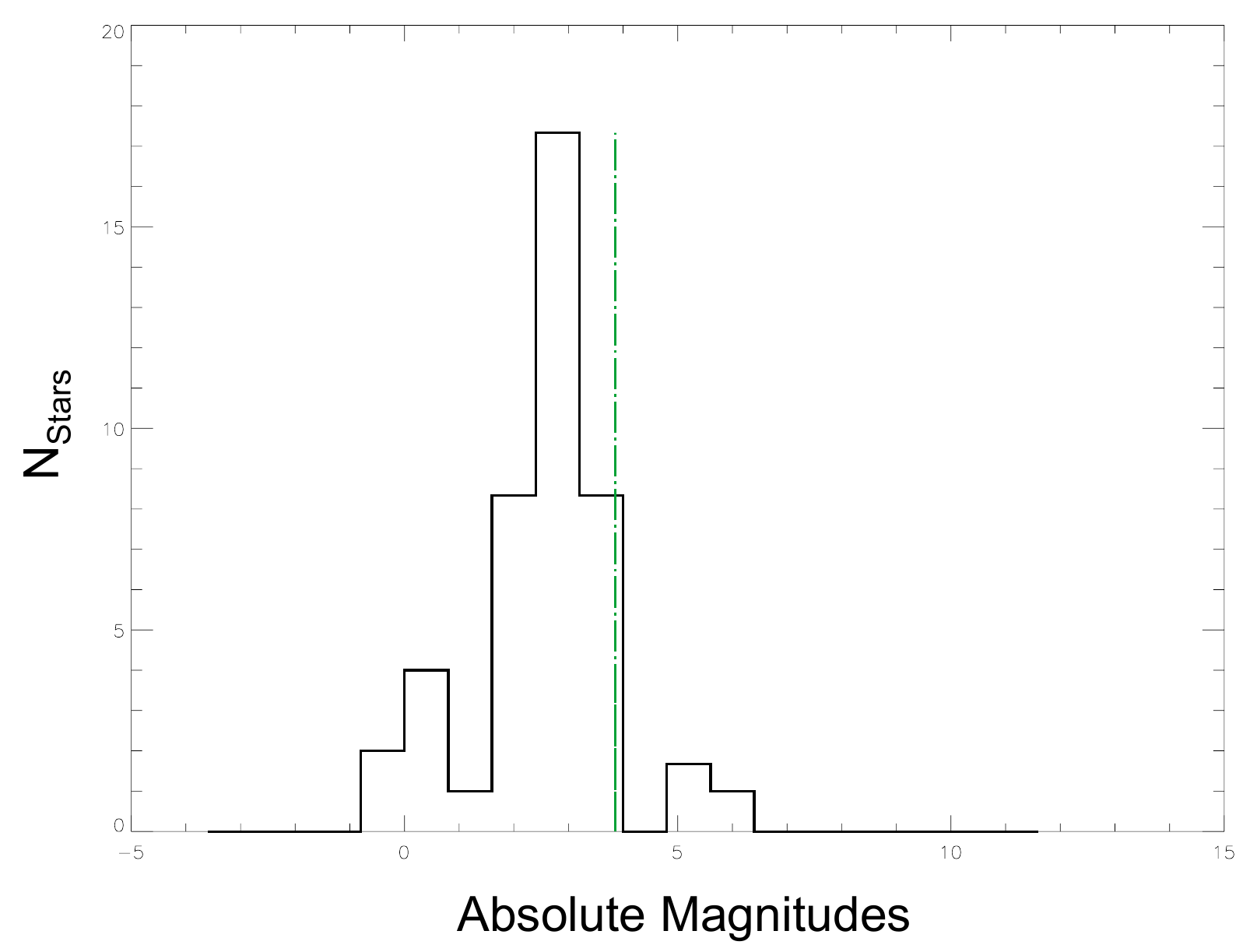} 
\includegraphics[angle=-0,origin=c,width=6cm]{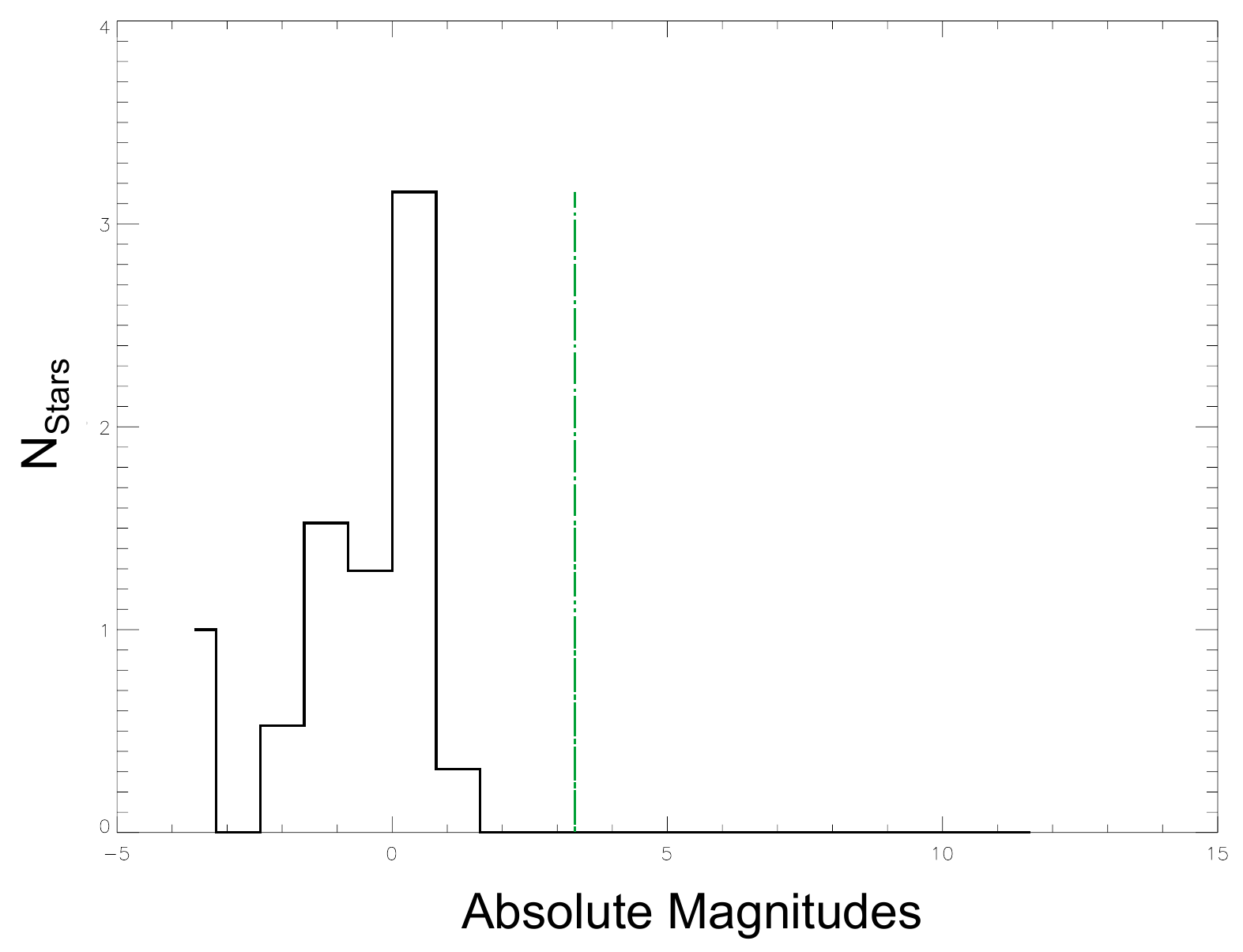} 
\includegraphics[angle=-0,origin=c,width=6cm]{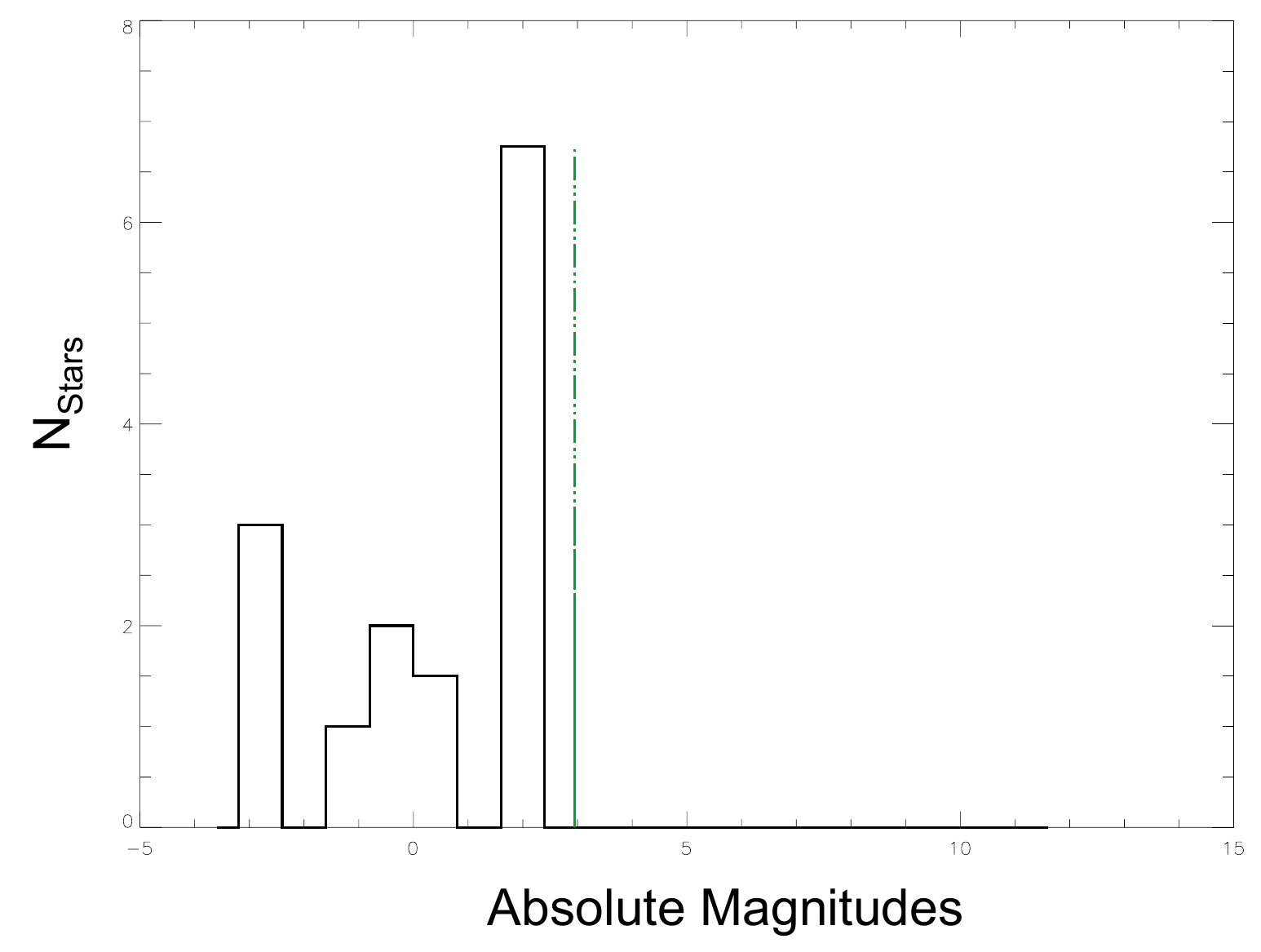}
}
\caption{Mol050 (left), Mol075 (center), Mol082 (right)}
\end{figure*}

\clearpage

\begin{figure*}[h]
\centering
\resizebox{\hsize}{!}
{
\includegraphics[angle=-0,origin=c,width=6cm]{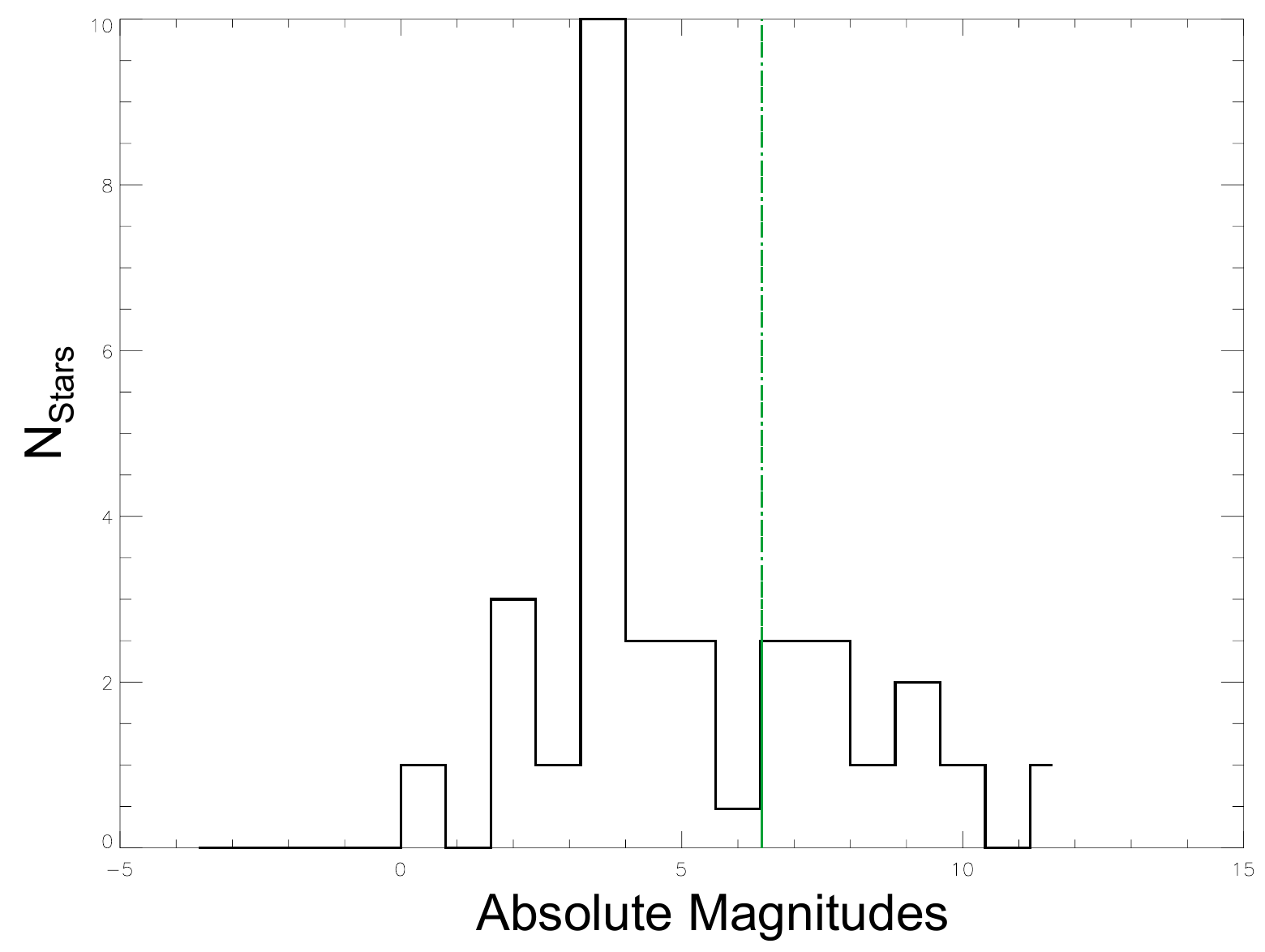} 
\includegraphics[angle=-0,origin=c,width=6cm]{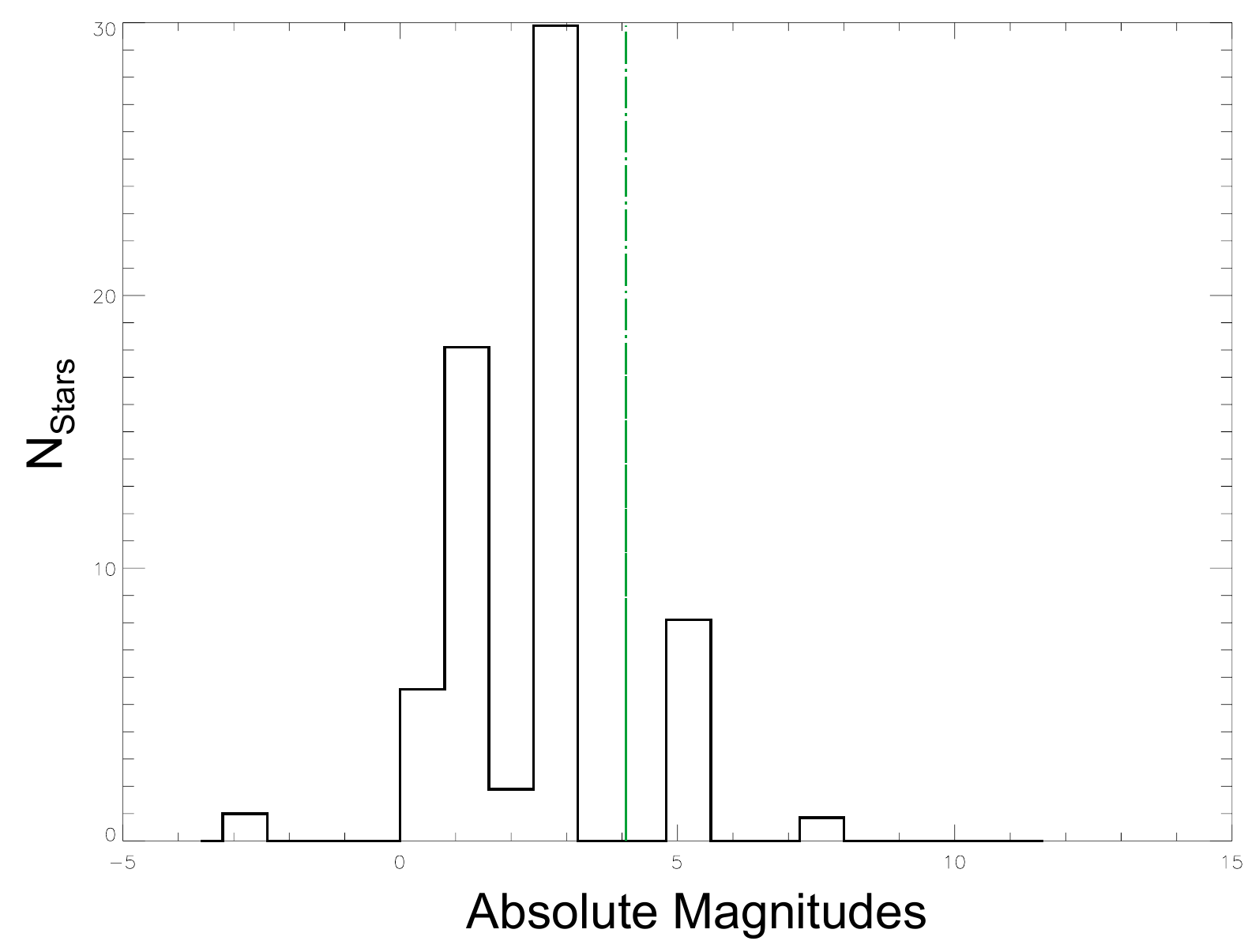} 
\includegraphics[angle=-0,origin=c,width=6cm]{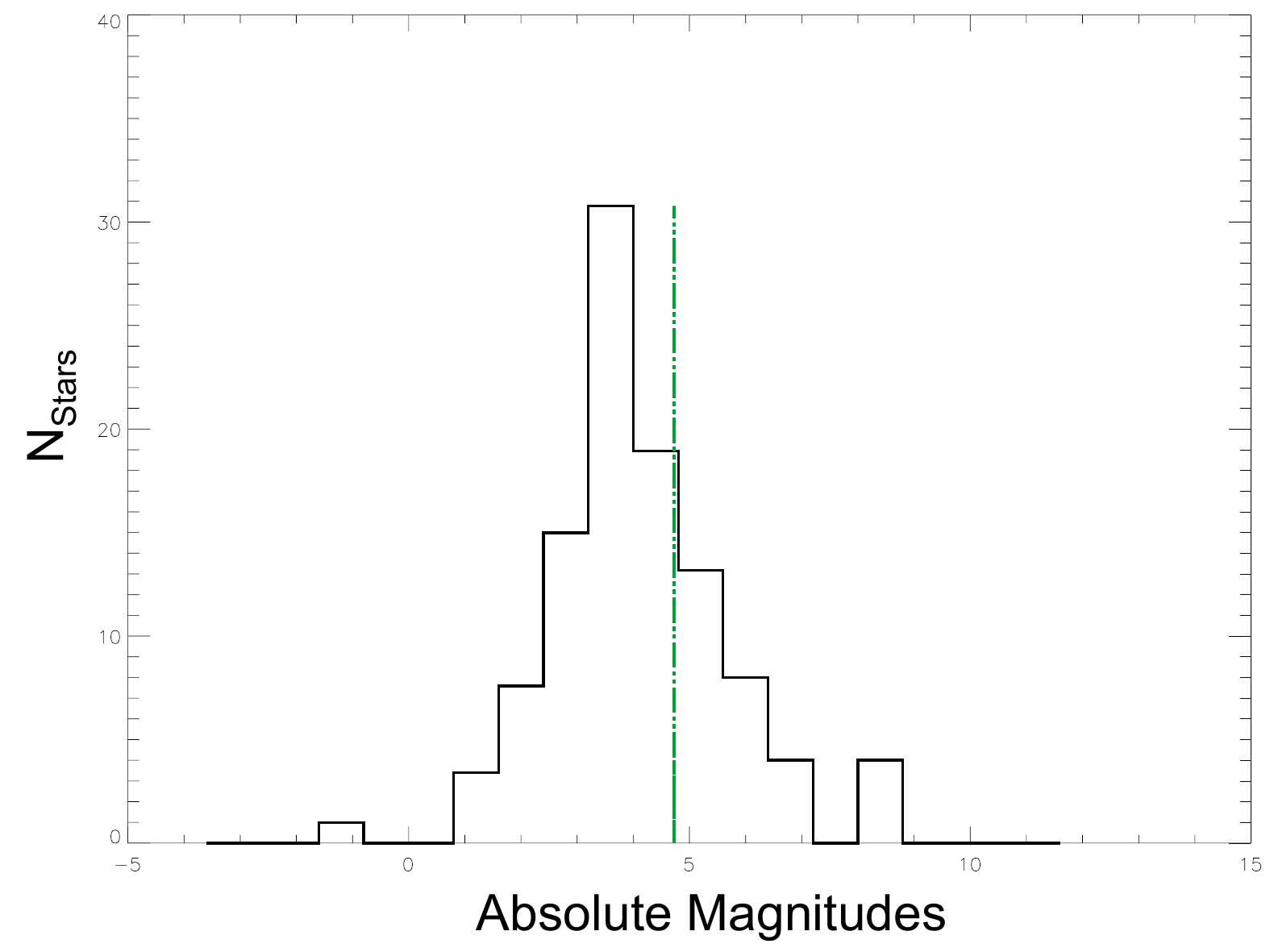}
}
\caption{Mol084 (left), Mol099 (center), Mol103 (right)}
\end{figure*}

\begin{figure*}[h]
\centering
\resizebox{\hsize}{!}
{
\includegraphics[angle=-0,origin=c,width=6cm]{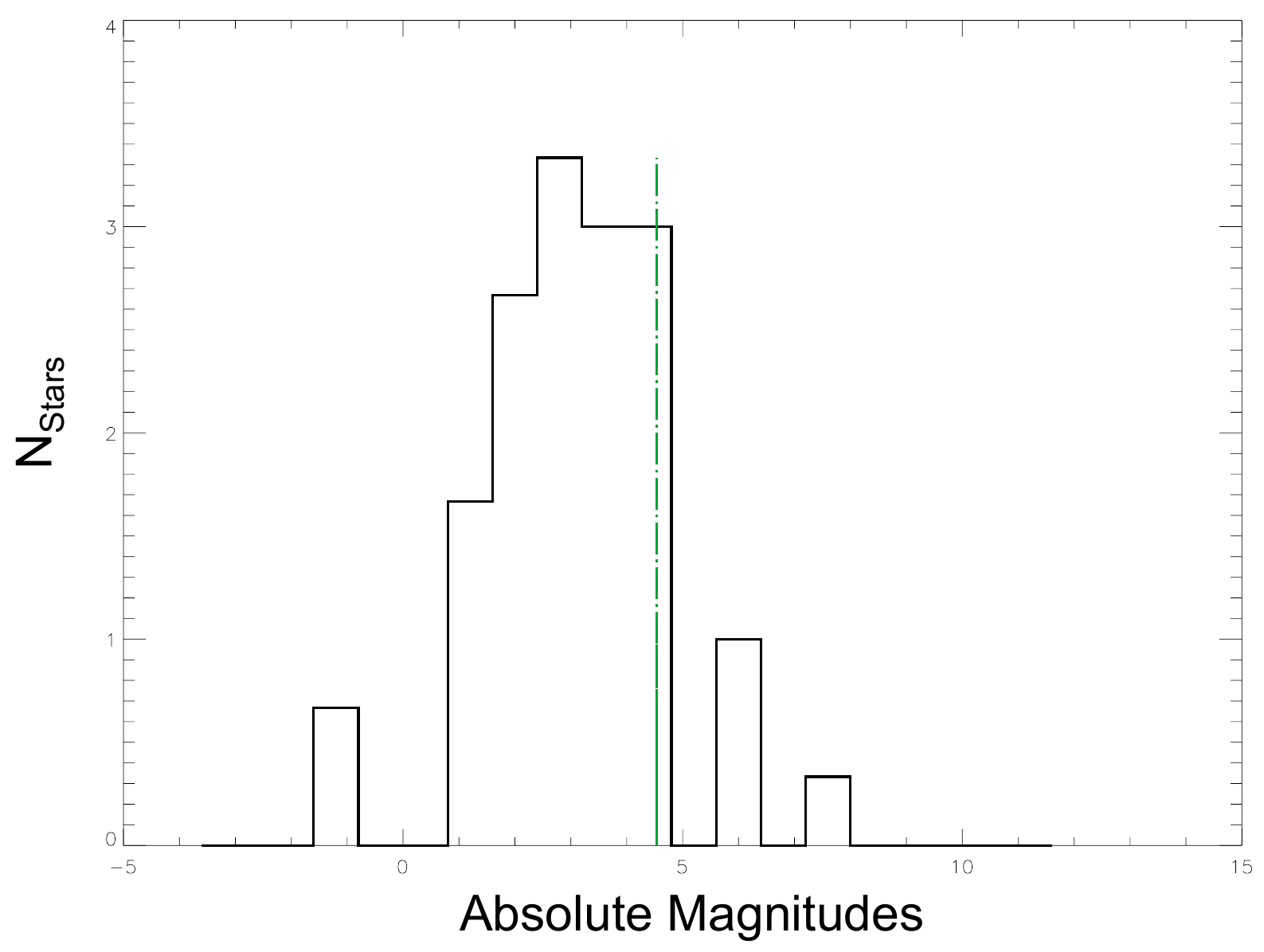} 
\includegraphics[angle=-0,origin=c,width=6cm]{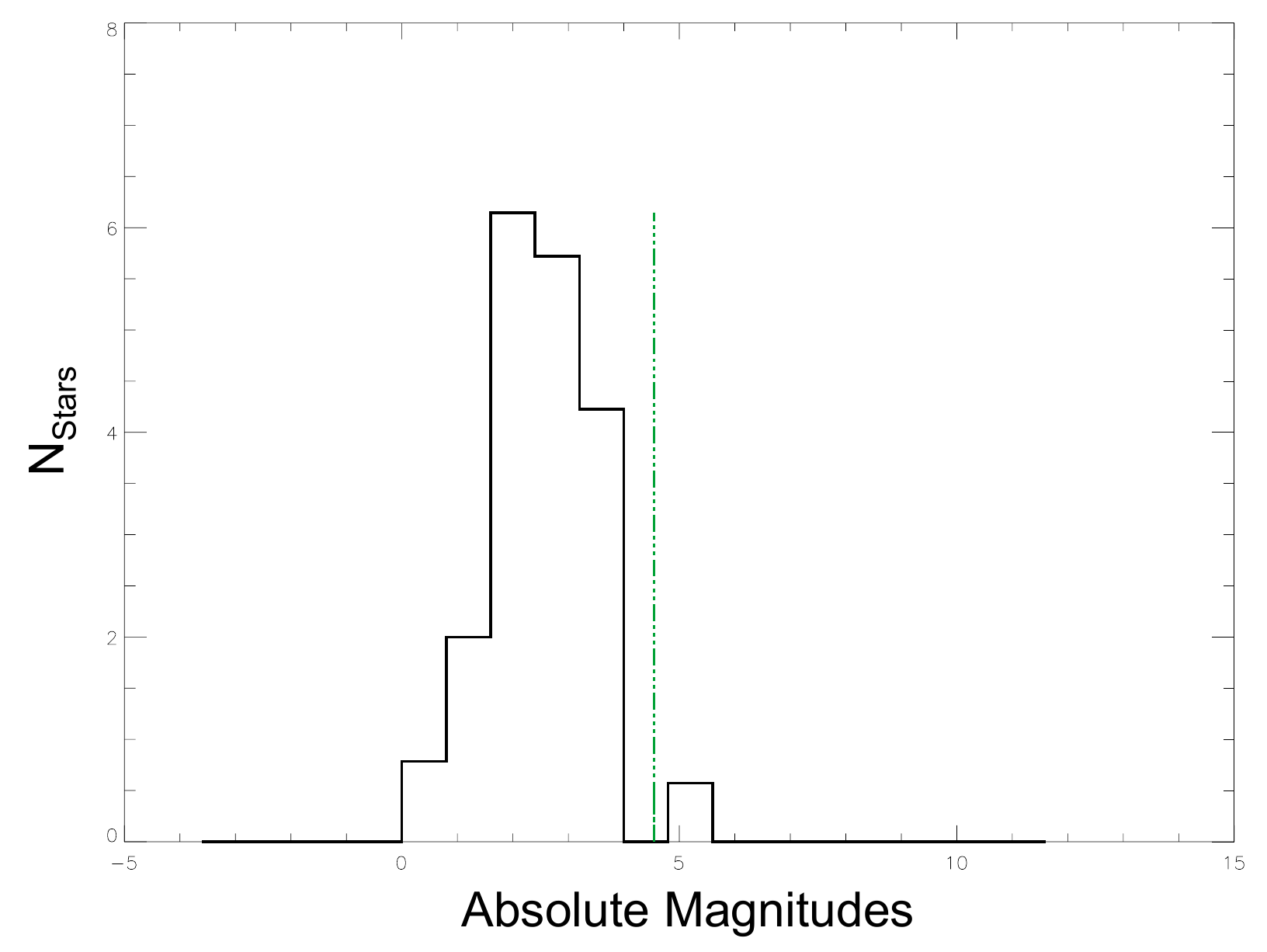} 
\includegraphics[angle=-0,origin=c,width=6cm]{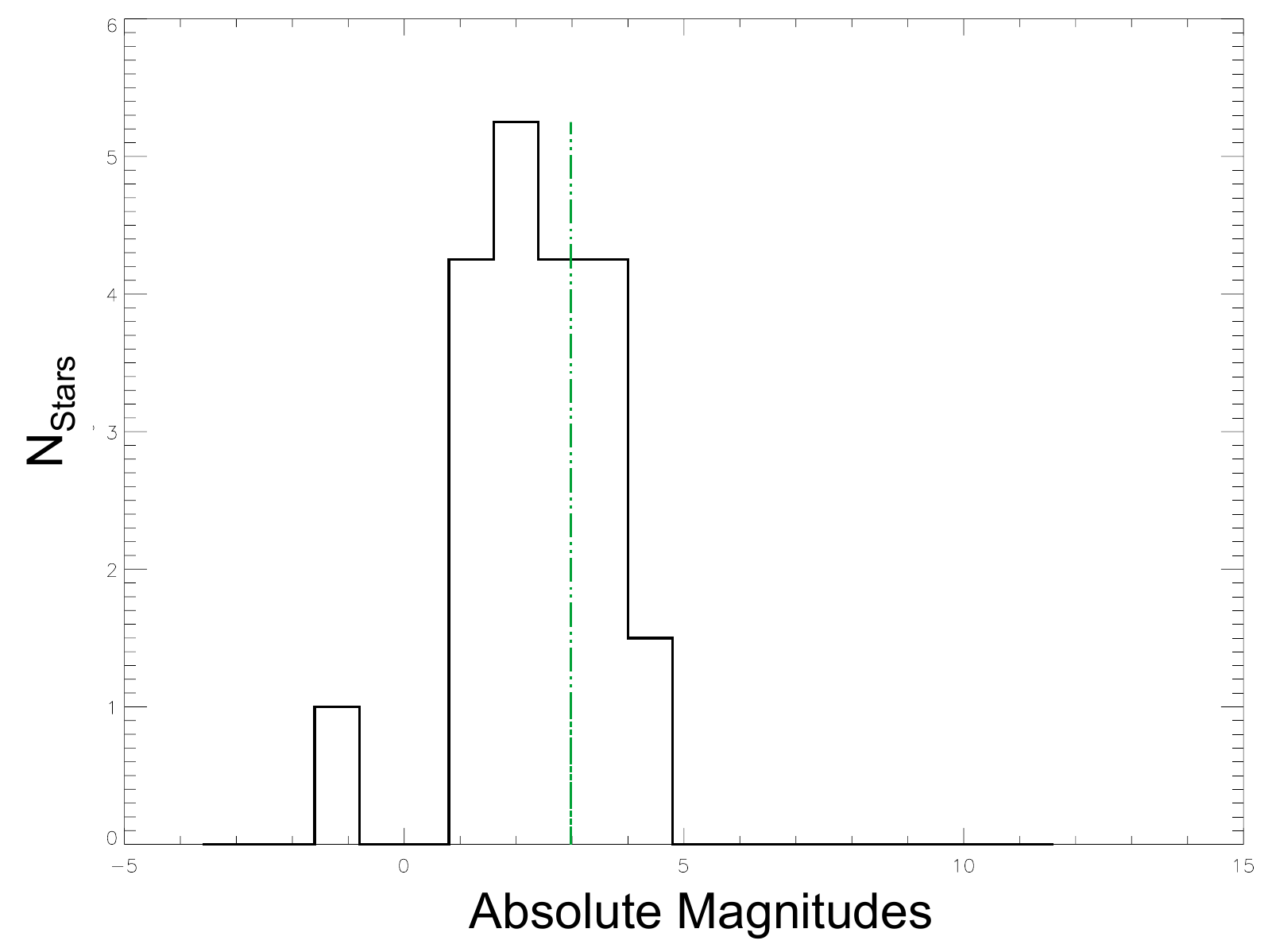}
}
\caption{Mol136 (left), Mol110 (center), Mol136 (right)}
\end{figure*}

\begin{figure*}[h]
\centering
\resizebox{\hsize}{!}
{
\includegraphics[angle=-0,origin=c,width=6cm]{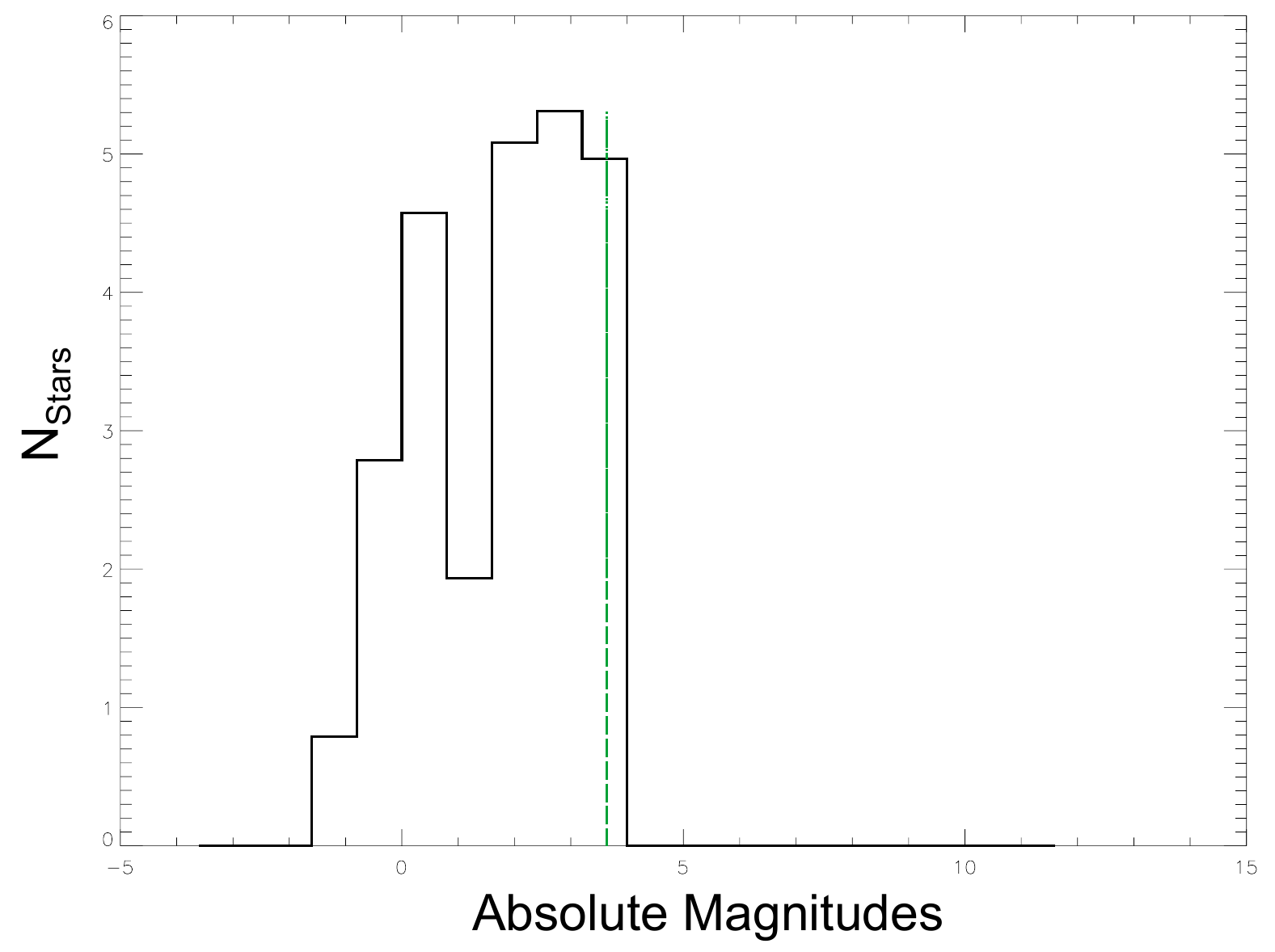} 
\includegraphics[angle=-0,origin=c,width=6cm]{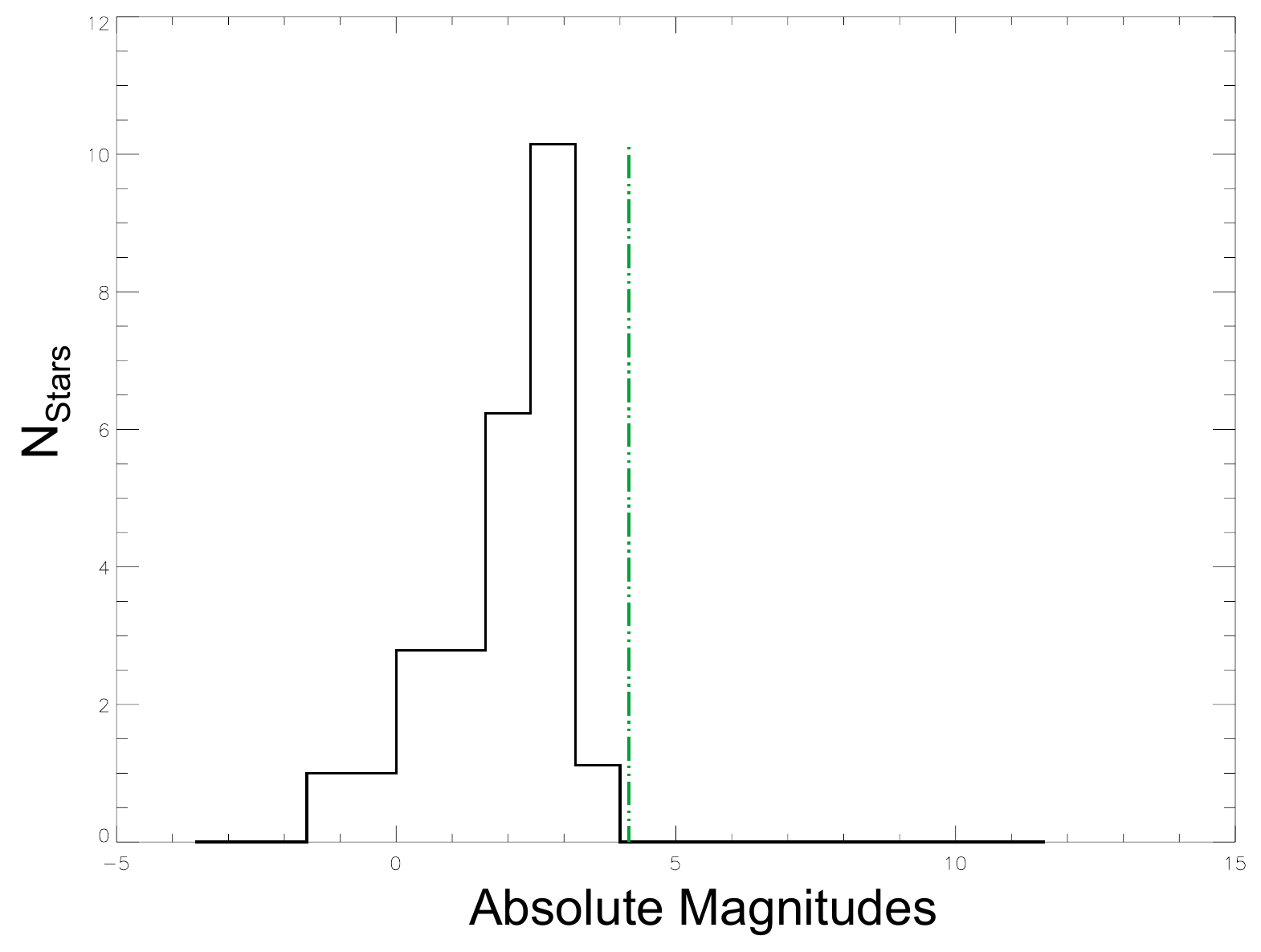} 
\includegraphics[angle=-0,origin=c,width=6cm]{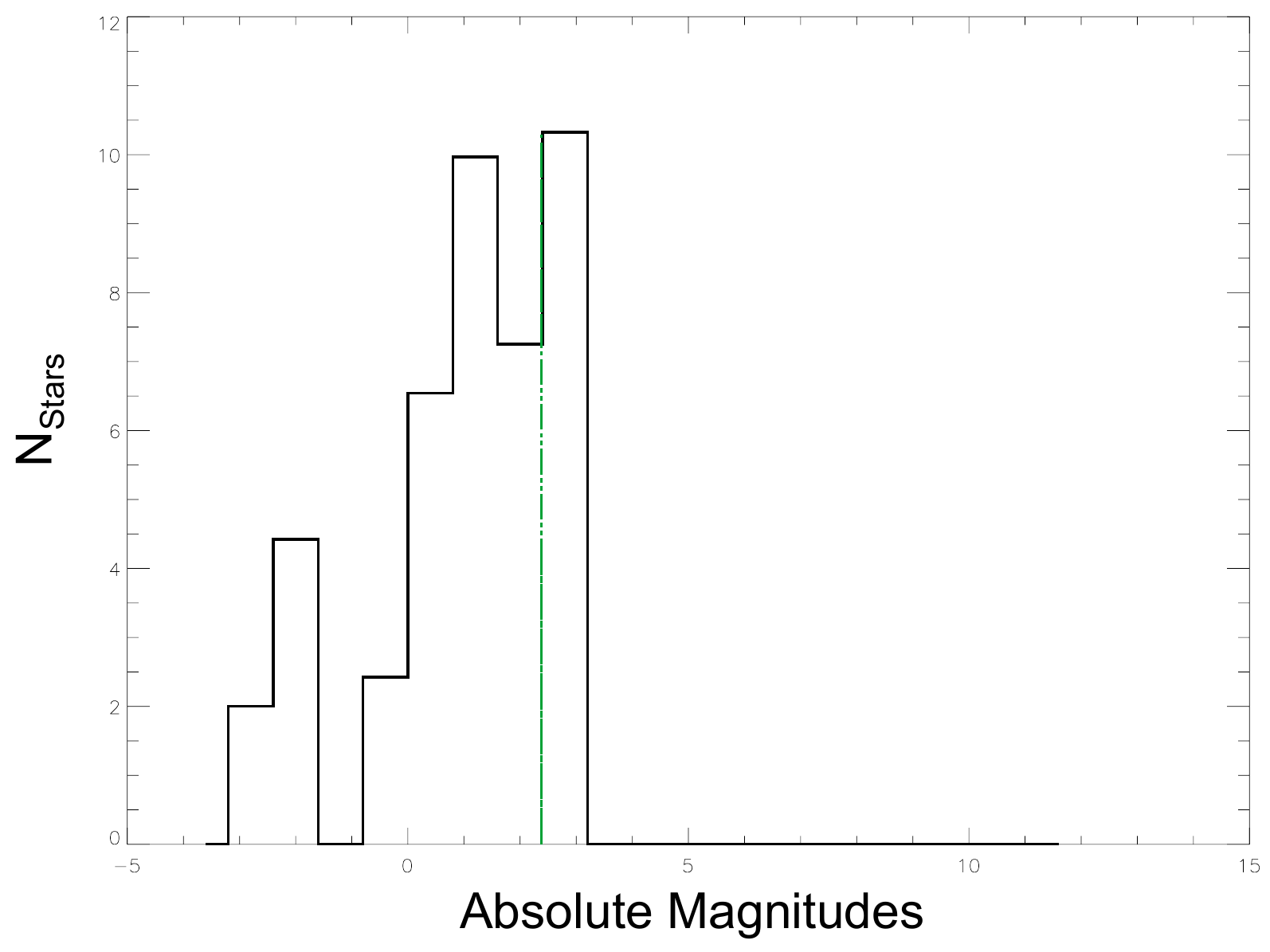}
}
\caption{Mol139 (left), Mol143 (center), Mol148 (right)}
\end{figure*}

\begin{figure*}[h]
\centering
\resizebox{\hsize}{!}
{
\includegraphics[angle=-0,origin=c,width=6cm]{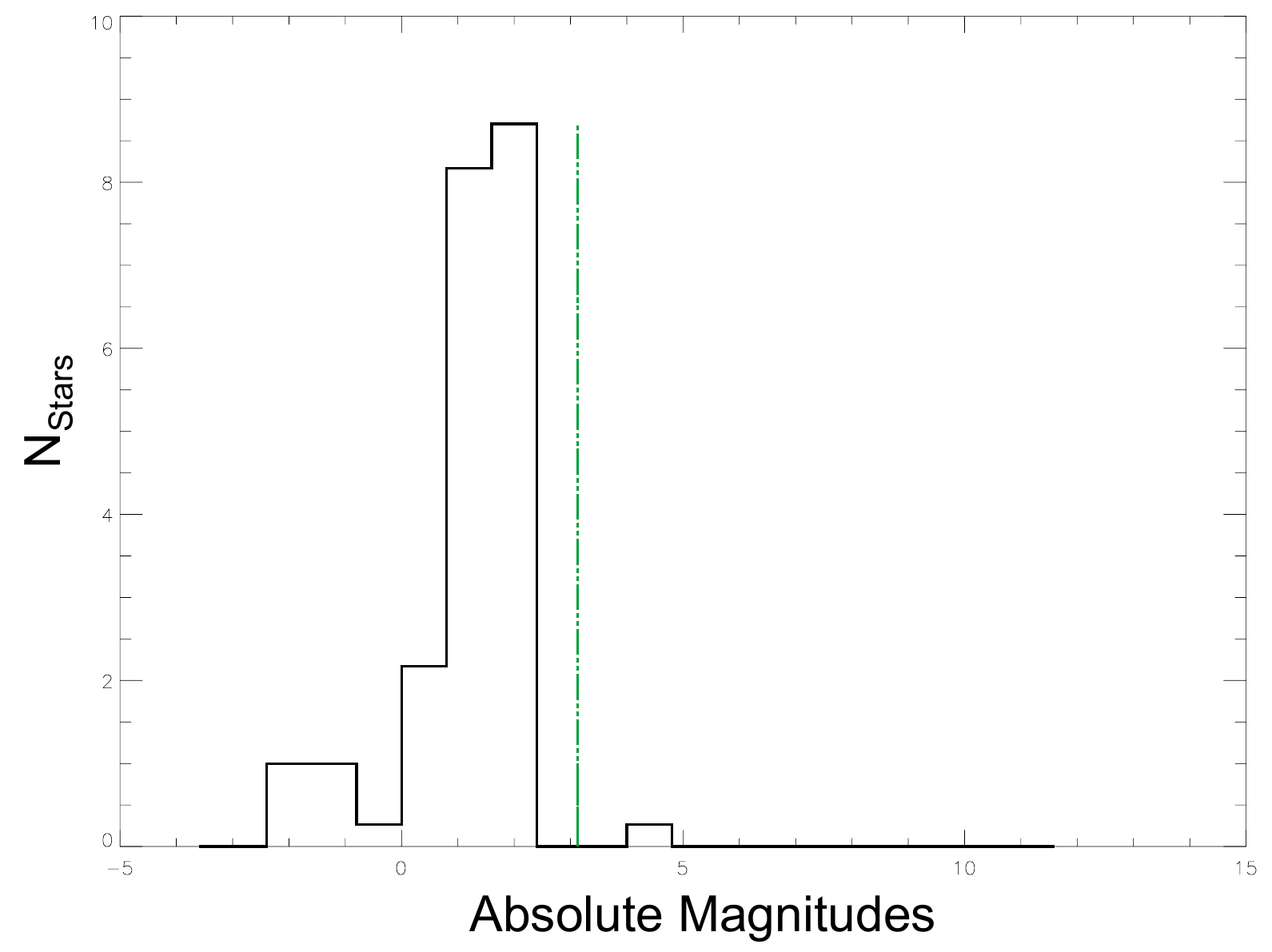} 
\includegraphics[angle=-0,origin=c,width=6cm]{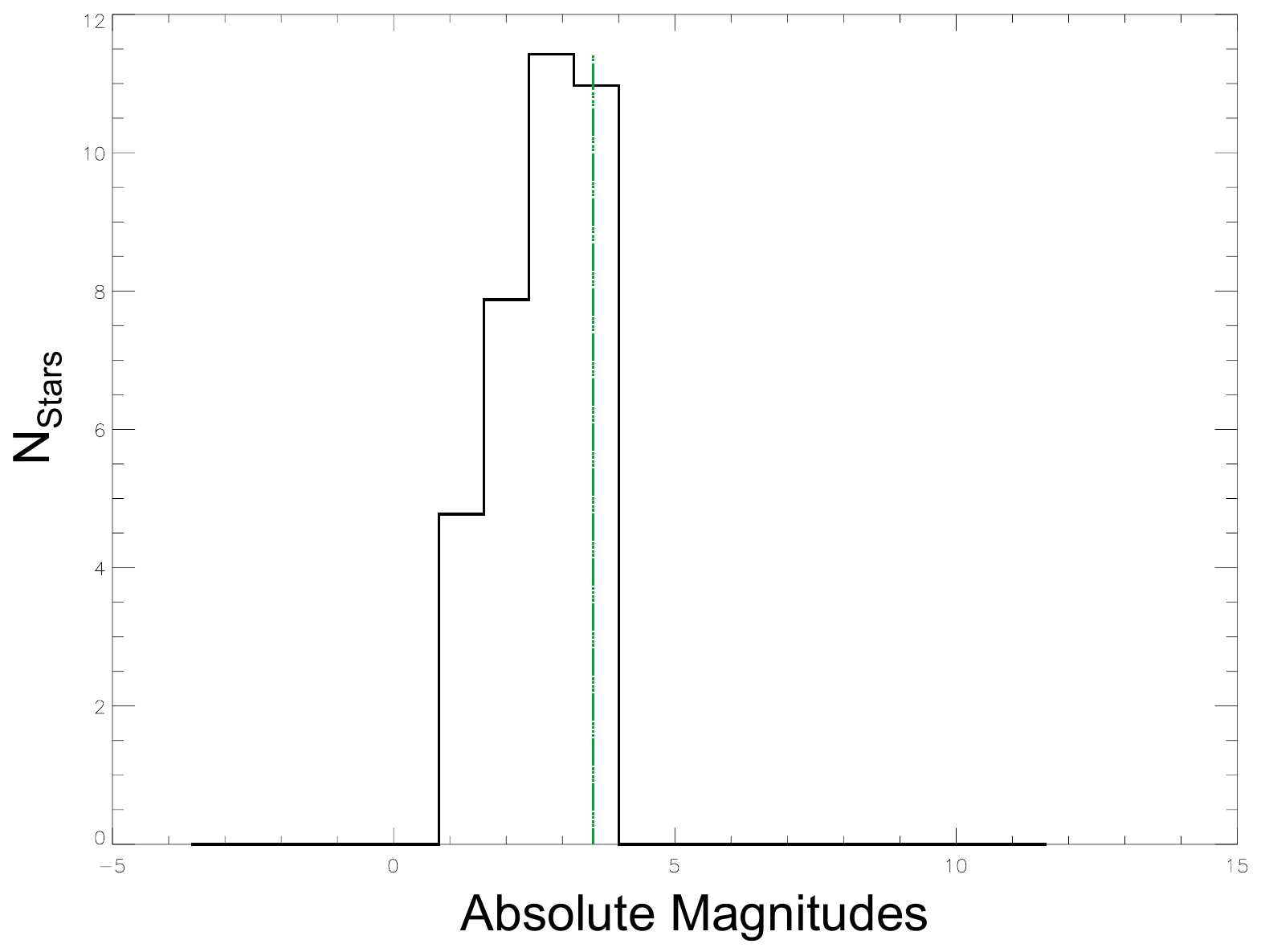}
}
\caption{Mol151 (left), Mol160 (right)}
\end{figure*}

\end{appendix}

\end{document}